%% file: invdec.tex
\documentclass[11pt,letter]{article}

\input{makros}

\begin{document}
\title{Structure Theorem and Isomorphism Test for Graphs with Excluded
  Topological Subgraphs\footnote{An extended abstract of the paper 
appeared in the proceedings of the 44th annual ACM symposium on Theory of computing (STOC 2012).}}
\author{Martin Grohe and Dániel Marx}
\maketitle

\begin{abstract}
  We generalize the structure theorem of Robertson and Seymour for
  graphs excluding a fixed graph $H$ as a minor to graphs excluding
  $H$ as a topological subgraph. We prove that for a fixed $H$, every
  graph excluding $H$ as a topological subgraph has a tree
  decomposition where each part is either ``almost embeddable'' to a
  fixed surface or has bounded degree with the exception of a bounded
  number of vertices. Furthermore, we prove that such a decomposition
  is computable by an algorithm that is fixed-parameter tractable with
  parameter $|H|$.

  We present two algorithmic applications of our structure theorem. To
  illustrate the mechanics of a ``typical'' application of the
  structure theorem, we show that on graphs excluding $H$ as a
  topological subgraph, \textsc{Partial Dominating Set} (find $k$
  vertices whose closed neighborhood has maximum size) can be solved
  in time $f(H,k)\cdot n^{O(1)}$ time. More significantly, we show
  that on graphs excluding $H$ as a topological subgraph,
  \textsc{Graph Isomorphism} can be solved in time $n^{f(H)}$.  This
  result unifies and generalizes two previously known important
  polynomial-time solvable cases of \textsc{Graph Isomorphism}:
  bounded-degree graphs \cite{luk82} and $H$-minor free graphs
  \cite{pon88}. The proof of this result needs a generalization of our
  structure theorem to the context of invariant treelike
  decomposition.
\end{abstract}

\input{intro}
\input{prelim}
\input{treedec}
\input{combined}
\input{tangles}
\input{local2}
\input{partial}
\input{treelike}
\input{canon}
\input{conclusions}


\end{document}

%% file: makros.tex
\usepackage[latin1]{inputenc}
\usepackage{amsmath}
\usepackage{amsthm}
\usepackage{amssymb}
\usepackage{MnSymbol}
\usepackage{graphicx}
\usepackage{ifthen}
\usepackage{calc}
\usepackage{cancel}
\usepackage{comment}
\usepackage{enumitem}
\usepackage{hyperref}
\usepackage{longtable}
\usepackage[margin=1in]{geometry}

\usepackage{tikz}
\usetikzlibrary{arrows}
\usetikzlibrary{shapes.geometric}
\usetikzlibrary{calc}

\usepackage{color}

\newcommand{\blue}{\color{blue}}

\definecolor{darkgreen}{rgb}{0,0.5,0.1}

\makeatletter
\renewcommand{\section}{\@startsection
  {section}%
  {1}%
  {0mm}%
  {-1\baselineskip}%
  {0.5\baselineskip}%
  {\normalfont\large\bfseries}%
}

\renewcommand{\subsubsection}{\@startsection
  {subsubsection}%
  {3}%
  {0mm}%
  {-1\baselineskip}%
  {0.3\baselineskip}%
  {\normalfont\itshape}%
}

\newsavebox{\tempbox}
\renewcommand{\@makecaption}[2]{
  \vspace{10pt}
  \sbox{\tempbox}{\textbf{#1.} #2}
  \ifthenelse{\lengthtest{\wd\tempbox > \linewidth}}{
    \textbf{#1.} #2\par
  }{
    \begin{center}
      \textbf{#1.} #2
    \end{center}
  }
}

\makeatother

\numberwithin{equation}{section}
\numberwithin{figure}{section}

\newtheoremstyle{mythm}
  {}
  {}
  {\itshape}
  {}
  {\bfseries}
  {.}
  {.5em}
  {\thmname{#1}~\thmnumber{#2}\ifthenelse{\equal{\thmnote{#3}}{}}{}{~(\thmnote{#3})}}

\newtheoremstyle{mydefn}
  {}
  {}
  {\upshape}
  {}
  {\bfseries}
  {.}
  {.5em}
  {\thmname{#1}~\thmnumber{#2}\ifthenelse{\equal{\thmnote{#3}}{}}{}{~(\thmnote{#3})}}

\newtheoremstyle{myremark}
  {}
  {}
  {\upshape}
  {}
  {\bfseries}
  {.}
  {.5em}
  {\thmname{#1}~\thmnumber{#2}\ifthenelse{\equal{\thmnote{#3}}{}}{}{~(\thmnote{#3})}}

\theoremstyle{mythm}
\newtheorem{theo}{Theorem}[section]
\newtheorem{lem}[theo]{Lemma}
\newtheorem{prop}[theo]{Proposition}
\newtheorem{cor}[theo]{Corollary}
\newtheorem{fact}[theo]{Fact}

\theoremstyle{mydefn}
\newtheorem{defn}[theo]{Definition}
\newtheorem{exa}[theo]{Example}

\theoremstyle{myremark}
\newtheorem{rem}[theo]{Remark}
\theoremstyle{mythm}

\newcounter{claimcounter}
\newenvironment{claim}[1][]{
  \renewcommand{\proof}{\smallskip\par\noindent\textit{Proof. }}
  \medskip\par\noindent%
  \ifthenelse{\equal{#1}{}}{%
    \setcounter{claimcounter}{0}\refstepcounter{claimcounter}\textit{Claim~\arabic{claimcounter}.}
  }{%
    \ifthenelse{\equal{#1}{resume}}{%
      \refstepcounter{claimcounter}\textit{Claim~\arabic{claimcounter}.}
    }{%
      \textit{Claim~#1.}
    }
  }
}{
  \par\medskip
}

\newcommand{\case}[1]{\par\medskip\noindent\textit{Case #1: }}

\newenvironment{cs}{
  \begin{description}
    \renewcommand{\case}[1]{\item[\itshape\mdseries Case ##1:]}
  }{
  \end{description}
}

\newlist{caselist}{description}{10}
\setlist[caselist]{font=\itshape\mdseries}

\newcommand{\uend}{\hfill$\lrcorner$}

\setenumerate[1]{label=(\arabic*)}
\newlist{eroman}{enumerate}{2}
\setlist[eroman,1]{label=(\roman*)}
\setlist[eroman,2]{label=(\alph*)}
\newlist{ealph}{enumerate}{1}
\setlist[ealph]{label=(\Alph*)}

\newcounter{nlistcounter}
\newenvironment{nlist}[1]{
  \renewcommand{\thenlistcounter}{\upshape(#1.\arabic{nlistcounter})}
  \begin{list}{\bfseries\thenlistcounter}{%
      \usecounter{nlistcounter}
      \setlength{\labelwidth}{1.5em}%
      \setlength{\leftmargin}{\labelwidth}%
      \addtolength{\leftmargin}{\labelsep}%
      \setlength{\listparindent}{0em}%
      \setlength{\topsep}{5pt}%
      \setlength{\itemsep}{5pt}%
      \setlength{\parsep}{0pt}%
    }
  }{
  \end{list}
}

\renewcommand{\fboxsep}{2mm}

\newsavebox{\fminibox}
\newlength{\fminilength}
\newenvironment{fminipage}[1][\linewidth]{
  \setlength{\fminilength}{#1-2\fboxsep-2\fboxrule}%
  \begin{lrbox}{\fminibox}\begin{minipage}{\fminilength}}{
    \end{minipage}\end{lrbox}\noindent\fbox{\usebox{\fminibox}}
}

\newlength{\pwidth}

\newenvironment{problem}[3]{
  \begin{center}
    \ifthenelse{\equal{#1}{}}%
    {\setlength{\pwidth}{\textwidth-2\parindent}}%
    {\setlength{\pwidth}{#1cm}}

    \begin{fminipage}[\pwidth]\upshape
      \ifthenelse{\equal{#3}{}}{}{{\scshape #3}}
      \begin{list}{}{
          \ifthenelse{\equal{#2}{}}%
          {\settowidth{\labelwidth}{\textit{Parameter:}}}%
          {\settowidth{\labelwidth}{\textit{#2:}}}
          
          \setlength{\leftmargin}{\labelwidth+\labelsep}
          \setlength{\itemsep}{0ex}
          \setlength{\parsep}{0ex}
          \setlength{\topsep}{0.2ex}
        }
      }{
      \end{list}
    \end{fminipage}
  \end{center}
}

\renewcommand{\max}{\operatorname{max}}
\renewcommand{\min}{\operatorname{min}}

\renewcommand{\phi}{\varphi}

\newcommand{\Bigmid}{\;\Big|\;}

\newcounter{rbcounter}
\newcommand{\torso}{\tau}
\newcommand{\mytorso}{\operatorname{torso}}

\newcommand{\ZZ}{\mathbb Z}
\newcommand{\NN}{\mathbb N}
\newcommand{\PN}{\mathbb N\setminus\{0\}}

\newcommand{\lex}{\le_{\textup{lex}}}
\newcommand{\slex}{<_{\textup{lex}}}

\renewcommand{\int}{\operatorname{int}}

\newcommand{\cone}{\gamma}
\newcommand{\comp}{\alpha}
\newcommand{\bag}{\beta}
\newcommand{\sep}{\sigma}

\newcommand{\CA}{\mathcal A}

\newcommand{\CC}{\mathcal C}
\newcommand{\CD}{\mathcal D}

\newcommand{\CI}{\mathcal I}

\newcommand{\CM}{\mathcal M}
\newcommand{\CN}{\mathcal N}

\newcommand{\CP}{\mathcal P}

\newcommand{\CW}{\mathcal W}

\newcommand{\CY}{\mathcal Y}

\newcommand{\FD}{{\mathbf D}}

\newcommand{\FS}{{\mathbf S}}

\newcommand{\hier}[1][]{\par\bigskip\hrule\mbox{}\\{HIER WEITER%
\ifthenelse{\equal{#1}{}}{}{\par\noindent#1}}\\\hrule\mbox{}\par\bigskip}

\newcommand{\defiff}{\;:\Leftrightarrow\;}

\newcommand{\Ka}{\mathfrak a}

\newcommand{\Kc}{\mathfrak c}

\newcommand{\KT}{\mathfrak T}

\newcommand{\todo}[1]{\par\noindent{\blue\itshape\textbf{To Do: }#1}\par}
\newcommand{\toproof}[1]{\todo{Proof}}

\newcommand{\separg}{S_G}
\newcommand{\boundary}{\partial}

\newcommand{\executeiffilenewer}[3]{%
\ifnum\pdfstrcmp{\pdffilemoddate{#1}}%
{\pdffilemoddate{#2}}>0%
{\immediate\write18{#3}}\fi%
}

\DeclareMathAlphabet{\mathcal}{OMS}{cmsy}{m}{n}

\renewcommand{\tilde}{\widetilde}
\renewcommand{\vec}{\bar}


%% file: intro.tex
\section{Introduction}
We say that a graph $H$ is a {\em minor} of $G$ if $H$ can be obtained
from $G$ by deleting vertices, deleting edges, and contracting
edges. A graph $G$ is {\em $H$-minor free} if $H$ is not a minor of
$G$.  Robertson and Seymour \cite{gm16} proved a structure theorem
for the class of $H$-minor-free graphs: roughly speaking, every $H$-minor
free graph can be decomposed in a way such that each part is ``almost
embeddable'' into a fixed surface. This structure theorem has
important algorithmic consequences: many natural computational
problems become easier when restricted to $H$-minor free graphs
\cite{DBLP:conf/stoc/DemaineHK11,DBLP:conf/soda/KawarabayashiDH09,DBLP:conf/soda/DornFT08,DBLP:journals/gc/KawarabayashiM07,DBLP:conf/stoc/KawarabayashiM06,DBLP:conf/focs/DemaineHK05,DBLP:journals/combinatorica/Grohe03}.
These algorithmic results can be thought of as far-reaching
generalizations of algorithms on planar graphs and bounded-genus
surfaces.

A more general way of defining restricted classes of graphs is to
exclude topological subgraphs instead of minors.  A graph $H$ is a
{\em topological subgraph} (or {\em topological minor}) of graph $G$
if a subdivision of $H$ is a subgraph of $G$. It is easy to see that
if $H$ is a topological subgraph of $G$, then $H$ is also a minor of
$G$. Thus the class of graphs excluding $H$ as a topological subgraph
is a more general class than $H$-minor free graphs.

One can ask if graphs excluding $H$ as a topological subgraph admit a
similar structure theorem as $H$-minor free graphs. However, graphs
excluding a topological subgraph can be much more general. For
example, no 3-regular graph can contain a subdivision of $K_5$ (as
$K_5$ is 4-regular). Therefore, the class of graphs excluding $K_5$ as
a topological subgraph includes in particular every 3-regular
graph. This suggests that it is unlikely that this class can be also
characterized by (almost) embeddability into surfaces. It is also
worth mentioning that graph classes that are closed under taking
minors can be characterised by finitely many excluded minors, or
equivalently, the minor-relation is a well quasi order; this is
Robertson and Seymour's famous Graph Minor Theorem~\cite{gm20}. It is
easy to show that the analogous result for classes closed under taking
topological subgraphs fails (see, for example, \cite{robsey85}). Thus
the topological-subgraph relation and the minor relation differ significantly.

Nevertheless, our first result is a structure theorem for graphs
excluding a graph $H$ as a topological subgraph. We prove that, in
some sense, only the bounded-degree graphs make this class more
general than $H$-minor free graphs. More precisely, we prove a
structure theorem that decomposes graphs excluding $H$ as a
topological subgraph into almost bounded-degree parts and into
$H'$-minor free parts (for some other graph $H'$). The
 $H'$-minor free parts can be further refined into almost-embeddable parts using the structure
theorem of Robertson and Seymour \cite{gm16}, to
obtain our main structural result (see
Corollary~\ref{cor:structure-emb} for the precise statement):
\begin{theo}[informal]\label{th:introstructure}
For every fixed graph $H$, every graph excluding $H$ as a topological subgraph has a tree decomposition where every torso
\begin{eroman}
\item either has bounded degree with the exception of a bounded number
  of vertices, or
\item almost embeddable into a surface of bounded genus.
\end{eroman}
Furthermore, such a decomposition can be computed in time $f(H)\cdot |V(G)|^{O(1)}$ for some computable function $f$.
\end{theo}

Our structure theorem allows us to lift problems that are tractable on
both bounded-degree graphs and on $H$-minor free graphs to the class
of graphs excluding $H$ as a topological subgraph. We demonstrate this
principle on the \textsc{Partial Dominating Set} problem (find $k$
vertices whose closed neighborhood is maximum). Following a bottom-up
dynamic programming approach, we solve the problem in each bag of the
tree decomposition (using the fact that the problem can be solved in
linear-time on both bounded-degree and on almost-embeddable graphs).
\begin{theo}\label{th:partialdom}
\textsc{Partial Dominating Set}
 can be solved in time $f(k,H)\cdot
n^{O(1)}$ when restricted to graphs excluding $H$ as a topological
subgraph.
\end{theo}
One could prove similar results for other basic problems such as
\textsc{Independent Set} or \textsc{Dominating Set}.  However, a
result of Dvorak et al.~\cite{DBLP:conf/focs/DvorakKT10} shows that
problems expressible in first-order logic can be solved in linear time
on classes of graphs having bounded expansion, and therefore on graphs
excluding $H$ as a topological subgraph. The problems
\textsc{Independent Set} and \textsc{Dominating Set} (for a fixed $k$)
can be expressed in first-order logic, thus the analogs of
Theorem~\ref{th:partialdom} for these problems follow from
\cite{DBLP:conf/focs/DvorakKT10}. On the other hand, \textsc{Partial
  Dominating Set} is not expressible in first-order logic, hence the
techniques of Dvorak et al.~\cite{DBLP:conf/focs/DvorakKT10} do not
apply to this problem.

The main algorithmic result of the paper concerns the \textsc{Graph
  Isomorphism} problem (given graphs $G_1$ and $G_2$, decide if they
are isomorphic). \textsc{Graph Isomorphism} is known to be
polynomial-time solvable for bounded-degree graphs
\cite{luk82,babluk83} and for $H$-minor free graphs
\cite{pon88,gro10+a}. In fact, for these classes of graphs, even the
more general canonization problem can be solved in polynomial time:
there is an algorithm labeling the vertices of the graph with positive
integers such that isomorphic graphs get isomorphic labelings.  It is
tempting to expect that our structure theorem together with a
bottom-up strategy give a canonization algorithm for graphs excluding
$H$ as a topological subgraph: in each bag, we use the canonization
algorithm either for bounded-degree graphs or $H$-minor free graphs
(after encoding somehow the canonized versions of the child bags,
which seems to be a technical problem only). However, this approach is
inherently doomed to failure: there is no guarantee that our
decomposition algorithm produces isomorphic decompositions for
isomorphic graphs. Therefore, even if two graphs are isomorphic, the
bottom-up canonization algorithm could be working on two completely
different decompositions and therefore could obtain different results
on the two graphs.

We overcome this difficulty by generalizing our structure theorem to
the context of treelike decompositions introduced by the first author in
\cite{gro08,gro10+a}. A treelike decomposition is similar to a tree
decomposition, but it is defined over a directed acyclic graph instead
of a rooted tree, and therefore it contains several tree
decompositions. The Invariant Decomposition Theorem
(Section~\ref{sec:treel-decomp}) generalizes the structure theorem by
giving an algorithm that computes a treelike decomposition in a way
that the decompositions obtained for isomorphic graphs are
isomorphic. Then the Lifting Lemma (Section~\ref{sec:canonization})
formalizes the bottom-up strategy informally described in the previous
paragraph: if we can compute treelike decompositions for a class of
graphs in an invariant way and we have a canonization algorithm for
the bags, then we have a canonization algorithm for this class of
graphs. Although the idea is simple, in order to encode the child
bags, we have to state this algorithmic result in a more general
form: instead of graphs, we have to work with weighted relational
structures. This makes the statement and proof of the Lifting Lemma
more technical.  Putting together these results, we obtain:
\begin{theo}\label{th:introiso}
  For every fixed graph $H$, \textsc{Graph Isomorphism} can be solved
  in polynomial-time restricted to graphs excluding $H$ as a
  topological subgraph.
\end{theo}

Actually, we not only obtain a polynomial time isomorphism test, but
also a polynomial time canonisation algorithm.  Our theorem
generalizes and unifies the results of Babai and
Luks~\cite{luk82,babluk83} on bounded-degree graphs and of
Ponomarenko~\cite{pon88} on $H$-minor free graphs. Let us remark that
Ponomarenko's result implies that there is a polynomial time
isomorphism test for all classes of graphs of bounded genus, which has
been proved earlier by Filotti and Mayer~\cite{filmay80} and
Miller~\cite{mil80}, and for all classes of graphs of bounded tree
width, which was also proved later (independently) by
Bodlaender~\cite{bod90}. Miller~\cite{mil83} gave a common
generalization of the bounded degree and bounded genus classes to
classes that he called $k$-contractible. These classes do not seem to
have a simple graph-theoretic characterization; they are defined in
terms of properties of the automorphism groups needed for the
algorithm. Excluding topological subgraphs, on the other hand, is a
natural graph theoretic restriction that generalizes both bounding the
degree and excluding minors and hence bounding the genus.


For the convenience of the reader, let us summarize how the different
results in the present paper depend on previous results in the
literature:
\begin{itemize}
\item The proof of the existence of the decomposition into $H$-minor
  free and almost bounded-degree parts is self-contained. The
  algorithm computing such a decomposition needs the minor testing
  algorithm of \cite{gm13} or \cite{DBLP:conf/stoc/KawarabayashiW10}.
\item The proof of the existence of the more refined decomposition
  into almost-embeddable and almost bounded-degree parts needs the
  graph structure theorem of Robertson and Seymour \cite{gm16}. The
  algorithm computing such a decomposition needs the algorithmic
  version of the structure theorem \cite{DBLP:conf/focs/DemaineHK05};
  to achieve $f(H)\cdot n^{O(1)}$ running time, a more recent stronger
  algorithmic result is needed \cite{kawwol11,grokawree13}.
\item The algorithm for \textsc{Partial Dominating Set} needs the more
  refined decomposition, hence it relies on
  \cite{gm13,kawwol11}. Additionally, it needs the fact proved in
  \cite{DBLP:journals/combinatorica/Grohe03} that almost-embeddable
  graphs have bounded local treewidth.
\item The result on \textsc{Graph Isomorphism} needs the minor testing
  algorithm of \cite{gm13} or \cite{DBLP:conf/stoc/KawarabayashiW10}
  to compute the treelike decomposition. Additionally, the
  canonization algorithms for bounded-degree graphs \cite{babluk83}
  and for $H$-minor free graphs (\cite{pon88} or \cite{gro10+a}) are
  needed.
\end{itemize}
Note that none of the results rely on the topological subgraph testing
algorithm of \cite{grohe-stoc2011-topminor} or need any substantial
result from the monograph \cite{gro10+a}.

The paper is organized as
follows. Sections~\ref{sec:preliminaries}--\ref{sec:tree-decompositions}
introduce the notation used in the
paper. Section~\ref{sec:local-glob-struct} states the structure
theorem and shows how it can be proved by appropriate local
decomposition lemmas. Section~\ref{sec:tangles} introduces the notion
of tangles, which is an important tool in the proofs of the local
decomposition lemmas in
Section~\ref{sec:local}. Section~\ref{sec:part-domin-set} uses the
structure theorem in an algorithm for \textsc{Partial Dominating
  Set}. Section~\ref{sec:treel-decomp} introduces treelike
decomposition and proves the Invariant Decomposition
Theorem. Section~\ref{sec:canonization} proves the Lifting Lemma for
canonizations, completing the proof of Theorem~\ref{th:introiso}.


%% file: prelim.tex
\section{Preliminaries}\label{sec:preliminaries}

$\ZZ$ and $\NN$ denote the sets of integers and nonnegative integers,
respectively.  For $m,n\in\ZZ$, we let $[m,n]:=\{\ell\in\ZZ\mid m\le\ell\le
n\}$ and $[n]:=[1,n]$. The power set of a set $S$ is denoted by $2^S$, and the
set of all $k$-element subsets of $S$ by $\binom{S}{k}$. For a mapping $f$
defined on $S$, we let $f(S):=\{f(s)\mid s\in S\}$. The cardinality of a set
$S$ is denoted by $|S|$.


Let $G$ be a graph. The \emph{order} of a graph $G$ is
$|G|:=|V(G)|$.  The set of all neighbors of a vertex $v\in V(G)$, called
the \emph{open neighborhood of $v$}, is denoted by $N^G(v)$. The \emph{closed
  neighborhood} of $v$ is the set $N^G[v]:=\{v\}\cup N^G(v)$. The
\emph{closed} and \emph{open neighborhood} of a subset $W\subseteq V(G)$ are
the sets $N^G[W]:= \bigcup_{w\in W}N^G[w]$ and $N^G(W):=N^G[W]\setminus W$,
respectively, and the \emph{closed} and \emph{open neighborhood} of a
subgraph $H\subseteq G$ are the sets $N^G[H]:=N^G[V(H)]$ and
$N^G(H):=N^G(V(H))$, respectively.  We omit the index ${}^G$ if $G$ is clear
from the context, and we do the same for similar notations introduced
later.
We let $\boundary^G(W)=|N^G(W)|$.


For every set $V$, we let $K[V]$ be
the complete graph with vertex set $V$, and for every $n\in\NN$, we let
$K_n:=K\big[[n]\big]$.

Let $G$ be a graph.  A graph $H$ is a {\em minor} of $G$ (denoted by
$H\preceq G$) if $H$ can be obtained from $G$ by deleting vertices,
deleting edges, and contracting edges.  Equivalently, we can define
$H\preceq G$ the following way. Two sets $S,T\subseteq V(G)$
\emph{touch} if either $S\cap T\neq\emptyset$ or there is an edge $vw\in
V(G)$ such that $v\in S$ and $w\in T$. It can be shown that $H\preceq
G$ if and only if there is a family $(I_w)_{w\in V(H)}$ of pairwise
disjoint connected subsets of $V(G)$ such that for every $u,v\in V(H)$
that are adjacent in $H$, the sets $I_u$ and $I_v$ touch in $G$.  We
call this family $I$ an {\em image} of $H$ in $G$ and the sets $I_w$
are the {\em branch sets} of the image. 

\begin{theo}[\cite{gm13,DBLP:conf/stoc/KawarabayashiW10}]\label{th:minortest}
 There is an $f(H)\cdot |V(G)|^3$ time algorithm (for some computable
  $f$) that finds an $H$-minor image in $G$, if it exists.
\end{theo}

A {\em subdivision} $H'$ of a graph $H$ is obtained by replacing each
edge of $H$ by a path of length at least 1. We say that $H$ is a {\em
  topological subgraph} (or {\em topological minor}) of $G$ and denote
it by $H\preceq_T G$ if a subdivision of $H$ is a subgraph of
$G$. Equivalently, $H$ is a topological subgraph of $G$ if $H$ can be
obtained from $G$ by deleting edges, deleting vertices, and dissolving
degree 2 vertices (which means deleting the vertex and making its two
neighbors adjacent).  For fixed $H$, it can be decided in cubic time
whether a graph $G$ contains a subdivision of $H$ (although we do not
need this result in the current paper):

\begin{theo}[\cite{grohe-stoc2011-topminor}]\label{th:subdivtest}
  There is an $f(H)\cdot |V(G)|^3$ time algorithm (for some computable
  $f$) that finds a subdivision of $H$ in $G$, if it exists.
\end{theo}

Let $D$ be a digraph. For every $t\in V(D)$, we let $N_+^D(t):=\{ u\in
V(D)\mid tu\in E(D)\}$. We call vertices of in-degree $0$ \emph{roots}
and vertices of out-degree $0$ \emph{leaves} of $D$.
The \emph{height} of an acyclic digraph $D$ is the length of the
longest path in $D$.

It will be convenient for us to view trees as being directed, unless
we explicitly call them undirected. Hence for us, a \emph{tree} is an
acyclic digraph $T$ that has a unique node $r(T)$ (the \emph{root}) such that
for every node $t$ there is a exactly one path from $r(T)$ to
$t$.\footnote{What we call ``directed tree'' here is somtimes called
  ``out branching''. Moreover, there is an obvious one-to-one
  correspondence between directed trees and rooted trees.}

For two graphs $A$ and $B$, the graph $A\cup B$ is defined by $V(A\cup
B)=V(A)\cup V(B)$ and $E(A\cup B)=E(A)\cup E(B)$.  Let $G$ be a graph.
A \emph{separation} of $G$ is a pair $(A,B)$ of subgraphs of $G$ such
that $A\cup B=G$ and $E(A\cap B)=\emptyset$. The \emph{order} of a
separation $(A,B)$ is $|V(A)\cap V(B)|$.


%% file: treedec.tex
\section{Tree Decompositions}\label{sec:tree-decompositions}
A \emph{tree decomposition} of a graph $G$ is a pair $(T,\bag)$, where $T$ is a
tree and $\bag:V(T)\to 2^{V(G)}$, such that for all nodes $v\in V(G)$ the
set $\{t\in V(T)\mid v\in\bag(t)\}$ is nonempty and connected in the
undirected tree underlying $T$, and for all edges $e\in E(G)$ there is a $t\in
V(T)$ such that $e\subseteq\bag(t)$. It will be convenient for us to
view the tree in a tree decomposition as being directed. Most readers will be familiar with this
definition, but it will be convenient for us to view tree decompositions
from a different perspective here. 

If $(T,\bag)$ is a tree decomposition of a graph $G$, then we define mappings
$\sep,\cone,\comp:V(T)\to2^{V(G)}$ by letting for all $t\in V(T)$
\begin{align}
  \label{eq:bagsep}
  \sep(t)&:=
  \begin{cases}
    \emptyset&\text{if $t$ is the root of $T$},\\
    \bag(t)\cap\bag(s)&\text{if $s$ is the parent of $t$ in $T$},
  \end{cases}\\
  \label{eq:bagcone}
  \cone(t)&:=\bigcup_{\text{$u$ is a descendant of $t$}}\bag(u),\\
  \label{eq:bagcomp}
  \comp(t)&:=\cone(t)\setminus\sep(t).
\end{align}
We call 
$\bag(t),\sep(t),\cone(t),\comp(t)$ the \emph{bag} at $t$,
\emph{separator} at $t$,
\emph{cone} at $t$, \emph{component} at $t$, respectively.
It is easy to verify that the following conditions hold:
\begin{nlist}{TD}
  \item\label{li:t1}
    $T$ is a tree.
  \item\label{li:t2} For all $t\in V(T)$ it holds that
    $\comp(t)\cap\sep(t)=\emptyset$ and
    $N^G(\comp(t))\subseteq\sep(t)$.

  \item\label{li:t3} For all $t\in V(T)$ and $u\in N_+^T(t)$ it holds that
    $\comp(u)\subseteq\comp(t)$ and $\cone(u)\subseteq\cone(t)$.
  \item\label{li:t4} For all $t\in V(T)$ and all distinct $u_1,u_2\in N_+^T(t)$ it holds that
  $\cone(u_1)\cap\cone(u_2)=\sep(u_1)\cap\sep(u_2)$.
  \item\label{li:t5} For the root $r$ of $T$ it holds that
    $\sep(r)=\emptyset$ and $\comp(r)=V(G)$.
\end{nlist}
Conversely, consider a triple $(T,\sep,\comp)$, where $T$ is a digraph and
$\sep,\comp:V(T)\to 2^{V(G)}$. We define $\cone,\bag:V(T)\to2^{V(G)}$ by
\begin{align}
  \label{eq:cone}
  \cone(t)&:=\sep(t)\cup\comp(t),\\
  \label{eq:bag}
  \bag(t)&:=\cone(t)\setminus\bigcup_{u\in N_+^T(t)}\comp(u)  
\end{align}
for all $t\in V(T)$. Then it is easy to prove that if
\ref{li:t1}--\ref{li:t5} are satisfied, then $(T,\bag)$ is a tree
decomposition (see \cite{gro10+a} for a proof). Thus we may also view
triples $(T,\sep,\comp)$ satisfying \ref{li:t1}--\ref{li:t5} as tree
decompositions. We jump back and forth between both versions of tree
decompositions, whichever is more convenient. The treelike
decompositions introduced in Section~\ref{sec:treel-decomp} need to be
defined as triples $(T,\sep,\comp)$, thus looking at tree
decompositions also this way in the first part of the paper makes the
transition between the two concepts smoother.

Let $(T,\bag)$ be a tree decomposition of a graph $G$. The
\emph{width} of $(T,\bag)$ is $\max\{|\bag(t)|-1\mid t\in
V(T)\}$, and the \emph{adhesion} of $(T,\bag)$ is
$\max\{|\sep(t)|\mid t\in V(T)\}$. The {\em tree width} of a graph $G$
is the minimum possible width of a tree decomposition of $G$. 
However, in the current paper, rather than minimizing tree width (i.e., minimizing the size of the bags), we 
are mostly interested in decompositions where the graph induced by each bag (plus some additional edges) is ``nice'' in a certain sense.
For every node $t\in V(T)$,
the \emph{torso} at $t$ is the graph
\begin{equation}
\label{eq:torso}
\tau(t):=G[\bag(t)]\cup K[\sep(t)]\cup\bigcup_{u\in N_+^T(t)}K[\sep(u)].
\end{equation}
That is, we take the graph induced by bag $\bag(t)$, turn $\sep(t)$
into a clique, and make vertices $x,y$ adjacent if they appear
together in the separator (or equivalently, the cone) of some child
$u$ of $t$.  For a class $\CA$ of graphs, $(T,\bag)$ is a tree
decomposition \emph{over} $\CA$ if all its torsos are in $\CA$.

A related notion is the {\em torso} of $G$ with respect to a set
$C\subseteq V(G)$, denoted by $\mytorso(G,C)$, which is defined as
graph on $C$ where $u,v\in V(G)$ are adjacent if there is a path $P$
in $G$ with endpoints $u$ and $v$ such that the internal vertices of
$P$ are disjoint from $C$. In other words,
\[
\mytorso(G,C):=G[C]\cup \bigcup_{\text{$X$ is a component of $G\setminus C$}}K[N^G(X)].
\]
It is easy to see that $\mytorso(G,\bag(t))\subseteq \tau(t)$. Equality
is not true in general: $G[\comp(u)]$ for some $u\in N_+^T(t)$ is not
necessarily connected, thus it is not necessarily true that $\sep(u)$
is $N^G(X)$ for some component $X$ of $G\setminus \bag(t)$.

%% file: combined.tex
\section{Local and Global Structure Theorems}
\label{sec:local-glob-struct}

The main structural result of the paper is a decomposition theorem for
graphs excluding a topological subgraph:
\begin{theo}[Global Structure Theorem]\label{theo:structure}
  For every $k\in\NN$, there exists constants $a(k)$, $b(k)$, $c(k)$,
  $d(k)$, $e(k)$, such that the following holds. Let $H$ be a graph on
  $k$ vertices. Then for every graph $G$ with $H\not\preceq_T G$ there is a tree
  decomposition $(T,\bag)$ of adhesion at most $a(k)$ such that for
  all $t\in V(T)$ one of the following three conditions is satisfied:
  \begin{eroman}
  \item\label{item:str-1} $|\bag(t)|\le b(k)$.
  \item\label{item:str-2} $\tau(t)$ has at most $c(k)$ vertices of degree larger than $d(k)$.
  \item\label{item:str-3} $K_{e(k)}\not\preceq\tau(t)$.
  \end{eroman}
  Furthermore, there is an algorithm that, given graphs $G,H$ of sizes
  $n,k$, respectively, in time $f(k)\cdot n^{O(1)}$ for some
  computable function $f$, computes either such a decomposition
  $(T,\beta)$ or a subdivision of $H$ in $G$.
\end{theo}
The reader could find it convenient to refer to the constants
$a,b,c,d,e$ as the bounds on the \underline{a}dhesion, \underline{b}ag
size, number of api\underline{c}es, maximum \underline{d}egree, and
\underline{e}xcluded clique.  We remark that all the constants are
polynomially large. Note that \ref{item:str-1} is redundant: by
choosing $d(k)$ or $e(k)$ sufficiently large, a bag satisfying
\ref{item:str-1} trivially satisfies \ref{item:str-2} and
\ref{item:str-3}. We state the result this way, because it shows the
high-level structure of the proof, which involves three decomposition
results corresponding to the three cases.

The proof of the Global Structure Theorem~\ref{theo:structure} builds
a tree decomposition step by step, iteratively decomposing the graph
locally in each step. The Local Structure Theorem describes the
``local'' structure of a graph, as seen from a single node of a tree
decomposition. We describe this local structure in terms of star
decompositions, to be defined next.  A \emph{star} is a tree of height
$1$. We usually call the root of a star its \emph{center} and the
leaves of a star its \emph{tips}.  A \emph{star decomposition} of a
graph $G$ is a tree decomposition $(T,\bag)$ where $T$ is a star. Note
that if $(T,\bag)$ is a star decomposition, then for every tip $t$ of
the star $T$ it holds that $\bag(t)=\cone(t)$.



\begin{theo}[Local Structure Theorem]\label{th:mainstar}
For every $k\in \NN$, there exists constants $a(k)$, $b(k)$, $c(k)$, $d(k)$, $e(k)$ such that the following
  holds.  There is an $f(k)\cdot |V(G)|^{O(1)}$ time algorithm
  that, given a graph $G$, a set $S\subseteq V(G)$ of size $\le a(k)$, and an integer $k$, 
\begin{enumerate}
\item \label{item:mainstar-1} either returns a subdivision of $K_k$ in $G$,
\item \label{item:mainstar-2} or computes a star decomposition $\Sigma_S=(T_S,\sep_S,\comp_S)$ of $G\cup K[S]$
  of adhesion $\le a(k)$ such that $S\subseteq \bag_S(s)$ for the
  center $s$, $\comp_S(t)\subset \comp_S(s)$ for every tip $t$, and one
  of the following three conditions is satisfied:
\begin{enumerate}
\item $|\bag_S(s)|\le b(k)$.
\item $\tau_S(s)$ does not contain a $K_{e(k)}$-minor.
\item At most $c(k)$ vertices of $\tau_S(s)$ have degree more than
  $d(k)$ in $\tau_S(s)$.
\end{enumerate}
\end{enumerate}

\end{theo}
The condition that $\comp_S(t)$ is a proper subset of $\comp_S(s)$ makes
sure that we make progress and compute a tree decomposition after a
finite number of applications of Theorem~\ref{th:mainstar}.  Note the
technical detail that $\Sigma_S$ in \ref{item:mainstar-2} is a
decomposition of $G\cup K[S]$ instead of $G$. As $G\cup K[S]$ has more
edges than $G$, this makes the statement slightly stronger (because it
makes harder to satisfy the requirements on $\tau_S(s)$).  The proof
of the Global Structure Theorem~\ref{theo:structure} needs this extra
condition, since the set $S$ will connect the graph to the part of the
tree decomposition already computed.  In \ref{item:mainstar-1},
however, the $K_k$-subdivision is found in $G$ (which is a slightly
stronger statement than finding it in $G\cup K[S]$).

The proof of the Global Structure Theorem~\ref{theo:structure} follows
from the Local Structure Theorem by a fairly simple induction (see below).
In Section~\ref{sec:three-local-decomp}, we show that Local Structure
Theorem~\ref{th:mainstar} can be proved by putting together three
decomposition lemmas. We prove these lemmas in
Sections~\ref{sec:tangles}--\ref{sec:local}. 
Let us remark that the Global
Structure Theorem can be seen as an instance of a general theorem due
to Robertson and Seymour~\cite[(11.3)]{gm10}, explaining how to construct a
tree decomposition whose torsos have a ``nice structure'' in graphs
with a ``nice local structure'', where the local structure is
described with respect to a tangle (see
Section~\ref{sec:tangles}). Our proof follows the ideas of Robertson and
Seymour's construction, but as Robertson and Seymour's theorem is not
algorithmic, and since there would be a large notational overhead, we
see no benefit in appealing to Robertson and Seymour's theorem here
and instead carry out our own version of the construction, which is
not very difficult anyway.
Not only here, but in several places throughout this paper we have to
carefully re-work results from Robertson and Seymour's structure
theory in order to make them algorithmic and, in addition, obtain
invariance results thate we need for the isomorphism test
later.

\begin{proof}[Proof of the Global Structure Theorem~\ref{theo:structure}]
  Let $a(k)$,
  $b(k)$, $c(k)$, $d(k)$, $e(k)$ as in the Local Structure Theorem~\ref{th:mainstar}.  Let $G$ be a graph. We shall describe the
  construction of a tree decomposition $(T,\bag)$ of $G$ satisfying
  all conditions asserted in the lemma. The construction may fail, but
  in that case it yields a subdivision of $H$ in $G$.

  We will built the tree $T$ inductively starting from the root. For every
  node $t$ we will define the set $N_+^T(t)$ of its children and sets
  $\sep(t),\comp(t)$ such that $|\sep(t)|\le a(k)$ and $N^G(\comp(t))\subseteq\sep(t)$. As usual, we define $\cone(t)$,
  $\bag(t)$, and $\tau(t)$ as in \eqref{eq:cone}, \eqref{eq:bag}, and
  \eqref{eq:torso}. In each step, we will prove that
  $\tau(t)$ satisfies one of (i), (ii), or (iii).

  We start with a root $r$ of $T$ and let $\sep(r):=\emptyset$ and
  $\comp(r):=V(G)$. For the inductive step, let $t$ be a node for
  which $\sep(t)$ and $\comp(t)$ are defined, but $N_+^T(t)$ is not yet
  defined. We let $G_t:=G[\cone(t)]$.  
  Let us run the algorithm of Theorem~\ref{th:mainstar} on $G_t$ (as
  $G$), $\sep(t)$ (as $S$), and $k$. If it returns a subdivision of
  $K_k$ in $G_t$, then we can clearly return a subdivision of $H$ in
  $G$ and we are done. Otherwise, it returns a star decomposition
  $\Sigma_t:=(T_t,\sep_t,\comp_t)$ of $G\cup K[\sep(t)]$ having adhesion at most $a(k)$; let $s_t$
  be the center of $T_t$.  We let $N_+^T(t):=V(T_t)\setminus\{s_t\}$ be
  the set of tips of $T_t$, where without loss of generality we assume
  that this set is disjoint from the tree $T$ constructed so far. For
  every $u\in N_+^T(t)$ we let $\sep(u):=\sep_t(u)$ and
  $\comp(u):=\comp_t(u)$.  Observe that we have
  $\bag(t)=\cone(t)\setminus\bigcup_{u\in
    N_+^T(t)}\comp(u)=\bag_t(s_t)$.  Furthermore, since $\Sigma_t$ is a
  decomposition of $G\cup K[\sep(t)]$ and $\sep(t)$ induces a clique
  in $G\cup K[\sep(t)]$, we have $\torso(t)=\torso_t(s_t)$.
  Thus one of the three cases of Theorem~\ref{th:mainstar} holds for
  the node $t$ as well.

  To see that $(T,\bag)$ is a tree decomposition, it is easiest to
  verify it satisfies \ref{li:t2}--\ref{li:t4}: it follows from the
  fact that the star decomposition $\Sigma_t$ used in each step of the
  construction does satisfy these conditions. Condition \ref{li:t1} is
  obvious and \ref{li:t5} follows because we start the construction
  with a node $t$ having $\comp(t)=V(G)$ and $\sep(t)=\emptyset$.
  Note that the bound $a(k)$ on the adhesion of $\Sigma_t$ implies the
  same bound on the adhesion of $(T,\bag)$.

  To see that the construction terminates, note that for all $t\in
  V(T)$, Theorem~\ref{th:mainstar} states that $\comp_t(u)\subset
  \comp_t(s_t)$ for every tip $u$ of $T_t$. This means that that
  $\comp(u)\subset \comp(t)$ holds for every $u\in N_+^T(t)$ and hence
  the height of the tree is at most $|V(G)|$. Moreover, $\comp(u_1)$
  and $\comp(u_2)$ are disjoint for two distinct children of node $t$
  and it follows that the total number of leaves can be bounded by
  $|V(G)|$. Thus the algorithm, excluding the calls to
  Theorem~\ref{th:mainstar}, runs in polynomial time. The claim on the
  running time follows from Theorem~\ref{th:mainstar}.
\end{proof}

\subsection{Almost Embeddable Graphs and a Refined Structure Theorem}
\label{sec:ae}
In this section, we combine our structure theorem with Robertson and Seymour's
structure theorem for graphs with excluded minors \cite{gm16}, which says that for graph
$H$, all graphs excluding $H$ as a minor have a tree decomposition into torsos
that are almost embeddable into some surface. 

We start by reviewing Robertson and Seymour's structure theorem. We
need first the definition of $(p,q,r,s)$-almost embeddable graphs (for
the current paper, the exact definition will not be important, thus
the reader can safely skip the details). We assume that the reader is
familiar with the basics of surface topology and graph embeddings. A
\emph{path decomposition} is a tree decomposition $(P,\bag)$ where $P$
is a path. For every $n\in\NN$, by $P^n$ we denote the path with
vertex set $[n]$ and edges $i(i+1)$ for all $i\in[n-1]$. A
\emph{$p$-ring} is a tuple $(R,v_1,\ldots,v_n)$, where $R$ is a graph
and $v_1,\ldots,v_n\in V(R)$ such that there is a path decomposition
$(P^n,\bag)$ of $R$ of width $p$ with $v_i\in\bag(i)$ for all
$i\in[n]$. A graph $G$ is \emph{$(p,q)$-almost embedded} in a surface
$\FS$ if there are graphs $G_0,G_1,\ldots,G_q$ and mutually disjoint
closed disks $\FD_1,\ldots,\FD_q\subseteq\FS$ such that:
\begin{eroman}
\item $G=\bigcup_{i=0}^qG_i$.
\item $G_0$ is embedded in $\FS$ and has a nonempty intersection with the
  interiors of the disks $\FD_1,\ldots,\FD_q$.
\item The graphs $G_1,\ldots,G_q$ are mutually disjoint.
\item For all $i\in[q]$ we have $E(G_0\cap G_i)=\emptyset$, and there are $n_i\in\NN$ and $v^i_1,\ldots,v^i_{n_i}\in
  V(G)$ such that $V(G_0\cap G_i)=\{v^i_1,\ldots,v^i_{n_i}\}$, and the
  vertices $v^i_1,\ldots,v^i_{n_i}$ appear in cyclic order on the boundary of
  the disk $\FD_i$.
\item For all $i\in[q]$ the tuple $(G_i,v^i_1,\ldots,v^i_{n_i})$ is a $p$-ring.
\end{eroman}
A graph $G$ is \emph{$(p,q,r,s)$-almost embeddable}\label{page:ae} if there is an {\em apex set}
$X\subseteq V(G)$ of size $|X|\le s$ such that $G\setminus X$ is isomorphic to
a graph that is $(p,q)$-almost embedded in a
surface of Euler genus $r$. 

\begin{theo}[\cite{gm16,kawwol11,grokawree13}]\label{theo:rsstructure}
  For every graph $H$ there are constants $p,q,r,s\in\NN$ such that every
  graph $G$ with $H\not\preceq G$ has a tree decomposition $(T,\bag)$ such that for all
  $t\in V(T)$ the torso $\tau(t)$ is $(p,q,r,s)$-almost embeddable.

  Furthermore, there is an algorithm that, given $G$ and $H$, in time
  $f(|H|)\cdot n^2$ for some
  computable function $f$, either finds an $H$-minor image in $G$, or computes
  such a tree decomposition and moreover, computes an apex set $Z_t$ of size
  at most $s$ for every $t\in V(T)$.
\end{theo}

As a corollary of this theorem and our structure theorem we get:

\begin{cor}\label{cor:structure-emb}
  For every graph $H$ there are constants $c,d,p,q,r,s\in\NN$ such that every
  graph $G$ with $H\not\preceq_T G$ has a tree decomposition $(T,\bag)$ such that for all
  $t\in V(T)$,
  \begin{eroman}
  \item either $\tau(t)$ is $(p,q,r,s)$-almost embeddable,
  \item or at most $c$ vertices of $\tau(t)$ have degree greater than $d$.
  \end{eroman}
  Furthermore, there is an algorithm that, given $G$ and $H$, in time
  $f(|H|)\cdot n^{O(1)}$ for some computable function $f$, either
  finds a subdivision of $H$ in $G$, or computes such a tree
  decomposition, and moreover computes an apex set $Z_t$ of size at
  most $s$ for every bag of the first type.
\end{cor}

\begin{proof}
  Let $G,H$ be a graphs such that $H\not\preceq_TG$. We let $k:=|H|$ and
  choose 
  constants $b,c,d,e$ (the adhesion $a$ is irrelevant here) according
  to the Global Structure Theorem~\ref{theo:structure}. Without loss
  of generality we may assume that $c\ge b$. Then $G$ has a
  tree decomposition $(T^1,\bag^1)$ into torsos $\tau^1(t)$ that
  either have at most $c$ vertices of degree greater than $d$ or
  exclude $K_e$ as a minor. 

We choose constants $p,q,r,s$ according to
  Theorem~\ref{theo:rsstructure} applied to $K_e$ (as $H$). 
 We refine the decomposition $(T^1,\bag^1)$ as follows: Let $t\in
 V(T^1)$ be a node such that $K_e\not\preceq\tau^1(t)$. Then by
 Theorem~\ref{theo:rsstructure}, we can find a decomposition
 $(T^2_t,\bag^2_t)$ of $\tau^1(t)$ into torsos that are $(p,q,r,s)$-almost
 embeddable. As $\sep^1(t)$ and $\sep^1(u)$ for all $u\in N_+^{T^1}(t)$
 are cliques in $\tau^1(t)$, there are nodes $x_t$ and $x_u$ such that
 $\sep^1(t)\subseteq\sep^2_t(x_t)$ and
 $\sep^1(u)\subseteq\sep^2_t(x_u)$. Without loss of generality we
 assume that $x_t$ is the root of $T^2_t$. We define a new decomposition by
 deleting $t$ from $T^1$, adding $T^2_t$, and adding edges from the
 parent of $t$ in $T^1$ to $x_t$ (if $t$ is the root of $T^1$, we omit
 this step) and from $x_u$ to $u$ for all $u\in N_+^D(t)$. All nodes in
 $t_1\in V(T^1)\setminus\{t\}$ keep their bags $\bag^1(t)$ in the new
 decomposition, and all nodes in $t_2\in V(T^2_t)$ keep their bags
 $\bag^2_t(t_2)$ as well. We carry out this construction for all $t\in
 V(T^1)$ such that $K_e\not\preceq\tau^1(t)$. All torsos of the
 resulting tree decomposition $(T,\bag)$ satisfy either (i) or (ii).

 Furthermore, the decomposition $(T,\bag)$ can be computed in time
 $f(|H|)\cdot n^{O(1)}$ because both $(T^1,\bag^1)$ and
 $(T^2_t,\bag^2_t)$ can.
\end{proof}

\subsection{The Three Local Decomposition Lemmas}
\label{sec:three-local-decomp}

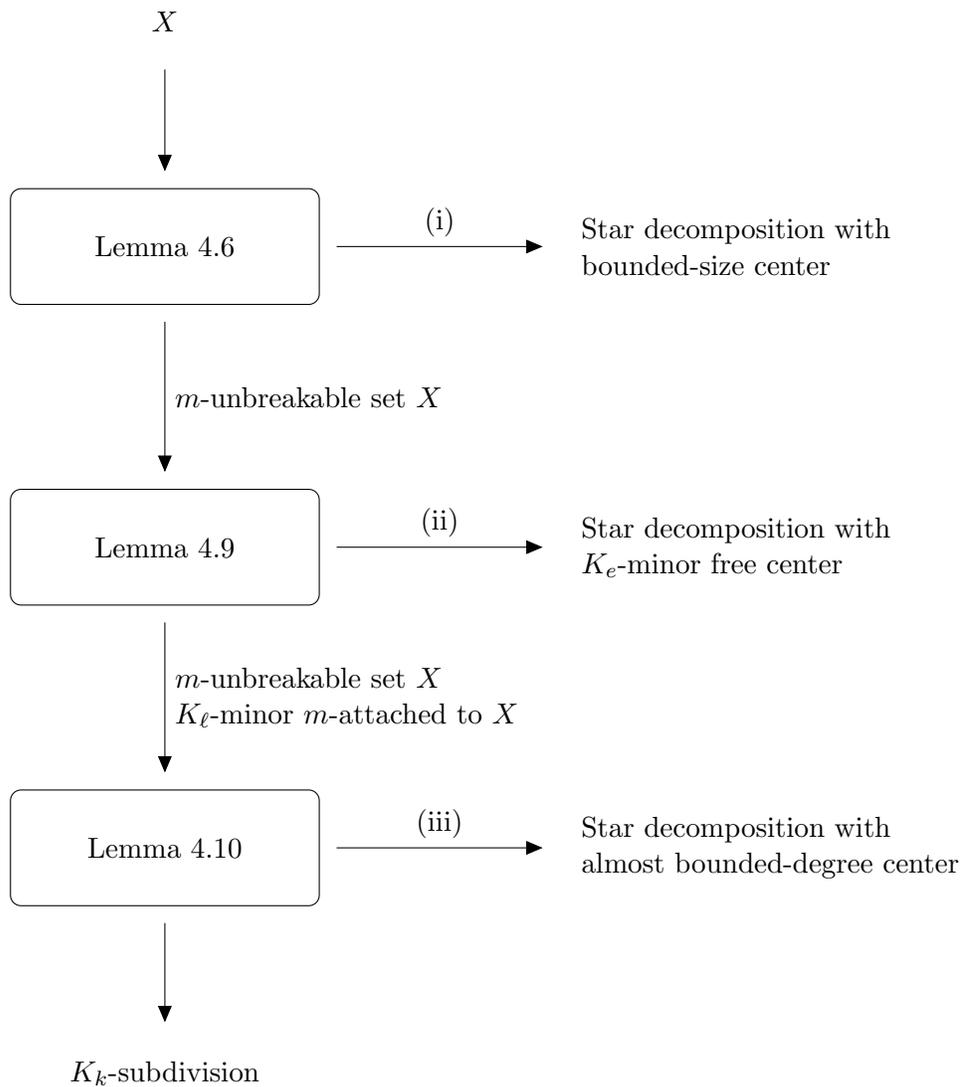
\begin{figure}
\tikzstyle{block} = [outer sep=0.6em, rectangle, draw,
    text width=10em,
 text centered, rounded corners, minimum height=4em, node distance=5.5cm]
\tikzstyle{endblock} = [outer sep=1em, rectangle,
    text width=10em,
 text centered, rounded corners, node distance=5.5cm]
\tikzstyle{rightblock} = [outer sep=1em, rectangle,
    text width=18em,
  rounded corners, node distance=5.5cm]
\tikzstyle{line} = [draw, -triangle 45]
\tikzstyle{arrownode} =[rectangle,text width=22em]
\begin{center}
{\centering
\begin{tikzpicture}[node distance = 2.5cm, auto]
    \node [endblock] (start) {$X$};
    \node [block, below of=start, node distance=3cm] (bounded) {Lemma~\ref{lem:compute-break}};
    \node [rightblock, right of=bounded, node distance=9cm] (bounded2) {Star decomposition with\\bounded-size center};
    \node [block, below of=bounded, node distance=4cm] (clique) {Lemma~\ref{lem:conncliquestar}};
    \node [rightblock, right of=clique, node distance=9cm] (clique2) {Star decomposition with\\$K_e$-minor free center};
    \node [block, below of=clique, node distance=4cm] (degree) {Lemma~\ref{lem:conndegstar}};
    \node [rightblock, right of=degree, node distance=9cm] (degree2) {Star decomposition with\\almost bounded-degree center};
    \node [endblock, below of=degree, node distance=3cm] (subdiv) {$K_k$-subdivision};

    \path [line] (start) -- node[arrownode] {} (bounded);
    \path [line] (bounded) -- node[arrownode] {$m$-unbreakable set $X$} (clique);
    \path [line] (bounded) -- node{\ref{item:str-1}} (bounded2);
    \path [line] (clique) -- node[arrownode] {$m$-unbreakable set $X$\\$K_\ell$-minor $m$-attached to $X$} (degree);
     \path [line] (clique) -- node{\ref{item:str-2}}  (clique2);
   \path [line] (degree) -- node[arrownode] {} (subdiv);
     \path [line] (degree) -- node{\ref{item:str-3}} (degree2);

\end{tikzpicture}
}
\end{center}
\caption{The three decomposition lemmas in the proof of Local Structure Theorem~\ref{th:mainstar}.}\label{fig:local}
\end{figure}

We prove the Local Structure Theorem~\ref{th:mainstar} by stacking
three decomposition lemmas on top of each other (see Figure~\ref{fig:local}). Each lemma provides
either a star decomposition corresponding to one of the three cases
\ref{item:str-1}--\ref{item:str-3} or an ``obstruction'' which can be
fed into the next lemma as input.

The first decomposition lemma either finds a star decomposition where
the center bag has bounded size or finds a ``highly connected'' set
in the following sense.
\begin{defn}\label{def:unbreakable}
  Let $G$ be a graph and $X\subseteq V(G)$.
  A separation $(A,B)$ of $G$ \emph{breaks} $X$ if $|(V(A)\cap X)\cup
  V(A\cap B)|<|X|$ and $|(V(B)\cap X)\cup
  V(A\cap B)|<|X|$.

  The set $X$ is \emph{$m$-unbreakable} if there is no
  separation $(A,B)$ of $G$ of order $<m$ that breaks $X$.
\end{defn}

The notion of an $m$-unbreakable set is closely related to that of an
$m$-linked set or a well-linked set
\cite{ree97,ree2003,chekhashe04}. We decided to present our
results in terms of the definition above, as it expresses most faithfully
our requirements in the proofs to follow.  To the best of our
knowledge, notions of this type were first used (implicitly) by
Robertson and Seymour \cite{gm10,gm13} in a similar context as
ours. The following lemma can be also traced back to \cite{gm13}.

There is a simple way of detecting if a set $X$ is $m$-unbreakable by
considering all possible ways of breaking $X$. Note that the running
time of the following algorithm is exponential in the size of the set,
but we will use it only on sets of bounded size.
\begin{lem}\label{lem:compute-break}
  There is an algorithm that, given a graph $G$ and a set $X\subseteq
  V(G)$ and $m\in\NN$, either computes a separation of $G$ of order
  $<m$ that breaks $X$ or correctly decides that $X$ is
  $m$-unbreakable. The running time of the algorithm is
  $3^{|X|}n^{O(1)}$.
\end{lem}

\begin{proof}
  The algorithm goes through all integers $0\le \lambda <m$ and
  partitions $(X_A,X_Q,X_B)$ of $X$ with $|X_A|+\lambda <|X|$ and
  $|X_B|+\lambda<|X|$. For each $\lambda$ and partition, we try to
  find a set $Y_Q\subseteq V(G)\setminus X_Q$ of size $<\lambda-|X_Q|$
  such that $Q:=X_Q\cup Y_Q$ separates $X_A$ from $X_B$, or in other
  words, $Y_Q$ separates $X_A$ from $X_B$ in $G\setminus X_Q$. Finding
  such a set can be done using standard polynomial-time minimum cut
  algorithms. If it succeeds to find such $(X_A,X_Q,X_B)$ and $Y_Q$,
  then it returns a separation $(A,B)$ with $V(A)\cap V(B)=Q$ such
  that $A$ contains all connected components of $G\setminus Q$ that
  have a nonempty intersection with $X_A$ and $B$ contains the
  remaining connected components of $G\setminus Q$. We have $|(V(A)\cap X)\cup
  V(A\cap B)|<|X_A|+|X_Q|+|Y_Q|\le |X_A|+\lambda<|X|$ and $|(V(B)\cap X)\cup
  V(A\cap B)|<|X|$ follows similarly, implying that $(A,B)$ breaks $X$. If the algorithm
  fails to find such a $\lambda$, $(X_A,X_Q,X_B)$, and $Y_Q$, then it correctly concludes
  that $X$ is $m$-unbreakable.
\end{proof}

It is not difficult to see that a large unbreakable set is an
obstruction for having small treewidth, that is, for having a tree
decomposition where every bag has small size. Therefore, it is not
surprising that the proof of the first local decomposition lemma is
very similar to algorithms finding tree decompositions.

\begin{lem}[Bounded-size star decomposition]\label{lem:mainbreak}
For every $m\in \NN$, there is a constant $b^*(m)$ such that the following
  holds.  There is an $f(m)\cdot |V(G)|^{O(1)}$ time algorithm
  that, given a graph $G$, an integer $m$, a set $X$ of size $\le 3m-2$, and an integer $k$, 
\begin{enumerate}
\item\label{item:bounded-1} either finds an $m$-unbreakable set $X'\supseteq X$  of size $3m-2$.
\item\label{item:bounded-2} or computes a star decomposition $\Sigma_X=(T_X,\sep_X,\comp_X)$
  of $G\cup K[X]$
 having adhesion $< 3m-2$ such
  that $X\subseteq \bag_X(s)$ and $|\bag_X(s)|\le b^*(m)$ for the center
  $s$ of $T_X$.
\end{enumerate}
\end{lem}
\begin{proof}
  Let $b^*(m)=4m-3$.  If $|V(G)|<3m-2$, then we can return a star
  decomposition consisting of a single center node $s$ with
  $\alpha(s)=V(G)$ and $\sep(s)=\emptyset$.  Otherwise, let $X'$ be an
  arbitrary superset of $X$ having size $3m-2$. Let us use the algorithm
  of Lemma~\ref{lem:compute-break} to test if $X'$ is $m$-unbreakable;
  if so, then we can return $X'$ and we are done. Otherwise, there is a
  separation $(A,B)$ of $G$ having order $<m$ such that $|(X'\cap
  V(A))\cup Q|,|(X'\cap V(B))\cup Q|<|X'|=3m-2$ for $Q:=V(A)\cap V(B)$.
  Let us construct a star decomposition
  $\Sigma_X=(T_X,\sep_X,\comp_X)$ with center $s$ and tips $t_A$,
  $t_B$. First, let $\comp(s)=V(G)$ and $\sep(s)=\emptyset$. Let
  $\comp(t_A)=V(A)\setminus (Q\cup X')$ and $\sep(t_A)=(X'\cap
  V(A))\cup Q$; it is clear that $|\sep(t_A)|< 3m-2$. Similarly, let
  $\comp(t_B)=V(B)\setminus (Q\cup X')$ and $\sep(t_B)=(X'\cap
  V(B))\cup Q$. It is straightforward to verify that this is indeed a
  star decomposition of $G\cup K[X]$ with adhesion $<3m-2$. Furthermore,
  $|\bag(s)|=|Q\cup X'|\le m-1+3m-2=b^*(m)$.
\end{proof}

The second local decomposition lemma takes an unbreakable set $X$ of
appropriate size, and either finds a star decomposition where the
torso of the center node excludes some minor or finds a large clique
minor. Furthermore, this clique minor has the additional property that
it is close to the unbreakable set $X$ in the following sense:
\begin{defn}
  Let $I$ be an $H$-minor image in $G$ and let $X$ be a set of
  vertices. We say that $I$ is {\em $m$-attached} to $X$ if there is
  no separation $(A,B)$ of order $<m$ such that $I(v)\subseteq
  V(A)\setminus V(B)$ for some $v\in V(H)$ and $|(V(B)\cap X)\cup
  V(A\cap B)|\ge |X|$.
\end{defn}
In particular, if $X$ is an $m$-unbreakable set and $I$ is
$m$-attached to $X$, then whenever $I(v)\subseteq V(A)\setminus V(B)$
for some $v\in V(H)$ and separation $(A,B)$ of order $<m$, then we
know that $|(V(A)\cap X)\cup V(A\cap B)|\ge |X|$. Thus in every
separation, $I$ is on the same side as the larger part of $X$.  (This
definition is similar to the notion of a tangle controlling a minor,
introduced by Robertson and Seymour \cite{gm16}.)

\begin{lem}[Excluded-minor star decomposition]\label{lem:conncliquestar}
  For every $\ell,m\in \NN$, there is a constant $e^*(\ell,m)$ such that
  the following holds.  There is an $f(\ell,m)\cdot |V(G)|^{O(1)}$ time
  algorithm that, given a graph $G$, integers $\ell$, $m$, and an
  $m$-unbreakable set $X$ of size $3m-2$
\begin{enumerate}
\item\label{item:conncliquestar-1} either finds a $K_\ell$-minor image $I$ in $G$ that is $m$-attached to $X$,
\item\label{item:conncliquestar-2} or computes a star decomposition $\Sigma_X=(T_X,\sep_X,\comp_X)$ of $G\cup K[X]$
having  adhesion $<|X|$ such
  that $X\subseteq \beta_X(s)$ and $\tau_X(s)$ does not contain a
  $K_{e^*(\ell,m)}$-minor for the center $s$ of $T_X$.
\end{enumerate}
Furthermore, suppose that the algorithm computes $\Sigma_X$ on input
$(G,X)$ and let $(G',X')$ be a pair such that there is an isomorphism $f$
from $G$ to $G'$ with $f(X)=X'$. Then the algorithm computes a star
decomposition $\Sigma'_{X'}$ on input $(G',X')$ and there exists an
isomorphism $g$ from $T_{X}$ to $T_{X'}$ such that for all $t\in
V(T_X)$ we have $\sep_{X'}(g(t))=f(\sep_{X}(t))$ and
$\comp_{X'}(g(t))=f(\comp_{X}(t))$.
\end{lem}
Lemma~\ref{lem:conncliquestar} states an invariance condition saying
that for isomorphic input the decomposition is isomorphic. This
condition is not required for the proof of the Global Structure
Theorem~\ref{theo:structure}, but will be essential for the proof of
the Invariant Decomposition Theorem~\ref{theo:idec} in
Section~\ref{sec:treel-decomp}. Note that Lemma~\ref{lem:mainbreak}
does not state such an invariance condition and in fact there does not
seem to be an obvious way of ensuring invariance (for example, already
the selection of $X'$ in the first step of the proof is completely
arbitrary and hence cannot be done in an invariant way). This is
precisely the reason why we need to use the more general treelike
decompositions in
Sections~\ref{sec:treel-decomp}--\ref{sec:canonization} if we want the
construction to be invariant.

The proof of Lemma~\ref{lem:conncliquestar} is deferred to
Section~\ref{sec:cliquestar}. The algorithm repeatedly finds
$K_\ell$-minor images and tests if they are $m$-attached to $S$. If
so, it returns it, otherwise there is a separator that we can use to
decrease the bag of the center in such a way that this particular
image is no longer in the torso of the center. Note that when we
exclude some vertices from the bag, then new cliques can appear in the
torso. The main technical challenge is to ensure that no new clique
minor images are created when decreasing the size of the bag.

The third and final decomposition lemma takes a clique minor image $I$ attached
to an unbreakable set $S$ and  finds either a star decomposition
where the torso of the center has ``almost bounded degree'' (that is,
bounded degree with the exception of a bounded number of vertices) or
a subdivision of a clique.  
\begin{lem}[Bounded-degree Star Decomposition]\label{lem:conndegstar}
  For every $k\in \NN$, there exist constants $c^*(k)$, $d^*(k)$,
  $m^*(k)$, $\ell^*(k)$ such that the following holds.  There is an
  $f(k)|V(G)|^{O(1)}$ time algorithm that given a graph $G$, integer
  $k$, an $m$-unbreakable set $X$ of size $3m-2$ (for $m:=m^*(k)$) and
  an image $I$ of $K_{\ell}$ that is $m$-attached to $X$ (for
  $\ell:=\ell^*(k)$),
\begin{enumerate}
\item either finds a subdivision of $K_k$ in $G$,
\item or computes a star decomposition
  $\Sigma_{X}=(T_{X},\sep_{X},\comp_{X})$ of $G\cup K[X]$ having adhesion
  $< |X|$ such that $X\subseteq \bag(s)$ and at most $c^*(k)$ vertices of
  $\torso(s)$ have degree greater than $d^*(k)$ in
  $\torso(s)$, where $s$ is the center of
  $T_X$.
\end{enumerate}
Furthermore, suppose that the algorithm computes $\Sigma_X$ on input
$(G,X)$ and let $(G',X')$ be a pair such that there is an isomorphism $f$
from $G$ to $G'$ with $f(X)=X'$. Then the algorithm computes a star
decomposition $\Sigma'_{X'}$ on input $(G',X')$ and there exists an
isomorphism $g$ from $T_{X}$ to $T_{X'}$ such that for all $t\in
V(T_X)$ we have $\sep_{X'}(g(t))=f(\sep_{X}(t))$ and
$\comp_{X'}(g(t))=f(\comp_{X}(t))$.
\end{lem}
The proof of Lemma~\ref{lem:conndegstar} is deferred to
Section~\ref{sec:degstar}. The main idea is that we are trying to
remove every high-degree vertex from the bag of the center using
appropriate separations. If there are at least $k$ high-degree
vertices that cannot be removed this way, then these vertices are close
to the clique minor image $I$, and we can use this fact to construct a
subdivision of a clique.

With the three local decomposition algorithms of
Lemmas~\ref{lem:mainbreak}--\ref{lem:conndegstar} at hand, we are
ready to prove Local Structure Theorem~\ref{th:mainstar}:

\begin{proof}[Proof of Local Structure Theorem~\ref{th:mainstar}]
  Let $c(k)=c^*(k)$, $d(k)=d^*(k)$, $\ell=\ell(k)=\ell^*(k)$,
  $m=m(k)=m^*(k)$ using the functions $c^*$, $d^*$, $\ell^*$, $m^*$ in
  Lemma~\ref{lem:conndegstar}. Let $e(k)=e^*(\ell,m)$ for the function
  $e^*$ in Lemma~\ref{lem:conncliquestar}. Let $b(k)=b^*(m)$ for the
  function $b^*(k)$ in Lemma~\ref{lem:mainbreak}.  Let $a(k)=3m-3$.
  Note that $b^*(m)\ge 3m-3$ in Lemma~\ref{lem:mainbreak}: otherwise,
  neither \ref{item:bounded-1} nor \ref{item:bounded-2} would be
  possible if $X=V(G)$ and $|X|=3m-3$. Thus we can assume $b(k)\ge
  a(k)$.

  If $S=V(G)$, then we can return a star decomposition consisting of a
  single center node $s$ with $\alpha(s)=V(G)$ and $\sep(s)=\emptyset$
  (here we use that $b(k)\ge a(k)\ge |S|$).  Otherwise, let $X:=S\cup \{v\}$ for
  an arbitrary vertex $v\not \in S$. Let us call the algorithm of
  Lemma~\ref{lem:mainbreak} on $G$, $X$, and $m$. If it returns a star
  decomposition $\Sigma_X=(T_X,\sep_X,\comp_X)$, then we return it and
  we are done.  Note that in this case $v\in X\subseteq \bag_X(s)$ for
  the root $s$ of $T_X$, thus $v\not\in \comp_X(t)$ for any tip $t$ of
  $T_X$, which means that the requirement $\comp_X(t)\subset
  \comp_X(s)$ indeed holds. Otherwise, let $X'$ be the $m$-unbreakable
  superset of $X$ returned by the algorithm. Let us call the algorithm
  of Lemma~\ref{lem:conncliquestar} with $G$, $\ell$, $m$, and
  $X'$. Again, if it returns a star decomposition, we are
  done. Otherwise, it returns a $K_\ell$-minor image $I$ that is
  $m$-attached to $X'$. Let us call the algorithm of
  Lemma~\ref{lem:conndegstar} with $G$, $k$, $X'$, and $I$. It returns
  either a $K_k$-subdivision or a star decomposition; we are done in
  both cases.
\end{proof}


%% file: tangles.tex
\section{Tangles}
\label{sec:tangles}

In the proofs of the local decomposition lemmas
(Section~\ref{sec:local}), we need to deal with separations that
separate some set from (the larger part of) an unbreakable
set. Robertson and Seymour \cite{gm10} defined the abstract notion of
{\em tangles,} which is a convenient tool for describing such
separations. While in principle our results could be described without
introducing tangles (in particular, we are not using any previous
results about tangles), we feel that they provide a convenient notation
for our purposes, and they make our results slightly more
general.

Let $m\in\PN$. A \emph{tangle of order $m$} in a graph $G$ is a set
$\KT$ of separations of $G$ of order $<m$ such that the following
axioms are satisfied:
\begin{nlist}{TA}
  \item\label{li:ta1}
    For every separation $(A,B)$ of $G$ of order $<m$,
    either $(A,B)\in\KT$ or $(B,A)\in\KT$.
  \item\label{li:ta2}
   For all $(A_1,B_1),(A_2,B_2),(A_3,B_3)\in\KT$ it holds that $A_1\cup
    A_2\cup A_3\neq G$.
  \item\label{li:ta3}
   For all $(A,B)\in\KT$ it holds that $V(A)\neq V(G)$.
\end{nlist}
Intuitively, one can think of each separation $(A,B)$ in the tangle
$\KT$ as having a ``small side'' $A$ and ``\underline{b}ig side'' $B$.
Axiom \ref{li:ta2} states that the ``small side'' is so small that not
even three of them can cover the whole graph.

In this section, we introduce basic concepts for dealing with
tangles in the algorithmic context we need later. The ideas are not
new, most of them already appear in \cite{gm10,ree97} in a
similar form. However, our exact definitions are sometimes different
(and therefore we use different terms),
and it will be important for us to work with these precise
definitions.


In this paper, we only consider tangles of a special form. These
tangles are defined by unbreakable sets (in the sense of Definition~\ref{def:unbreakable}).
\begin{lem}\label{lem:unbreakabletangle}
  Let $X$ be an $m$-unbreakable set of size at least $(3m-2)$ in graph
  $G$. Let $\KT$ contain every separation of order $<m$ such that
  $|(X\cap V(B))\cup V(A\cap B)|\ge |X|$. Then $\KT$ is a
  tangle of order $m$ in $G$ (and we call it the tangle of order $m$
  defined by the set $X$). Furthermore, for every separation
  $(A,B)\in\KT$ it holds that $|V(A)\cap X|\le|V(A\cap B)|<m$.
\end{lem}
\begin{proof}
  Let us first observe that for every separation $(A,B)\in\KT$ with
  $Q:=V(A\cap B)$ we have $|V(A)\cap X|\le |Q|\le m-1$:
    otherwise, we would have
\[
|(X\cap V(B))\cup V(A\cap B)|=|(X\cap (V(B)\setminus V(A)))\cup Q|= |X|-|V(A)\cap X|+|Q|< |X|-|Q|+|Q|=|X|,
\]
contradicting the assumption that $(A,B)\in\KT$. In particular, this
makes it impossible that $V(A)=V(G)$, proving \ref{li:ta3}.

To see that $\KT$ satisfies \ref{li:ta2}, let
$(A_1,B_1),(A_2,B_2),(A_3,B_3)\in \KT$. By our observation in
the previous paragraph, we have $|V(A_i)\cap X|\le m-1$ for $i=1,2,3$, thus
 \[
 \big|\big(V(A_1)\cap X\big)\cup \big(V(A_2)\cap X\big)\cup \big(V(A_3)\cap X\big)\big|\le 3m-3<|X|.
 \]
 Therefore, $X\not\subseteq V(A_1\cup A_2\cup A_3)$, implying
 \ref{li:ta2}.

Finally, \ref{li:ta1} follows immediately from the fact that $X$ is $m$-unbreakable.
\end{proof}



The size of a tangle (even of small order) can be exponential in the
size of the graph. Observe that specifying the vertex set $V(A)\cap
V(B)$ is not sufficient for describing the separation $(A,B)$. For
example, a star with $n$ leaves have at least $2^n$ separations of
order 1. Therefore, when stating algorithmic results that take a graph
and a tangle as input, we have to state how the tangle is
represented. To obtain maximum generality of the results, we assume
that the tangle is given by an oracle. We define two type of
oracles. The first type simply answers if a separation $(A,B)$ is a
member of the tangle. However, in applications we often need to find a
separation of small order in the tangle that separates two given sets
$S$ and $T$. The min-cut oracle answers queries of this type. Note
that there are more than one natural way of defining such oracles, in
particular, we might want to allow or forbid the separator $V(A)\cap
V(B)$ to intersect $S$ and/or $T$. We define the min-cut oracle in a
way that includes all these possibilities: the query contains a set
$F$ of forbidden vertices and we require the separator to be disjoint
from $F$.

\begin{defn}\label{def:oracle}
Let $\KT$ be a tangle of order $k$ in a graph $G$.
\begin{enumerate}
\item An {\em oracle} for $\KT$ answers in constant time whether a given separation $(A,B)$ is in $\KT$. 
\item Given sets $S,T,F\subseteq V(G)$ and an integer $\lambda<k$, a {\em min cut oracle} for $\KT$ returns in constant time either a separation $(A,B)\in\KT$ of order at most $\lambda$ such that $S\subseteq V(A)$, $T\subseteq V(B)$, and $V(A)\cap V(B)\cap F=\emptyset$, or ``no'' if no such separation exists.
\end{enumerate}
\end{defn}




For tangles defined by unbreakable sets it is easy to
 implement both type of oracles:
\begin{lem}\label{lem:connectedtangleoracle}
Let $X$ be an $m$-unbreakable set of size at least $3m-2$ in a graph $G$ and let $\KT$ be the tangle of order $m$ defined by $X$.
\begin{enumerate}
\item  The oracle for $\KT$ can be implemented in polynomial time.
\item  The min cut oracle for $\KT$ can be implemented in time $2^{|X|}\cdot |V(G)|^{O(1)}$.
\end{enumerate}
\end{lem}

\begin{proof}
To implement the oracle, all we have to do is to check if $|X\cap V(B\setminus
  A))\cup V(A\cap B)|\ge |X|$, which can be clearly done in polynomial time.

  To implement the min cut oracle, observe that the answer for a query
  $S$, $T$, $F$, $\lambda$ is yes if and only if there is a separation
  $(A,B)$ with separator $Q=V(A)\cap V(B)$ satisfying
\begin{enumerate}
\item $S\subseteq V(A)$,
\item $T\subseteq V(B)$,
\item $Q\cap F=\emptyset$,
\item $|Q|\le \lambda$, and
\item there is a set $X'\subseteq X$ of size at least $|X|-|Q|$ with $X'\subseteq V(B)\setminus V(A)$. 
\end{enumerate}
In this case, $(A,B)\in\KT$ and satisfies the requirements. To find
such an $(A,B)$, we guess the size of $|Q|$ (at most $\lambda+1$
possibilities) and the set $X'$ (at most $2^{|X|}$ possibilities). For
each such guess, we check if there is a $Q$ of the given size that is
disjoint from $F\cup X'$ and separates $S$ from $T\cup X'$. This can be
checked using standard minimum cut algorithms in polynomial time. If
we find such a set $Q$ for at least one of the guesses, then we can
return a separation $(A,B)\in\KT$ satisfying the requirements. If
there is no such $Q$ for any of the guesses, then the answer is no.
\end{proof}

\begin{rem}
  Let us mention that for all tangles, and not only tangles defined by
  unbreakable sets, we can implement a min cut oracle using just a
  plain oracle in time $2^{O(k)}n^{O(1)}$, where $n$ is the order of
  the graph and $k$ the order of the tangle. This can be proved using
  ``important separators'' (introduced
  in~\cite{marx-separation-full}). As we do not need this result in
  the present article, we omit the proof.
\end{rem}

\subsection{Boundaries and separations}
In this section, we summarize some useful properties of boundaries of
sets and their relations to tangles. These facts will be used
extensively in Section~\ref{sec:local}.

Recall that $\boundary(X)=|N(X)|$. The following lemma states that the function
$\boundary$ satisfies the submodular inequality and a variant of the
posimodular inequality:
\begin{lem}\label{lem:subandposimodular}
Let $G$ be a graph and $X,Y\subseteq V(G)$. 
\begin{enumerate}
\item $\boundary(X)+\boundary(Y)\ge \boundary(X\cap Y)+\boundary(X\cup Y)$.
\item $\boundary(X)+\boundary(Y)\ge \boundary(X\setminus N[Y])+\boundary(Y\setminus N[X])$.
\end{enumerate}
\end{lem}
\begin{proof}
Both statements can be proved by checking that the contribution of
each vertex to the right-hand side is at most the contribution to the
left-hand side. This can be verified by a simple case
analysis. Figure~\ref{fig:submodular}
may help.

\begin{figure}
  \centering
  \input{submodular}
  \caption{Proof of Lemma~\ref{lem:subandposimodular}; note that there are only edges between regions whose boundaries have at least one point in common}
  \label{fig:submodular}
\end{figure}
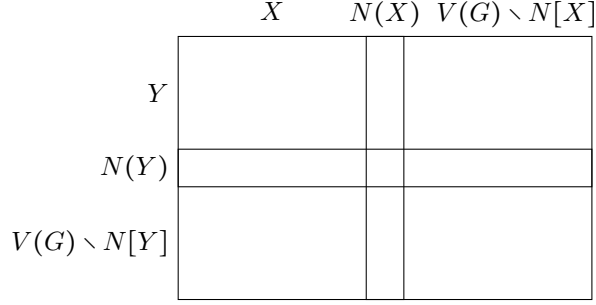

  (1) Any vertex that contributes to one of the terms on the
  right-hand side (i.e., appears in $N(X\cap Y)$ or in $N(X\cup
  Y)$) has to appear either in $N(X)$ or in $N(Y)$, and therefore
  contributes at least one to the left-hand side. Furthermore, if a
  vertex $v$ appears in both $N(X\cap Y)$ and $N(X\cup Y)$, then
  it is easy to check that $v\in N(X)$ and $v\in N(Y)$.

  (2) If $v\in N(X\setminus N[Y])$, then either $v\in
  N(X)$, or $v\in N[Y]$. Note that there is no edge between
  $X\setminus N[Y]$ and $Y$, thus in the latter case $v\in
  N(Y)$ holds. Similarly, $v\in N(Y\setminus N[X])$
  implies that either $v\in N(Y)$ or $v\in N(X)$. Finally, we
  claim that if $v\in N(X\setminus  N[Y])$ and $v\in
  N(Y\setminus N[X])$ both hold, then $v\in N(X)$ and
  $v\in N(Y)$. Suppose that $v\not\in N(X)$; by $v\in
  N(X\setminus N[Y])$, this is only possible if $v\in
  X$. Every neighbor of $v$ is in $N[X]$, thus $v$ has no
  neighbor in $Y\setminus N[X]$, contradicting the assumption that
  $v\in N(Y\setminus N[X])$.
\end{proof}

We often work with separations that separate a subset of vertices from
the rest of the graph:
\begin{defn}\label{def:separg}
  Let $G$ be a graph and $X\subseteq V(G)$. Then we define the
  separation $\separg(X)=(A,B)$ by $A=G[N[X]]$,
  $V(B)=V(G)\setminus X$, $E(B)=E(G)\setminus E(A)$.
\end{defn}
Note that the order of $\separg(X)$ is exactly $\boundary(X)$.

The following observation, together with
Lemma~\ref{lem:subandposimodular}, will allow us to use uncrossing
arguments in Section~\ref{sec:local}:
\begin{lem}\label{lem:tanglesepprop}
Let $\KT$ be a tangle of order $m$ in graph $G$ and let $X,Y\subseteq V(G)$ be sets such that $\separg(X),\separg(Y)\in \KT$.
\begin{enumerate}
\item For every $X'\subseteq X$, if $\separg(X')$ is of order $<m$, then $\separg(X')\in \KT$.
\item If $\separg(X\cap Y)$ is of order $<m$, then $\separg(X\cap Y)\in\KT$.
\item If $\separg(X\cup Y)$ is of order $<m$, then $\separg(X\cup Y)\in\KT$.
\end{enumerate}
\end{lem}
\begin{proof}
  (1) Let $\separg(X)=(A,B)$ and $\separg(X')=(A',B')$; note that
  $B'\supseteq B$. If $(A',B')\not\in \KT$ and has order $<m$, then
  $(B',A')\in \KT$ by \ref{li:ta1}. But then $A\cup B'\supseteq A\cup B=G$ and
  \ref{li:ta2} is violated.

(2) Follows from (1).

(3) Let $\separg(X)=(A_X,B_X)$, $\separg(Y)=(A_Y,B_Y)$, and
$\separg(X\cup Y)=(A_{X\cup Y},B_{X\cup Y})$. If $\separg(X\cup
Y)\not\in \KT$ and has order $<m$, then $(B_{X\cup Y},A_{X\cup Y})\in
\KT$ by \ref{li:ta1}.  We claim that $A_X\cup A_Y\cup B_{X\cup Y}=G$,
violating \ref{li:ta2}. Consider an edge $e\in E(G)$. If $e$ has an
endpoint in $X$, then $e\in E(G[N[X]])=E(A_X)$. Similarly, if $e$
has an endpoint in $Y$, then $e\in E(G[N[Y]])=E(A_Y)$. Finally, if
$e$ does not have an endpoint in $X\cup Y$, then $e\not\in E(G[
N[X\cup Y]])=E(A_{X\cup Y})$, implying that $e\in E(B_{X\cup
  Y})=E(G)\setminus E(A_{X\cup Y})$. We can conclude that
$E(G)=E(A_X)\cup E(A_Y)\cup E(B_{X\cup Y})$. A similar argument shows
that $V(G)=V(A_X)\cup V(A_Y)\cup V(B_{X\cup Y})$.
\end{proof}

We say that a separation $(A,B)$ {\em removes} a set $X\subseteq V(G)$ if
$X\subseteq V(A)\setminus V(B)$.  Note that $\separg(W)$ removes $X$
if and only if $X\subseteq W$. It follows from
Lemmas~\ref{lem:subandposimodular} and \ref{lem:tanglesepprop} that
for every set $X$, there is a unique ``closest minimum cut'' of the
tangle that removes $X$:
\begin{lem}\label{lem:uniqueminsep}
  Let $\KT$ be a tangle of order $m$ in a graph $G$. Suppose that
  there is an $(A,B)\in\KT$ with $X\subseteq V(A\setminus B)$.  Then
  there is a unique $W(X)\subseteq V(G)$ such that
\begin{enumerate}
\item $X\subseteq  W(X)$,
\item $\separg(W(X))\in \KT$,
\item among sets satisfying (1) and (2), the order of $\separg(W(X))$ is minimum possible, and
\item among sets satisfying (1)--(3), $|W(X)|$ is minimum possible.
\end{enumerate}
Furthermore, given a min cut oracle for $\KT$, this unique minimal set can be found in polynomial time.
\end{lem}
\begin{proof}
  Let $m_0<m$ be the minimum possible order of a separation
  $\separg(W)\in\KT$ over all $W$ containing $X$ (the set
  $V(A\setminus B)$ shows that at least one such $W$ exists). To prove
  the uniqueness of $W(X)$, we show a stronger statement: there is
  such a $W(X)$ with the property that $W(X)\subseteq W$ for every
  $W\supseteq X$ with $\separg(W)\in \KT$ and $\boundary(W)=m_0$.  To
  prove this statement, suppose that $W_1,W_2\supseteq X$ are sets
  such that $\separg(W_1),\separg(W_2)\in \KT$ both have order $m_0$.
  By Lemma~\ref{lem:subandposimodular}(1),
\[
2m_0=\boundary(W_1)+\boundary(W_2)\ge \boundary(W_1\cap W_2)+\boundary(W_1\cup W_2).
\]
Observe that $W_1\cap W_2$ and $W_1\cup W_2$ both contain $X$. If
$\boundary(W_1\cup W_2)<m_0$, then $\separg(W_1\cup W_2)\in \KT$ by
Lemma~\ref{lem:tanglesepprop}(3), contradicting the minimality of the
order of $\separg(W_1)$ and $\separg(W_2)$. If $\boundary(W_1\cup
W_2)\ge m_0$, then $\boundary(W_1\cap W_2)\le m_0$. By
Lemma~\ref{lem:tanglesepprop}(2), $\separg(W_1\cap W_2)\in \KT$, and its
order is not larger than the order of $\separg(W_1)$ and
$\separg(W_2)$. Thus the intersection of the two sets is also a set
satisfying the requirements.  It follows that the common intersection
of every $W_i\supseteq X$ such that $\boundary(W_i)=m_0$ and
$\separg(W_i)\in \KT$ is the required minimal set $W(X)$.

To find this unique set $W(X)$, we let $S:=X$, initially define
$T=\emptyset$, and use the min cut oracle to check if there is a
separation $(A,B)$ of order at most $\lambda$ with $X\subseteq V(A)$,
$T\subseteq V(B)$, and $V(A)\cap V(B)$ disjoint from $F:=X$.  Let us
fix the smallest $\lambda$ for which the answer is yes: then the min
cut oracle returns a separation $(A,B)\in\KT$, such that
$W:=V(A)\setminus V(B)$ satisfies the first three properties above. To
ensure that the last property holds as well, we pick a vertex $v\in
W$, and call the min cut oracle to check if there is a separation
$(A',B')\in \KT$ of order $\lambda$ such that $X\subseteq V(A')$,
$T\cup\{v\}\subseteq V(B')$, and $V(A')\cap V(B')$ disjoint from
$X$. If there is such a separation, then we include $v$ in $T$, and
repeat this process with the new separation $(A',B')$.  As the size of
$T$ strictly increases, eventually we arrive at a set $W$ such that
including any vertex $v\in W$ into $T$ increases the minimum cut size
to above $\lambda$. We have seen that this set $W$ contains the unique
minimal set $W(X)$ defined above. Furthermore, $W=W(X)$ has to hold:
otherwise, including a vertex $v\in W\setminus W(X)$ into $T$ would
not increase the minimum cut size.
\end{proof}

In the following, we will denote by $W(X)$ the unique set defined by Lemma~\ref{lem:uniqueminsep}. Note that if there is no separation $(A,B)\in \KT$ with $X\subseteq V(A\setminus B)$, then $W(X)$ is undefined.

\begin{prop}\label{prop:connectedmin}
Let $X\subseteq V(G)$ such that $W(X)$ is defined and $G[X]$ is
connected. Then $G[W(X)]$ is connected.
\end{prop}
\begin{proof}
  Suppose that $G[W(X)]$ is not connected. Since $X\subseteq W(X)$,
  there is a component $C$ of $G[W(X)]$ containing $C$. By
  Lemma~\ref{lem:tanglesepprop}(1), $\separg(W(X))\in \KT$ implies
  $\separg(C)\in \KT$. Clearly, $N(C)\subseteq N(W(X))$, thus
  $\boundary(C)\le \boundary(W(X))$. Together with $C\subset W(X)$,
  this contradicts the minimality of $W(X)$.
\end{proof}


%% file: submodular.tex
\begin{tikzpicture}
  \draw (0,0) rectangle (5.5,3.5);
  \draw (2.5,0) rectangle (3,3.5);
  \draw (0,1.5) rectangle (5.5,2);
  \small
  \path (1.25,3.6) node[anchor=south] {$X$} (2.75,3.5) node[anchor=south] {$N(X)$} (4.5,3.5) node[anchor=south]
  {$V(G)\setminus N[X]$} (0,2.75) node[anchor=east] {$Y$} (0,1.75) node[anchor=east]
  {$N(Y)$} (0,0.75) node[anchor=east] {$V(G)\setminus N[Y]$} (1.75,1.75);

\end{tikzpicture}


%% file: local2.tex
\section{Proofs of the Local Decomposition Lemmas}
\label{sec:local}

This section completes the proof of Global Structure
Theorem~\ref{theo:structure} by proving
Lemmas~\ref{lem:conncliquestar} and \ref{lem:conndegstar}. First, in
Section~\ref{sec:using-clique-minor} we describe a useful tool (taken
from \cite{gm13}): using a clique minor as a ``crossbar switch'' to
connect a set of vertices. The proofs of
Lemmas~\ref{lem:conncliquestar} and \ref{lem:conndegstar} are
contained in Sections~\ref{sec:cliquestar} and \ref{sec:degstar},
respectively. Note that the proofs in this section contain somewhat
more work than what is  strictly necessary for the proof of the Global
Structure Theorem~\ref{theo:structure}: the proof of the invariance
conditions in Lemmas~\ref{lem:conncliquestar} and
\ref{lem:conndegstar} require extra arguments. These invariance
conditions are not needed for the Global Structure Theorem, but they
will be crucial for the invariance of the treelike decompositions in
Section~\ref{sec:treel-decomp} and therefore for the results of
Section~\ref{sec:canonization} on isomorphism and canonization.

We prove variants of Lemmas~\ref{lem:conncliquestar} and
\ref{lem:conndegstar} stated in terms of tangles instead of
unbreakable sets (Lemmas~\ref{lem:tanglecliquestar} and
\ref{lem:tangledegstar}, respectively); the proofs of
Lemmas~\ref{lem:conncliquestar} and \ref{lem:conndegstar} follows
easily from these variants. The statements involving tangles need the
following definitions:
\begin{defn}
  Let $\KT,\KT'$ be tangles in graphs $G,G'$, respectively.
  An \emph{isomorphism} from $(G,\KT)$ to $(G',\KT')$ is an isomorphism $f$
  from $G$ to $G'$ such that for all $(A,B)\in\KT$ we have
  $(f(A),f(B))\in\KT'$. 
\end{defn} 

\begin{defn}\label{def:tanglerespect}
  Let $\Sigma=(T,\bag)$ be a star decomposition of graph $G$ and let
  $\KT$ be a tangle of $G$. We say that $\Sigma$ {\em respects $\KT$}
  if for every tip $t$ of $T$ the separation $(A,B)$ with
  $A=G[\cone(t)]$ and $V(B)=V(G)\setminus\comp(t)$ is in $\KT$. In
  particular, this implies $\separg(\comp(t))\in \KT$ and $|\sep(t)|$
  is less than the order of $\KT$.
\end{defn}

\subsection{Using a clique minor}
\label{sec:using-clique-minor}

The following lemma follows from \cite[(5.3)]{gm13}:
\begin{lem}[\cite{gm13}]\label{lem:rs}
  For every $r\in \NN$, there is a constant $t(r)=O(r^2)$ such that
  the following holds.  Let $G$ be a graph and $R\subseteq V(G)$ with
  $|R|=r$. Let $t\ge t(r)$ and let $(B_i)_{i\in[t]}$ be an image of a
  $K_t$-minor in $G$. Suppose that there is no separation $(G_1,G_2)$
  of $G$ of order $<|R|$ with $R\subseteq V(G_1)$ and $B_b\cap
  V(G_1)=\emptyset$ for some $b\in[t]$. Then there is a
  $K_{|R|}$-minor image in $G$ such that every branch set contains
  exactly one vertex of $R$ and such an image can be found in
  polynomial time.
\end{lem}

In fact, \cite[(5.3)]{gm13} gives a better constant $t(r)=O(r)$. For
completeness, we give a short and self-contained proof of
Lemma~\ref{lem:rs} here (which achieves only $r(r)=O(r^2)$). We need the following lemma first:
\begin{lem}\label{lem:rspaths}
  Let $G$ be a graph and $R\subseteq V(G)$. Let $\ell>|R|$, and let
$(B_i)_{i\in[\ell]}$ be an image of a
  $K_\ell$-minor of $G$. Suppose that there is no separation
  $(G_1,G_2)$ of $G$ of order $<|R|$ with $R\subseteq V(G_1)$ and
  $B_b\cap V(G_1)=\emptyset$ for some $b\in[\ell]$. Then there are
  pairwise vertex-disjoint paths $P_1$, $\dots$, $P_{|R|}$ such that
  the following conditions are satisfied.
\begin{enumerate}
\item The first endpoint of each $P_i$ is a vertex of $R$.
\item The second endpoints of the $P_i$'s are in distinct branch sets $B_i$.
\item The $|R|$ paths intersect exactly $|R|$ branch sets.
\end{enumerate}
\end{lem}
\begin{proof}
  Let $b_i$ be an arbitrary vertex of $B_i$. By Menger's Theorem, we
  get that either there are $|R|$ vertex-disjoint paths $P_1$,
  $\dots$, $P_{|R|}$ such that $P_i$ connects $R$ and $b_i$, or there
  is a separation $(G_1,G_2)$ of order $<|R|$ such that $R\subseteq
  V(G_1)$ and every $b_i$ with $1\le i \le |R|$ is
  in $G_2$. Now $|V(G_1)\cap V(G_2)|<|R|$ means that there is a
  $B_i$ with $1\le i \le |R|$ that is disjoint from $V(G_1)\cap V(G_2)$. However, since
  $B_i$ is connected and $b_i\in V(G_2)$, this would imply that
  $B_i$ is fully contained in $V(G_2)\setminus V(G_1)$, contradicting
  the assumption of the lemma.  Therefore, there exists paths $P_1$,
  $\dots$, $P_{|R|}$ that satisfy the first two conditions. In order
  to ensure that the third condition holds, pick a path $P_i$ and, if
  possible, shorten it such that its new second endpoint is in a
  branch set different from where the second endpoints of the other
  paths are. If it is not possible to shorten any of the paths this
  way, then no path visits a branch set other than where the endpoints
  of the other paths are. In other words, the paths visit exactly
  $|R|$ branch sets.
\end{proof}

\begin{proof}[Proof of Lemma~\ref{lem:rs}]
  Let $t(r)=r^2+r$.  Let $I^0=(B_i)_{i\in[t]}$. For $i=1,\dots,r$, we define
  $B^i$ and $I^i$ as follows. We use Lemma~\ref{lem:rspaths} to obtain
  a collection $\CP_i$ of paths between $R$ and the branch sets in
  $I^{i-1}$. Let $B^i$ be the set of branch sets that the paths in
  $\CP_i$ intersect and let $I^{i}$ be obtained by removing $B^i$ from
  $I^{i-1}$. As $t\ge t(r)=r^2+r$, $I^{r}$ is well defined and has at least $r$
  branch sets.

  Let $G'$ obtained from $G$ by contracting each branch set of $I^{r}$
  into a single vertex. Let $v_1$, $\dots$, $v_{r}$ be contracted versions
  of the $r$ branch sets of $I^{r}$. Suppose that there are vertex-disjoint paths $P_1$, $\dots$,
  $P_{r}$ in $G'$ such that $P_i$ connects $R$ and $v_i$. Let $S_i$
  be the union of $V(P_i)$ and the uncontracted version of $v_i$ in
  $G$. Clearly, the $S_i$'s are vertex-disjoint and they form the
  required $K_{r}$-minor image.

  Suppose that there are no $r$ vertex-disjoint paths connecting $R$ and
  $\{v_1,\dots,v_{r}\}$ in $G'$. By Menger's Theorem, this means
  that $G'$ has a separation $(G'_1,G'_2)$ of order $<r$ such that
  $R\subseteq V(G'_1)$ and $\{v_1,\dots,v_{r}\}\subseteq
  V(G'_2)$. Therefore, there is a $v_j$ in $V(G'_2)\setminus V(G'_1)$.
  For every $1\le i \le r$, the branch sets in $B^i$ and the paths
  in $\CP_i$ are disjoint from the branch sets in $I^{r}$, hence
  these sets and paths still appear in the contracted graph $G'$.  As the
  sets $B^1$, $\dots$, $B^{r}$ are disjoint, there is a $1\le i \le
  r$ such that $V(G'_1)\cap V(G'_2)$ is disjoint from the $r$ branch
  sets in $B^i$. This means that these branch sets are fully contained
  in $V(G'_2)\setminus V(G'_1)$: they cannot be contained in
  $V(G'_1)\setminus V(G'_2)$, since they are adjacent to $v_j\in V(G'_2)\setminus V(G'_1)$. Therefore,
  each of the $r$ vertex-disjoint paths in $\CP^i$ has to go through
  $V(G'_1)\cap V(G'_2)$, which is of size $<r$, a contradiction.
\end{proof}

\subsection{Star decomposition with clique-minor free center}
\label{sec:cliquestar}
We prove Lemma~\ref{lem:conncliquestar} in this section. First we
prove a variant of the lemma stated in terms of tangles
(Lemma~\ref{lem:tanglecliquestar}) and then deduce
Lemma~\ref{lem:conncliquestar} it at the end of the section.

Recall that a separation $(A,B)$ removes a set $X$ if $X\subseteq
V(A)\setminus V(B)$.  We say that a separation $(A,B)$ {\em removes}
the $H$-minor image $I=(I_w)_{w\in V(H)}$ if it removes one of the
branch sets, that is, $I_w\subseteq V(A)\setminus V(B)$ for some $w\in
V(H)$. A tangle $\KT$ in $G$ \emph{removes an $H$-minor image $I$} if
$I$ is removed by some $(A,B)\in\KT$ with order $<|H|$.  The following
lemma is analogous to Lemma~\ref{lem:uniqueminsep}: for every clique
minor, there is a unique ``closest minimum separation'' that removes
it.
\begin{lem}\label{lem:uniqueminremove}
  Let $\KT$ be a tangle of order $m$ in a graph $G$ and let $e>2m$.  For every image
  $I$ of $K_e$ in $G$ removed by $\KT$, there is a unique
  $W(I)\subseteq V(G)$ such that
\begin{enumerate}
\item $\separg(W(I))$ removes $I$,
\item $\separg(W(I))\in\KT$,
\item among sets satisfying (1) and (2), the order of $\separg(W(I))$ is minimum possible, and
\item among sets satisfying (1)--(3), $|W(I)|$ is minimum possible.
\end{enumerate}
Furthermore, $G[W(I)]$ is connected and there is a polynomial-time algorithm that, given $G$,
$m$, $I$, and a min cut oracle for $\KT$, either finds $W(I)$ or
concludes that $\KT$ does not remove $I$.
\end{lem}
\begin{proof}
  As $\KT$ removes $I$, there has to be at least one separation
  $(A,B)\in \KT$ that removes $I$. Thus the set $W=V(A)\setminus V(B)$ is one
  such set. To prove the uniqueness, suppose that there are two
  distinct minimal sets $X$ and $Y$.  By Lemma~\ref{lem:subandposimodular}(1),
  either $\boundary(X\cap Y)\le \boundary(X)$ or $\boundary(X\cup
  Y)<\boundary(Y)$.

  Suppose first that $\boundary(X\cap Y)\le \boundary(X)<m$. By
  Lemma~\ref{lem:tanglesepprop}(2), $\separg(X\cap Y)\in \KT$.  We
  claim that $\separg(X\cap Y)$ removes $I$. As both $\separg(X)$ and
  $\separg(Y)$ remove $I$, there are vertices $x,y\in V(K_e)$ such
  that $V(I_x)\subseteq X$ and $V(I_y)\subseteq Y$. Since
  $\boundary(X),\boundary(Y)<m$ and $e>2m$, there is a vertex $z\in
  K_e$ such that $V(I_z)$ is disjoint from $N^G(X)\cup N^G(Y)$. As
  $K_e$ is a clique, a vertex of $V(I_z)$ has to be adjacent to
  $V(I_x)\subseteq X$, which is only possible if this vertex is also
  in $X$ (since it cannot be in $N^G(X)$). It follows that $V(I_z)$ is
  fully contained in $X$. A symmetrical argument shows that
  $V(I_z)\subseteq Y$. Thus $V(I_z)\subseteq X\cap Y$, i.e.,
  $\separg(X\cap Y)$ removes $I$. Therefore, $X\cap Y\subset X$ and
  $\boundary(X\cap Y)\le \boundary(X)$ contradicts the minimality of
  $X$.

  Suppose now that $\boundary(X\cup Y)<\boundary(Y)<m$. By
  Lemma~\ref{lem:tanglesepprop}(3), $\separg(X\cup Y)\in
  \KT$. Clearly, $\separg(X\cup Y)$ removes $I$ (as any branch set
  contained in $X$ or $Y$ is also contained in $X\cup Y$). Therefore,
  $S_G(X\cup Y)$ contradicts the minimality of $S_G(Y)$.

  To check if an image $I$ is removed by $\KT$, we use the algorithm
  of Lemma~\ref{lem:uniqueminsep} to compute the set $W(I_v)$ for
  every $v\in V(K_e)$ (if such a set exists).  If $\KT$ removes $I$,
  then at least one of these sets should exist. Furthermore, if $\KT$
  removes $I$, then it should be clear that $W(I)$ is equal to one of
  these sets $W(I_v)$: if $W(I)$ contains $I_v$, then it cannot be
  different from $W(I_v)$ (as it would contradict the minimality and
  uniqueness of either $W(I)$ or $W(I_v)$). As $W(I_v)$ is connected
  by Prop.~\ref{prop:connectedmin}, it follows that $W(I)$ is
  connected as well.
\end{proof}

A simple uncrossing argument shows that the minimum separations
defined in Lemma~\ref{lem:uniqueminremove} cannot properly intersect
one other:
\begin{lem}\label{lem:cliqueuncross}
  Let $\KT$ be a tangle of order $m$ in a graph $G$ and let
  $e>2m$. Let $I^x$ and $I^y$ be two $K_e$-minor images removed by
  $\KT$. Then either
\begin{enumerate}
\item $W(I^x)\subseteq W(I^y)$,
\item $W(I^x)\supseteq
  W(I^y)$, or 
\item $W(I^x)$ and $W(I^y)$ are disjoint and do not touch.
\end{enumerate}
\end{lem}
\begin{proof}
  Let $X:=W(I^x)$ and $Y:=W(I^y)$ and suppose that none of the three
  possibilities hold. Assume first that $I^x$ has a branch set fully
  contained in $X\cap Y\subset X$. If $\boundary(X\cap Y)\le \boundary(X)<m$, then $\separg(X\cap Y)\in\KT$ by
  Lemma~\ref{lem:tanglesepprop}(2) and $\separg(X\cap Y)$ removes $I^x$, contradicting the 
  minimality of $W(I^x)$. Thus we can assume that $\boundary(X\cap Y)>\boundary(X)$. By
  Lemma~\ref{lem:subandposimodular}, it follows that $\boundary(X\cup
  Y)<\boundary(Y)<m$. Therefore, $\separg(X\cup Y)$ is in $\KT$ by
  Lemma~\ref{lem:tanglesepprop}(3) and it clearly removes $I^y$ (since
  $\separg(Y)$ already does), contradicting the minimality of
  $Y=W(I^y)$.

We have proved that $I^x$ has no branch set fully contained in
  $X\cap Y$, and a symmetrical argument shows that $I^y$ has no such branch set either.
 By
  Lemma~\ref{lem:subandposimodular}(2), either $\boundary(X)\ge
  \boundary(X\setminus N[Y])$ or $\boundary(Y)\ge \boundary(Y\setminus
  N[X])$. Assume without loss of generality the first
  case.  Consider a branch set $I^x_1$ of $I^x$ fully contained in
  $X$ (such a set exists, as $\separg(X)$ removes $I^x$) and a branch
  set $I^x_2$ disjoint from $N^G(X)\cup N^G(Y)$ (since $e>2m$, there has to
  be such a set). The branch set $I^x_2$ has a vertex adjacent to $I^x_1\subseteq
  X$. Since $I^x_2$ is disjoint from $N^G(X)$, this is only possible if
  $I^x_2$ is fully contained in $X$. Moreover, we assumed that $I^x_2$ is disjoint
  from $N^G(Y)$ and it is not fully contained in $X\cap Y$, thus $I^x_2$
  is fully contained in $X\setminus N^G[Y]$, that is, the separation $\separg(X\setminus N^G[Y])$
  removes $I^x$.  Note that $X\setminus N^G[Y]$ is a proper subset of
  $X$, otherwise $X$ and $Y$ are disjoint and do not touch.
  Lemma~\ref{lem:tanglesepprop}(1) implies that $\separg(X\setminus
  N[Y])\in \KT$, and therefore $X\setminus N[Y]\subset X$ violates the
  minimality of $X=W(I^x)$.
\end{proof}

Another useful property of the definition of minimum separation in
Lemma~\ref{lem:uniqueminremove} is that if $\separg(W(I))=(A,B)$, then the
clique minor $I$ allows us to connect vertices of $V(A)\cap V(B)$ with
each other using paths in $A$ in an arbitrary way. We use the following definition to
state this property:
\begin{defn}
  We say that a separation $(A,B)$ of order $m$ is {\em generic} if
  there is a $K_m$-minor image in $A$ such that each branch set contains
  exactly one vertex of $V(A)\cap V(B)$.  Such an image is called a
  {\em witness}.
\end{defn}

\begin{lem}\label{lem:sepgeneric}
  Let $\KT$ be a tangle of order $m$ in a graph $G$ and let
  $e>t(m)+m$
  for the function $t$ of Lemma~\ref{lem:rs}.  For every image $I$ of
  $K_e$ in $G$ removed by $\KT$, the separation $\separg(W(I))$ is
  generic. Furthermore, given $I$ and a min cut oracle for $\KT$, a
  witness can be found in polynomial time.
\end{lem}
\begin{proof}
  Let $\separg(W(I))=(A,B)$ and $R=V(A)\cap V(B)$. By definition,
  $(A,B)$ removes $I$, thus at least one branch set of $I$ is
  contained in $V(A)\setminus V(B)$ and at most $|R|<m$ branch sets intersect
  $R$. Thus at least $t(m)$ branch sets are fully contained in
  $V(A)\setminus V(B)$. Therefore, $A$ contains a $K_{t(m)}$-minor image $I'$. We
  verify that the conditions of Lemma~\ref{lem:rs} hold for graph $A$
  and set $R$. Suppose that there is a separation $(G_1,G_2)$ of order
  $<|R|$ with $R\subseteq G_1$ and $I'_w\subseteq V(G_2)\setminus
  V(G_1)$ for some branch set $I'_w$ of $I'$ (which is also a branch set of $I$).  Let $X'=V(G_2)\setminus
  V(G_1)\subset V(A)\setminus V(B)=W(I)$. It follows that
  $\separg(X')$ has order $<|R|$ (which is the order of $(A,B)$) and
  is in $\KT$ by Lemma~\ref{lem:tanglesepprop}(1). However, $\separg(X')$
  also removes $I$, contradicting the minimality of $W(I)$.  We can
  conclude that $A$ and $R$ satisfy the conditions of
  Lemma~\ref{lem:rs}, and the existence of the required $K_{|R|}$-minor image follows.
\end{proof}

It follows from Lemma~\ref{lem:sepgeneric} that if $W(I)=(A,B)$, then
removing $V(A)\setminus V(B)$ and replacing $V(A)\cap V(B)$ with the
clique $K[V(A)\cap V(B)]$ does not create any new clique minor images in $B$
(because the edges in the clique $K[V(A)\cap V(B)]$ can be simulated
by connections in $A$ in the original graph). Repeated application of
this observation shows that after removing all the clique minor
images, we get a bag whose torso does not contain clique minors of the
given size.
\begin{lem}\label{lem:separcliquefind}
  Let $\KT$ be a tangle of order $m$ in a graph $G$ and let $e>t(m)+m$
  for the function $t$ of Lemma~\ref{lem:rs}. Let $I^1$, $\dots$,
  $I^p$ be $K_e$-minor images removed by $\KT$. Let $W=\bigcup_{i=1}^p
  W(I^i)$ and let $G'=\mytorso(G,V(G)\setminus W)$. The graph $G'$ has
  a $K_e$-minor $I'$ if and only if $G$ has a $K_e$-minor image $I$
  not removed by any $\separg(W(I^i))$. Furthermore, given a min cut
  oracle for $\KT$ and such a $K_e$-minor image $I$, one can compute a $K_e$-minor image $I'$ in $G'$
  in polynomial time and vice versa.
\end{lem}
\begin{proof}
  We can assume that the sets $W(I^1)$, $\dots$, $W(I^m)$ are pairwise
  incomparable (because if $W(I^i)\subseteq W(I^j)$, then omitting
  $I^i$ from this collection does not change $W$), thus by
  Lemma~\ref{lem:cliqueuncross}, we can assume that these sets are
  pairwise disjoint and do not touch.  This means that
  $R_i=N^G(W(I^i))$ is a subset of $V(G)\setminus W$ and induces a
  clique in $G'$.  By Lemma~\ref{lem:uniqueminremove}, each
  $G[W(I^i)]$ is connected. Thus $G'=\mytorso(G,V(G)\setminus W)$ is
  exactly the union of $G\setminus W$ with a clique on each
  $R_i$.

  Let $I'$ be the image of a $K_e$-minor in $G'$. Note that this is
  not necessarily a $K_e$-minor image in $G\setminus W$ as $G'$ has
  edges that $G\setminus W$ does not have. However, we can use the
  subgraph inside $G[W(I^i)]$ to simulate these edges. By
  Lemma~\ref{lem:sepgeneric}, every $\separg(W(I^i))$ is generic and
  we can obtain the corresponding clique minor images.  This means
  that for each $R_i$, there is a set of $r$ pairwise disjoint and
  touching connected subgraphs in $G[N^G[W(I^i)]]$. Using these
  connected sets, we can extend each $I'_w$ of $G'$ into a connected
  set $I_w$ of $G$ and obtain a $K_e$-minor image $I$ in $G$.

  For the reverse direction, let $I$ be a $K_e$-minor image in $G$ not
  removed by any $\separg(W(I^i))$. Let $I'$ be defined by
  $I'_w=G'[V(I_w)\setminus W]$ for every $w\in V(K_e)$. Note that
  $V(I'_w)\neq\emptyset$: this would be only possible if
  $V(I_w)\subseteq W(I^i)$ for some $1\le i \le p$, which would imply
  that $\separg(W(I^i))$ removes $I$. We claim that $I'$ is a
  $K_e$-minor image. The connectedness of $I'_w$ is easy to see: any
  path with internal vertices in $W(I^i)$ can be replaced by an edge
  in $R_i$ (as $R_i$ induces a clique in $G'$). To see that $I'_w$ and
  $I'_u$ touch for every $w,u\in V(K_e)$, consider an edge $e$ between
  $I_w$ and $I_u$ in $G$. If both endpoints of $e$ are in $W(I^i)\cup
  R_i$, then $I'_w$ and $I'_u$ both intersect $R_i$, hence they
  touch. Otherwise, $e$ is an edge of $G\setminus W$, implying that it
  is also an edge of $G'$.
\end{proof}

Now we state and prove a version of Lemma~\ref{lem:conncliquestar} in
terms of tangles. The following lemma is closely related to Theorem
(3.1) of \cite{gm16}, but in addition is algorithmic and has an invariance
statement that we need for our isomorphism test later.

\begin{lem}\label{lem:tanglecliquestar}
For every $\ell,m\in \NN$, there is a constant $e'(\ell,m)$ such that the following holds.
  There is an $f(\ell,m)\cdot |V(G)|^{O(1)}$ time algorithm that, given
  a graph $G$, $\ell$, $m$,  a min cut oracle for a tangle $\KT$ of order $m$,  either
\begin{enumerate}
\item finds a $K_\ell$-minor image $I$ not removed by $\KT$, or
\item computes a $\KT$-respecting star decomposition $\Sigma_\KT=(T_\KT,\sep_\KT,\comp_\KT)$ with center $s$ such
  that $\tau_\KT(s)$ does not contain a $K_{e'(\ell,m)}$-minor.
\end{enumerate}

Furthermore, if the algorithm returns $\Sigma_{\KT}$ for $(G,\KT)$ and
$\KT'$ is another tangle of order $m$ in a graph $G'$, and $f$ is an isomorphism from
$(G,\KT)$ to $(G',\KT')$, then the algorithm returns a star
decomposition $\Sigma_{\KT'}$ for $(G',\KT')$ such that there is an
isomorphism $g$ from $T_{\KT}$ to $T_{\KT'}$ such that for all $t\in
V(T_\KT)$ we have $\sep_{\KT'}(g(t))=f(\sep_{\KT}(t))$ and
$\comp_{\KT'}(g(t))=f(\comp_{\KT}(t))$.
\end{lem}
\begin{proof}
  Let $e=e'(\ell,m)=\max(\ell,t(m)+m+1)$ for the function $t$ in
  Lemma~\ref{lem:rs}.  We show first that if $\KT$ removes every
  $K_\ell$-minor image (and therefore every $K_e$-minor image as $e\ge \ell$), then
  there exists a star decomposition satisfying the requirements.
  Suppose that $\KT$ removes every $K_\ell$-minor image, implying that
  $W(I)$ from Lemma~\ref{lem:uniqueminremove} is defined for every $K_e$-minor image $I$.
Let $\CW$ contain the inclusionwise-maximal sets in $\{W(I) \mid \text{$I$ is a $K_e$-minor image}\}$.
 Let us pick a representative $K_e$-minor image for every $W\in \CW$: let $p=|\CW|$ and let $I^1$,
  $\dots$, $I^p$ be a list of $K_e$-minor images such that $\{W(I^i)\mid 1\le i \le p\}=\CW$ (this implies that $W(I^i)\neq W(I^j)$ for $i\neq j$).  By Lemma~\ref{lem:cliqueuncross},
  $W(I^i)$ and $W(I^j)$ do not touch for $i\neq
  j$. Let $W=\bigcup_{i=1}^p W(I^i)$. We construct a star
  decomposition $\Sigma_\KT=(T_\KT,\sep_\KT,\comp_\KT)$ with center
  $s$ and $p$ tips $t_i$ ($1\le i \le p$).  We set
  $\comp_\KT(s)=V(G)$, $\sep_\KT(s)=\emptyset$,
   $\comp_\KT(t_i)=W(I^i)$, and $\sep_\KT(t_i)=N^G(W(I^i))$.

It easy to verify that $\Sigma_\KT$ is a tree decomposition:
\begin{claim}\label{claim:topcl1}
$\Sigma_\KT$  satisfies properties \ref{li:t1}--\ref{li:t5}.
\end{claim}
The definition of $W(I^i)$ implies that $\separg(W(I^i))\in \KT$ for every $1\le i \le p$. Therefore,
\begin{claim}[resume]\label{claim:topcl5}
$\Sigma_\KT$ respects $\KT$.
\end{claim}

\begin{claim}[resume]\label{claim:topcl2}
$G'=\tau(s)=\mytorso(G,V(G)\setminus W)$ does not
  contain a $K_e$.
\proof
If $G'$ contains a $K_e$-minor, then
  Lemma~\ref{lem:separcliquefind} implies that there is a $K_e$-minor
  image $I$ in $G$ not removed by any of the separations
  $\separg(W(I^i))$.  However, this contradicts the assumption that
  $I^i$, $\dots$, $I^p$ is the list of all images for which $W(I^i)$
  is inclusionwise maximal. \footnote{Note to typesetter: we are following here the usual convention that the $\qedsymbol$ symbol marks the end of a proof and the $\lrcorner$ symbol marks the end of the proof of a claim embedded in a longer proof. In our experience, this significantly improves the readability of longer proofs.}
\uend
\end{claim}

Algorithmically, we can find the set $W$ defined above as follows. We
construct collections $\CI^{(0)}\subset \CI^{(1)} \subset \dots$ of
$K_e$-minor images, each of which is removed by $\KT$. We start with
$\CI^{(0)}=\emptyset$. Given $\CI^{(j)}$, we construct $\CI^{(j+1)}$ as
follows. Let $W^{(j)}=\bigcup_{I\in \CI^{(j)}}W(I)$ and
$G^{(j)}=\mytorso(G,V(G)\setminus W^{(j)})$. We test if $G^{(j)}$ has
a $K_e$-minor (using the algorithm of Theorem~\ref{th:minortest}). By
Lemma~\ref{lem:separcliquefind}, if there is a $K_e$-minor model $I'$
in $G^{(j)}$, then there is a corresponding $K_e$-minor model
$I^{(j)}$ of $G$ which is not removed by $\separg(W(I))$ for any $I\in
\CI^{(j)}$. Let us use the algorithm of
Lemma~\ref{lem:uniqueminremove} to compute the set $W(I^{(j)})$. If
the algorithm returns that $W(I^{(j)})$ is not defined, that is,
$I^{(j)}$ is not removed by $\KT$, then we can stop and return
$I^{(j)}$ (or more precisely, as $e\ge \ell$, a restriction of
$I^{(j)}$ to a $K_\ell$-minor) and we are done. Otherwise, let us
obtain $\CI^{(j+1)}$ from $\CI^{(j)}$ by inserting $I^{(j)}$. Let us
observe that $W(I^{(j)})\not\subseteq W^{(j)}$: by
Lemma~\ref{lem:cliqueuncross}, $W(I^{(j)})\subseteq W^{(j)}$ is only possible if
$W(I^{(j)})\subseteq W(I)$ for some $I\in \CI^{(j)}$, but this means
that $\separg(W(I))$ already removes $I^{(j)}$, a contradiction. It follows
that $W^{(j)}\subset W^{(j+1)}$. After including $I^{(j)}$ into
$\CI^{(j+1)}$, we repeat this procedure until we arrive to a $j$ such
that $G^{(j)}$ has no $K_e$-minor.

  As the size of $W^{(j)}$ strictly increases in each step, the
  process described above stops in at most $|V(G)|$ steps with a
  $G^{(j)}$ that does not contain a $K_e$-minor. 

\begin{claim}[resume]\label{claim:topcl3}
  $W^{(j)}=W$.
\proof
Suppose that $W^{(j)}\neq W$, i.e., there is an image
  $I^*$ such that $W(I^*)\not\subseteq W^{(j)}$. Since $G^{(j)}$ has
  no $K_e$-minor, by Lemma~\ref{lem:separcliquefind}, there is an
  $I\in \CI^{(j)}$ such that $\separg(W(I))$ removes $I^*$. As
  $\separg(W(I))$ and $\separg(W(I^*))$ both remove $I^*$, the sets
  $W(I)$ and $W(I^*)$ both contain a branch set of $I^*$, hence it is
  not possible that the two sets are disjoint and do not
  touch. Therefore, by Lemma~\ref{lem:cliqueuncross}, one of the two
  sets is contained in the other. From $W(I^*)\not\subseteq W^{(j)}$,
  we know that $W(I^*)\subseteq W(I)$ is not possible, hence we have
  $W(I)\subset W(I^*)$, implying that $\separg(W(I^*))$ removes $I$ as
  well. Now $\boundary(W(I^*))<\boundary(W(I))$ would contradict the
  minimality of $W(I)$ and $\boundary(W(I^*))\ge \boundary(W(I))$
  would contradict the minimality of $W(I^*)$ (as $|W(I)|<|W(I^*)|$). Thus we have proved
  that $W^{(j)}$ obtained by this procedure is indeed the set $W$ defined at the beginning of the
  proof.
\uend
\end{claim}

What remains to be proven is the invariance condition.  Suppose that
$\KT'$ is another tangle of order $m$ in a graph $G'$. Let $f$ be an
isomorphism from $(G,\KT)$ to $(G',\KT')$.  Let $I=(I_v)_{v\in
  V(K_\ell)}$ be a $K_e$-minor image in $G$ and let
$I'=(f(I_v))_{v\in V(K_e)}$ be the corresponding $K_e$-minor
image in $G'$.  Let $W(I)$ and $W'(I')$ be the set given by
Lemma~\ref{lem:uniqueminremove} on $I$ and $I'$, respectively.
\begin{claim}[resume]\label{claim:topcl4}
$W'(I')=f(W(I))$.
\proof The definition of the
set $W(I)$ depends only on the branch sets of $I$, the tangle $\KT$ and the graph-theoretical
properties of $G$ (size of the boundaries of certain sets etc.) and
all these properties are preserved by $f$.  \uend
\end{claim}

Therefore, if $\{W(I^1),\dots, W(I^p)\}$ is the collection of
inclusionwise maximal sets appearing in the definition of $W$ for
$(G,\KT)$, then exactly $\{f(W(I^1)),\dots, f(W(I^p))\}$ is the
collection of sets appearing in the definition of $W'$. If follows
that for every $t_i$, there is a $g(t_i)$ such that
$\comp_{\KT}(t_i)=W(I^i)$ and
$\comp_{\KT'}(g(t_i))=f(W(I^i))$. Moreover,
$\sep_{\KT}(t_i)=N^G(W(I^i))$ and
$\sep_{\KT'}(g(t_i))=N^{G'}(f(W(I^i)))=f(N^{G}(W(I^i)))$ follows, as
required. Setting $g(s)=s'$ (where $s'$ is the center of the
decomposition of $G'$) completes the definition of $g$.
\end{proof}

Finally, we can prove Lemma~\ref{lem:conncliquestar} by invoking
Lemma~\ref{lem:tanglecliquestar} on the tangle defined by the
unbreakable set $X$:
\begin{proof}[Proof of Lemma~\ref{lem:conncliquestar}]
  Let $e^*(\ell,m)=e'(\ell,m)+3m-2$ for the function $e'$ in
  Lemma~\ref{lem:tanglecliquestar}.  Let $\KT$ be the tangle of order
  $m$ defined by the $m$-unbreakable set $X$;
  Lemma~\ref{lem:unbreakabletangle} provides an implementation of the
  min-cut oracle for $\KT$. Let us call the algorithm of
  Lemma~\ref{lem:tanglecliquestar} with $G$, $\KT$, $\ell$, and $m$.
  If it returns a $K_\ell$-minor image $I$ not removed by $\KT$, then
  this is equivalent to saying that $I$ is $m$-attached to $X$. Thus
  we can return $I$ and we are done. Otherwise, the algorithm of
  Lemma~\ref{lem:tanglecliquestar} returns a $\KT$-respecting star
  decomposition $\Sigma_\KT=(T_\KT,\sep_\KT,\comp_\KT)$ of $G$.  The
  star decomposition $\Sigma_\KT$ almost satisfies the requirements of
  Lemma~\ref{lem:conncliquestar}, except that $X$ is not necessarily
  contained in $\bag_\KT(s)$ for the center $s$. To move $X$ to the
  center, we construct a star decomposition a star decomposition
  $\Sigma_X=(T_X,\sep_X,\comp_X)$ as follows. First, let $T_X=T_\KT$
  and for the center $s$ of $T_X$, let $\comp_X(s)=V(G)$ and
  $\sep_X(s)=\emptyset$. For every tip $t$ of $T_X$, we let
  $\comp_X(t)=\comp_\KT(t)\setminus X$ and $\sep_X(t)=\sep_\KT(t)\cup
  (X\cap \comp_{\KT}(t))$. It is straightforward to verify that
  $\Sigma_X$ is also a star decomposition of $G$, and in fact it is
  star decomposition even for the supergraph $G\cup K[X]$ (since
  $X\subseteq \beta_X(s)$). Note that $\torso_\KT(s)\setminus
  X=\torso_X(s)\setminus X$ (because the two bags differ only in the
  vertices of $X$ and all the extra edges of $G\cup K[X]$ are incident
  to $X$). As $\torso_{\KT}(s)$ has no $K_{e'(\ell,m)}$-minor, this
  means that $\torso_X(s)$ cannot have a clique minor of order
  $e'(\ell,m)+|X|=e^*(\ell,m)$, as required.  Furthermore, as
  $\Sigma_\KT$ is $\KT$-respecting, it follows that
  $|\sep_{\KT}(t)|<m$ and $\separg(\comp_{\KT}(t))\in\KT$ for every
  tip $t$. By Lemma~\ref{lem:unbreakabletangle}, this also means that
  $|\comp_\KT(t)\cap X|\le m-1$ and therefore $|\sep_X(t)|\le
  m-1+m-1<|X|$. Thus the adhesion of $\Sigma_X$ is less than $|X|$, as
  required.\footnote{This is the point (and the analogous argument in
    the proof of Lemma~\ref{lem:conndegstar}) where it becomes motived
    why we used the tangle $\KT$ defined by the unbreakable set $X$.
    If we have no bound on $|\comp_\KT(t)\cap X|$, then moving $X$ to
    the center can increase the adhesion by up to $|X|=3m-2$, which
    means that the bound on the adhesion would be larger than
    $|X|$. Therefore, the repeated application of this lemma in the
    proof of Global Structure Theorem~\ref{theo:structure} would
    increase the adhesion in each step. In all the arguments in the
    section, we were careful enough to use only separations that are
    in $\KT$, and therefore we have the bound that the component of
    each child of $t$ contains at most $m-1$ vertices of $X$. } The
  invariance condition follows easily from the invariance condition of
  Lemma~\ref{lem:tanglecliquestar}: if $f$ is an isomorphism from $G$
  to $G'$ with $f(X)=X'$ and $\KT$ and $\KT'$ are the tangles defined
  by the unbreakable sets $X$ and $X'$, respectively, then $f$ is an
  isomorphism from $(G,\KT)$ to $(G',\KT')$.
\end{proof}

\subsection{Star decomposition with a bounded-degree center}
\label{sec:degstar}

The proof of Lemma~\ref{lem:conndegstar} has the same high-level
strategy as the proof of Lemma~\ref{lem:conncliquestar} in
Section~\ref{sec:cliquestar}: we identify those parts of the graph
that we want to exclude from the bag of the center (this time, the
high-degree vertices) and we use an uncrossing argument to show that
all of them can be removed more or less independently from each
other. The uncrossing argument is somewhat more involved due to the
technicality that a high-degree vertex can be part of the separator
removing some other high-degree vertex.



First we need the following lemma, which shows that all but at most
$k$ high-degree vertices can be removed by separations in the tangle,
or we can find a $K_k$-subdivision. Recall from
Lemma~\ref{lem:uniqueminsep} that if we have tangle $\KT$ in a graph
and a vertex $v$ such that $v\in V(A\setminus B)$ for some separation
$(A,B)\in\KT$, then $W(\{v\})$ is the ``closest minimum separation''
that removes $v$ from the tangle. If there is no $(A,B)\in\KT$ such
that $v\in V(A\setminus B)$ then $W(\{v\})$ is undefined. The set $Z$
in the following lemma consist of all vertices that cannot be removed
from the tangle $\KT$ by a separation of order less than $k(k-1)$.

\begin{lem}\label{lem:tangledeg2}
  For every $k\in\NN$, there is a constant $\ell'(k)$ such that the
  following holds.  For a graph $G$, integer $k\in\NN$, tangle $\KT$
  of order at least $k(k-1)$, and an image $I$ of $K_{\ell'(k)}$ not
  removed by $\KT$, we define $Z$ to be the set of all vertices $v\in
  V(G)$ such that $v$ has
  degree at least $k$ and either $W(\{v\})$ is undefined or
  $\boundary(W(\{v\}))\ge k(k-1)$. If $|Z|\ge k$, then given $G$, $k$,
  a min-cut oracle for $\KT$, and $I$, a subdivision of $K_k$ in $G$
  can be found in polynomial time.
\end{lem}
\begin{proof}
  Let $\ell=\ell'(k)=t(k(k-1))$ for the function $t$ in Lemma~\ref{lem:rs}.
  We show that if $|Z|\ge k$, then we can find a subdivision of $K_k$
  in $G$. Let $Z_0$ be a subset of $Z$ of size exactly $k$.  Let $G'$
  be the graph obtained from $G$ by extending each vertex $z\in Z_0$
  into a clique $K_z$ of $k-1$ vertices: for every $z\in Z_0$, we
  introduce $k-2$ new vertices that are adjacent to each other, to
  vertex $z$, and to every neighbor of $z$. The clique $K_z$ contains
  $z$ and these $k-2$ new vertices.  Let $R:=\bigcup_{z\in Z_0}K_z$.

  Let $I_1$, $\dots$, $I_\ell$ be the branch sets in the $K_\ell$
  minor image $I$.  Let us show first that the conditions of
  Lemma~\ref{lem:rs} hold for $R$ in $G'$.  Suppose for contradiction
  that $(A',B')$ is a separation of $G'$ of order less than
  $|R|=k(k-1)$ with $R\subseteq V(A')$ and $I_{b}\subseteq
  V(B')\setminus V(A')$ for some $b\in[\ell]$.  Let $Q':=V(A')\cap
  V(B')$ be the separator.  Without loss of generality, we may assume
  that for all $z\in Z_0$, either $K_z\cap Q'=\emptyset$ or
  $K_z\subseteq Q'$.  Let $A:=A'\setminus(R\setminus Z_0)$ and
  $B:=B'\setminus(R\setminus Z_0)$ (i.e., we remove from $Q'$ the
  extra vertices that were introduced in the definition of $G'$). Then
  $(A,B)$ is a separation of $G$; let $Q=V(A)\cap V(B)$ be the
  separator. Now it is clear that $|Q|\le |Q'|<k(k-1)$. Furthermore,
  there has to be a vertex $z\in Z_0$ which is not in $Q$: otherwise,
  $Z_0\subseteq Q$ implies that the size of $Q'$ in $G'$ is at least
  $k(k-1)$. Therefore, $(A,B)$ is a separation of order $<k(k-1)$ with
  $z\in V(A)\setminus V(B)$. This separation is in $\KT$: otherwise,
  $(B,A)\in \KT$ by \ref{li:ta1} (here we use that the order of $\KT$
  is at least $k(k-1)$) and $I_b\subseteq V(B)\setminus V(A)$ means that $\KT$
  removes $I$, contradicting our assumption on $I$. It follows that
  $(A,B)\in \KT$ is a separation of order $<k(k-1)$ with $z\in
  V(A)\setminus V(B)$, contradicting $z\in Z$ and the definition of
  $Z$.  Thus we can conclude that there is no such separation
  $(G'_1,G'_2)$ of $G'$, and the conditions of Lemma~\ref{lem:rs} hold
  for $Z$ and $G'$.

  Lemma~\ref{lem:rs} gives us a $K_{k(k-1)}$-minor image in $G'$, that is, for
  every $q\in R$, a connected set $I_q$ such that these sets are pairwise
  disjoint and touch.  Consider a partition of $R$ into $\binom{k}{2}$
  classes, each of size 2, such that for every pair $z_1,z_2\in Z_0$
  of distinct vertices, there is a class of the partition containing a
  vertex of $\hat z_1\in K_{z_1}$ and a vertex of $\hat z_2 \in
  K_{z_2}$.  (As the size of each $K_{z}$ is exactly $k-1$, such a
  partition is possible.)  We define a path $P'_{\{z_1,z_2\}}\subseteq
  I_{\hat z_1}\cup I_{\hat z_2}$ connecting $\hat z_1$ and $\hat z_2$;
  let $\CP'$ be the collection of these $\binom{k}{2}$ paths.  For
  each such path $P'_{\{z_1,z_2\}}\in \CP'$ of $G'$, there is a
  corresponding path $P_{\{z_1,z_2\}}$ in $G$: whenever
  $P'_{\{z_1,z_2\}}$ contains a vertex of some $K_z$, then we replace
  it by $z$. Let $\CP$ be the collection of these $\binom{k}{2}$ paths
  in $G$. As the paths $\CP'$ are pairwise disjoint, the corresponding
  paths in $\CP$ can intersect only in $Z_0$. Therefore, we have $k$
  vertices $Z_0$ and a collection of $\binom{k}{2}$ internally
  pairwise disjoint paths that connect every pair of vertices in
  $Z_0$. In other words, we have formed a $K_k$ topological minor
  image in $G$, which we can return.
 \end{proof}

 The following lemma is a version of Lemma~\ref{lem:conndegstar} stated in terms
 of tangles:
\begin{lem}\label{lem:tangledegstar}
  For every $k\in\NN$ there are constants $m'(k),d'(k),\ell'(k)$ such that the following holds. There is a
  polynomial-time algorithm that, given a graph $G$, an integer $k$, min cut oracle for a
  tangle $\KT$ of order $m'(k)$, and 
an image $I$ of $K_{\ell'(k)}$ not removed by $\KT$
either
\begin{enumerate}
\item finds a subdivision of $K_k$ in $G$, or
\item  computes a $\KT$-respecting star decomposition
  $\Sigma_{\KT}=(T_{\KT},\sep_{\KT},\comp_{\KT})$ of $G$ with center $s$ such that
  at most $k$ vertices of $\tau_\KT(s)$ have degree more than $d'(k)$.
\end{enumerate}

Furthermore, if the algorithm returns $\Sigma_{\KT}$ for $(G,\KT)$ and
$\KT'$ is another tangle of order $m$ in a graph $G'$, and $f$ is an isomorphism from
$(G,\KT)$ to $(G',\KT')$, then the algorithm returns a star
decomposition $\Sigma_{\KT'}$ for $(G',\KT')$ such that there is an
isomorphism $g$ from $T_{\KT}$ to $T_{\KT'}$ such that for all $t\in
V(T_\KT)$ we have $\sep_{\KT'}(g(t))=f(\sep_{\KT}(t))$ and
$\comp_{\KT'}(g(t))=f(\comp_{\KT}(t))$.
 \end{lem}

\begin{proof}
  Let $\ell'(k)$ be as in Lemma~\ref{lem:tangledeg2}.  We will define
  later (in Claim~\ref{claim:topdeg6}) a constant $a$ depending on
  $k$; let $m'(k)=\max\{k(k-1),a+1\}$. Let $Z$ contain a vertex $v\in
  V(G)$ if $v$ has degree at least $k$ and either $W(\{v\})$ is
  undefined or $\boundary(W(\{v\}))\ge k(k-1)$
   (the algorithm of
  Lemma~\ref{lem:uniqueminsep} can be used to check this condition).
  If $|Z|\ge k$, then we can use the algorithm of
  Lemma~\ref{lem:tangledeg2} to return a subdivision of $K_k$ in $G$,
  and we are done.

  Otherwise, let $L\subseteq V(G)$ be the set of vertices not in $Z$ having
  degree at least $k$.  For every $v\in L$, let us use
  the algorithm of Lemma~\ref{lem:uniqueminsep} to compute the unique
  minimal set $W_v=W(\{v\})$ (as $v\not\in Z$, such a set
  exists). By Prop.~\ref{prop:connectedmin}, $G[W_v]$ is connected.

  Let $\CW$ contain
  the inclusionwise-maximal sets in $\{W_v\mid v\in L\}$; i.e.,
  $W_v\in \CW$ if and only if there is no $u\in L$ with $W_v\subset
  W_u$. Note that we define $\CW$ such that it does not contain duplicate sets.

\begin{claim}\label{claim:topdeg1}
  Every $b\in V(G)$ appears in $O(k^2)$ members of $\CW$.  \proof For
  every $W\in \CW$, let us choose a representative $v\in L$ with
  $W=W_v$; let $M\subseteq L$ be the set of selected
  representatives. We define a directed graph $\overrightarrow H$ on
  $M$ where $\overrightarrow{vu}\in E(\overrightarrow H)$ if and only
  if $u\in N^G(W_v)$. Note that $|N^G(W_v)|< k(k-1)$ implies that the
  outdegree of $v$ is at most $k(k-1)-1$. This further implies that
  the maximum clique size in the undirected graph $H$ underlying
  $\overrightarrow H$ is at most $2k(k-1)-1$: the average degree of
  every subgraph of $H$ is at most $2k(k-1)-2$.

We show that the representatives of the sets in $\CW$ containing $b$
form a clique in $H$, thus by the argument in the previous paragraph,
there can be at most $2k(k-1)-1$ sets in $\CW$ containing $b$.
Consider two distinct vertices $u,v\in M$ with $b\in W_u$ and $b\in
W_v$. We claim that $u$ and $v$ are adjacent in the undirected graph
$H$. Otherwise, $u\not\in N^G(W_v)$ and $v\not\in N^G(W_u)$. We
consider the following cases:
    \begin{cs}
      \case1 $u\in W_u\cap W_v$. By
      Lemma~\ref{lem:subandposimodular}(1), we have two possibilities:
      \begin{enumerate}
      \item $\boundary(W_u\cup
        W_v)< \boundary(W_v)$.  In this case $W_u\cup W_v$ contradicts
        the minimality of $W_v$ (note that by Lemma~\ref{lem:tanglesepprop}(3),
        $\separg(W_u\cup W_v)\in\KT$).
      \item $\boundary(W_u\cup
        W_v)\ge \boundary(W_v)$ and $\boundary(W_u\cap W_v)\le \boundary(W_u)$.  In this case,
        $W_u\cap W_v$ contradicts the minimality of $W_u$ (by
        Lemma~\ref{lem:tanglesepprop}(2), $\separg(W_u\cap
        W_v)\in\KT$).
      \end{enumerate}
      \case2 $v\in W_u\cap W_v$. Similar to case 1.

      \case3 $u\in W_u\setminus W_v$ and $v\in W_v\setminus W_u$. Let
      $W'_u:=W_u\setminus N^G[W_v]$ and $W'_v:=W_v\setminus
      N^G[W_u]$. Note that $b\in W_u\cap W_v$ implies that
      $W'_u\subset W_u$ and $W'_v\subset W_v$. Furthermore, the assumptions $u\in
      W_u\setminus W_v$ and $u\not\in N^G(W_v)$ imply that $u\in
      W'_u$, and we have $v\in W'_v$ in a similar way. By
      Lemma~\ref{lem:subandposimodular}(2), either $\boundary(W'_u)\le
      \boundary(W_u)$ or $\boundary(W'_v)\le \boundary(W_v)$. If, say,
      $\boundary(W'_u)\le \boundary(W_u)$, then $\separg(W'_u)\in\KT$ follows by
      Lemma~\ref{lem:tanglesepprop}(1), contradicting the minimality
      of $W_u$.
    \end{cs}
    Therefore, the vertices $u$ of $M$ for which $b\in W_u$ form a
    clique in $H$, thus there are less than $2k(k-1)$ such vertices.
    \uend
\end{claim}

We define
\[
B:=(V(G)\setminus \bigcup_{W\in \CW} W)\cup \bigcup_{W\in \CW}N^G(W)
\]
(see Figure~\ref{fig:kisolated}).
\begin{figure}
\begin{center}
{\small \def\svgwidth{0.5\linewidth}%
\executeiffilenewer{kisolated.svg}{kisolated.pdf}%
{inkscape -z -D --file=kisolated.svg %
--export-pdf=kisolated.pdf --export-latex}%
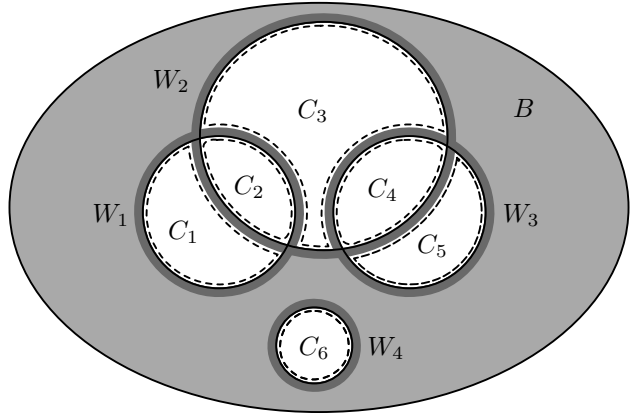%
}
\end{center}
\caption{Definition of the set $B$ in Lemma~\ref{lem:tangledegstar}. The four solid circles represent the sets $W_1$, $W_2$, $W_3$, $W_4$ contained in $\CW$. The dark gray area contains the boundaries of these sets. Set $B$ (light and dark gray area) is defined to be the union of these boundaries and the area outside these sets. The six regions $C_1$, $\dots$, $C_6$ with dashed outline are the components of $G\setminus B$.}\label{fig:kisolated}
\end{figure}

\begin{claim}[resume]\label{claim:topdeg2}
For every $W\in \CW$, $|N^G[W]\cap B|=O(k^6)$.
\proof 
Let us fix a $W\in \CW$.
We bound first the number of sets
$Y\in\CW$ such that $N^G(Y)$ intersects $W$.  As $G[Y]$ is connected
and $Y$ is not contained in $W$ (by the definition of $\CW$), $Y$ has
to contain a vertex $b\in N^G(W)$. By Claim~\ref{claim:topdeg1}, there
are at most $O(k^2)$ sets in $\CW$ containing a particular $b\in N^G(W)$. Together with
$|N^G(W)|<k(k-1)$, this gives a total bound of $O(k^4)$ on the number
of sets $Y\in\CW$ for which $N^G(Y)$ intersects $W$.  As
$|N^G(Y)|<k(k-1)$ for every $Y\in\CW$, this means that $W$ contains at
most $O(k^6)$ vertices of $B$.  Additionally, $N^G(W)$ can contain at
most $|N^G(W)|<k(k-1)$ vertices of $B$, and the claim follows.  \uend
\end{claim}

Let $C_1$, $\dots$, $C_p$ be the connected components of $G\setminus
B$.  We construct a star decomposition
$\Sigma_\KT=(T_\KT,\sep_{\KT},\comp_\KT)$ with center $s$ and $p$ tips
$t_i$ ($1\le i \le p$).  We set $\comp_\KT(s)=V(G)$,
$\sep_\KT(s)=\emptyset$, $\comp_\KT(t_i)=C_i$, and
$\sep_\KT(t_i)=N^G(C_i)$.

It easy easy to verify that $\Sigma_\KT$ is a tree decomposition:
\begin{claim}[resume]\label{claim:topdeg3}
$\Sigma_\KT$  satisfies properties \ref{li:t1}--\ref{li:t5}.
\end{claim}

The following claim implies a bound on the adhesion of $\Sigma_\KT$:
\begin{claim}[resume]\label{claim:topdeg6}
There is a constant $a=O(k^6)$ such that $|\sep_\KT(t_i)|\le a$ for every $1\le i \le p$.
\proof  Observe that $\comp_\KT(t_i)$ is disjoint from $B$ and therefore it
  has to be fully contained in $W$ for some $W\in\CW$: vertices
  outside every $W\in \CW$ are in $B$ and $N^G(W)\subseteq B$ for
  every $W\in \CW$, thus $\comp_\KT(t_i)$ has to be contained in a
  single $W\in \CW$.  Furthermore,
$\comp_\KT(t_i)\subseteq W$ implies  $\sep_\KT(t_i)\subseteq N^G[W]\cap B$.
By Claim~\ref{claim:topdeg2}, $|N^G[W]\cap B|=O(k^6)$, and we have the
required bound on $|\sep_\KT(t)|$.  \uend
\end{claim}

Using the bound on the adhesion, it is easy to show that $\Sigma_\KT$ respects $\KT$:
\begin{claim}[resume]\label{claim:topdeg8}
  $\separg(\comp_\KT(t_i))\in \KT$ for every $1\le i \le p$.  \proof
As in the previous claim, $\comp_\KT(t_i)\subseteq W$ for some $W\in \CW$.
The order of $\separg(\comp_\KT(t_i))$ is exactly
  $|\sep_\KT(t)|$, which is at most $a$ by
  Claim~\ref{claim:topdeg6}. As the order of $\KT$ is $m'(k)> a$ and
  $\comp_\KT(t)\subseteq W$, $\separg(W)\in\KT$ hold,
  Lemma~\ref{lem:tanglesepprop}(1) implies that
  $\separg(\comp_\KT(t_i))\in\KT$ holds as well.  \uend
\end{claim}

The following claim proves the bound on the maximum degree:
\begin{claim}[resume]\label{claim:topdeg7}
There is a constant $d'(k)=O(k^7)$ such that every vertex $v\not \in Z$ has degree at most $d'(k)$ in $\tau(s)$,
\proof  Let us observe first that for every $W\in \CW$,
the graph $\tau(s)$ has no edge between $W\cap B$ and $B\setminus
N^G[W]$. To see this, recall that, for every $1\le i \le p$,
$\sigma(t_i)=N^G(\comp_\KT(t_i))$, $G[\comp_\KT(t_i)]$ is connected,
and $\comp_\KT(t_i)\cap B=\emptyset$. As $N^G(W)\subseteq B$, it
follows that $\comp_\KT(t_i)$ cannot have a neighbor both inside $W$
and outside $N^G[W]$. Therefore, $\sigma(t_i)$ is either a subset
$N^G[W]$ or disjoint from $W$. This means that in the definition of
$\tau(s)$, there is no clique that introduces an edge between a vertex
in $W$ and a vertex outside $N^G[W]$.

Consider a $u \in B\setminus Z$.
\begin{cs}
  \case1 $u\in W$ for some $W\in \CW$. By our observation above, every
  neighbor of $u$ in $\tau(t)\setminus Z$ is contained in $N^G[W]$. Therefore, Claim~\ref{claim:topdeg2} gives a bound of
  $O(k^6)$ on the degree of $u$ in $\tau(t)$.

  \case2 $u\not\in W$ for any $W\in \CW$. As $u\not\in Z$, this is only possible if the
  degree of $u$ is at most $k$ in $G$. Therefore, $u$ is adjacent to
  at most $k$ components of $G\setminus B$.  Each new clique in
  $\tau(t)$ corresponds to the neighborhood of such a component. Thus
  $u$ is part of at most $k$ cliques introduced in the definition of
  $\tau(t)$. The size of each clique can be bounded by the adhesion of
  $\Sigma_\KT$, which is at most $a$ by Claim~\ref{claim:topdeg6}. Therefore,
  $k$ receives at most $k\cdot O(k^6)$ new edges.
\uend
\end{cs}
\end{claim}

What remains to be proven is the invariance condition.
Suppose that $\KT'$ is another tangle of order $k$ in a graph
 $G'$. Let  $f$ be an
 isomorphism from $(G,\KT)$ to $(G',\KT')$. Let $B$ and $B'$ be the sets computed by the algorithm on $(G,\KT)$ and $(G',\KT')$, respectively.

\begin{claim}[resume]\label{claim:topdeg9}
$B'=f(B)$.
\proof Let $\CW$ and $\CW'$ be the two collection of sets constructed
by the algorithm on $(G,\KT)$ and $(G',\KT')$, respectively. Let us
observe that $W\in \CW$ if and only if $f(W)\in \CW'$: the definition
of $\CW$ depends only on the definition of the sets $W_v$, which
depends only on the tangle $\KT$ and the graph-theoretic properties of
$G$, all of which are preserved by the isomorphism $f$. Taking into
account that the definition of $B$ depends only on the sets in $\CW$
and their neighborhoods in $G$, we can deduce $B'=f(B)$. \uend
\end{claim}

As $B'=f(B)$, for every component of $G\setminus B$, there is a
corresponding component of $G'\setminus B'$. Let $C'_1$, $\dots$,
$C'_p$ be the components of $G'$, as enumerated by running the
algorithm on $(G',\KT')$, and let $s'$, $t'_1$, $\dots$, $t'_p$ be
the nodes of the constructed star decomposition.  Let us define
$g(s)=s'$ and let $g(t_i)=t'_j$ such that $f(C_i)=C'_j$.
\begin{claim}[resume]\label{claim:topdeg10}
For all $t\in
 V(T_\KT)$ we have $\sep_{\KT'}(g(t))=f(\sep_{\KT}(t))$ and
 $\comp_{\KT'}(g(t))=f(\comp_{\KT}(t))$.

\proof The statement immediately follows from the fact that
$\comp_{\KT}(t_i)=C_i$ and $\comp_{\KT'}(g(t_i))=f(C_i)$ by definition
of $G$, and hence $\sep_{\KT}(t_i)=N^G(C_i)$ and
$\sep_{\KT'}(g(t_i))=N^{G'}(f(C_i))=f(N^G(C_i))$.  \qedhere
\end{claim}
\end{proof}

Finally, we can prove Lemma~\ref{lem:conndegstar} by invoking
Lemma~\ref{lem:tangledegstar} on the tangle defined by the unbreakable
set $X$:
\begin{proof}[Proof of Lemma~\ref{lem:conndegstar}]
  Let $c^*(k)=k+3m-2$, $d^*(k)=d'(k)+3m-2$, $\ell=\ell^*(k)=\ell'(k)$,
  $m=m^*(k)=m'(k)$ for the functions $d'$, $\ell'$, $m'$ in
  Lemma~\ref{lem:tangledegstar}.  Let $\KT$ be the tangle of order $m$
  defined by the $m$-unbreakable set $X$ (see Lemma~\ref{lem:unbreakabletangle});
  Lemma~\ref{lem:connectedtangleoracle} provides an implementation of the
  min-cut oracle for $\KT$. As $I$ is $m$-attached to $X$, tangle $\KT$ does not remove $I$. Let us call the algorithm of
  Lemma~\ref{lem:tangledegstar} with $G$, $k$, $\KT$, and $I$.
  If it returns a subdivision of $K_k$ in $G$, then we are
  done. Otherwise, the algorithm of Lemma~\ref{lem:tangledegstar}
  returns a $\KT$-respecting star decomposition
  $\Sigma_\KT=(T_\KT,\sep_\KT,\comp_\KT)$ of $G$.
The
  star decomposition $\Sigma_\KT$ almost satisfies the requirements of
  Lemma~\ref{lem:conndegstar}, except that $X$ is not necessarily
  contained in $\bag_\KT(s)$ for the center $s$. To move $X$ to the
  center, we construct a star
  decomposition $\Sigma_X=(T_X,\sep_X,\comp_X)$ as follows. First, let
  $T_X=T_\KT$ and for the center $s$ of $T_X$, let $\comp_X(s)=V(G)$
  and $\sep_X(s)=\emptyset$. For every tip $t$ of $T_X$, we let
  $\comp_X(t)=\comp_\KT(t)\setminus X$ and $\sep_X(t)=\sep_\KT(t)\cup
  (X\cap \comp_{\KT}(t))$. It is straightforward to verify that
  $\Sigma_X$ is also a star decomposition of $G$, and in fact it is
  star decomposition even for the supergraph $G\cup K[X]$ (since
  $X\subseteq \beta_X(s)$).  As $\torso_\KT(s)\setminus
  X=\torso_X(s)\setminus X$, and $\torso_\KT(s)$ contains at most $k$
  vertices of degree higher than $d^*(k)$, we have that $\torso_X(s)$
  contains at most $k+|X|=c^*(k)$ vertices of degree higher than
  $d^*(k)$.
  The bound $<|X|$ on the adhesion and the invariance requirement can
  be proved the same way as in Lemma~\ref{lem:conncliquestar}.
\end{proof}


%% file: kisolated.pdf_tex
\begingroup%
  \makeatletter%
  \providecommand\color[2][]{%
    \errmessage{(Inkscape) Color is used for the text in Inkscape, but the package 'color.sty' is not loaded}%
    \renewcommand\color[2][]{}%
  }%
  \providecommand\transparent[1]{%
    \errmessage{(Inkscape) Transparency is used (non-zero) for the text in Inkscape, but the package 'transparent.sty' is not loaded}%
    \renewcommand\transparent[1]{}%
  }%
  \providecommand\rotatebox[2]{#2}%
  \ifx\svgwidth\undefined%
    \setlength{\unitlength}{260.8bp}%
    \ifx\svgscale\undefined%
      \relax%
    \else%
      \setlength{\unitlength}{\unitlength * \real{\svgscale}}%
    \fi%
  \else%
    \setlength{\unitlength}{\svgwidth}%
  \fi%
  \global\let\svgwidth\undefined%
  \global\let\svgscale\undefined%
  \makeatother%
  \begin{picture}(1,0.66257669)%
    \put(0,0){\includegraphics[width=\unitlength]{kisolated.pdf}}%
    \put(0.8297546,0.47703318){\color[rgb]{0,0,0}\makebox(0,0)[b]{\smash{${\Large B}$}}}%
    \put(0.28198949,0.27983777){\color[rgb]{0,0,0}\makebox(0,0)[b]{\smash{$C_1$}}}%
    \put(0.38781771,0.35323825){\color[rgb]{0,0,0}\makebox(0,0)[b]{\smash{$C_2$}}}%
    \put(0.4901402,0.47484209){\color[rgb]{0,0,0}\makebox(0,0)[b]{\smash{$C_3$}}}%
    \put(0.60276074,0.34885617){\color[rgb]{0,0,0}\makebox(0,0)[b]{\smash{$C_4$}}}%
    \put(0.68295354,0.25792717){\color[rgb]{0,0,0}\makebox(0,0)[b]{\smash{$C_5$}}}%
    \put(0.49233129,0.0935976){\color[rgb]{0,0,0}\makebox(0,0)[b]{\smash{$C_6$}}}%
    \put(0.19478528,0.30832153){\color[rgb]{0,0,0}\makebox(0,0)[rb]{\smash{$W_1$}}}%
    \put(0.26226994,0.52304545){\color[rgb]{0,0,0}\makebox(0,0)[b]{\smash{$W_2$}}}%
    \put(0.57822086,0.0935976){\color[rgb]{0,0,0}\makebox(0,0)[lb]{\smash{$W_4$}}}%
    \put(0.79601227,0.30832153){\color[rgb]{0,0,0}\makebox(0,0)[lb]{\smash{$W_3$}}}%
  \end{picture}%
\endgroup%

%% file: partial.tex
\section{Partial Dominating Set}
\label{sec:part-domin-set}
The goal of this section is to prove that \textsc{Partial Dominating
  Set} (find $k$ vertices whose closed neighborhood has maximum size)
can be solved in time $f(H,k)\cdot n^{O(1)}$ on graphs excluding $H$
as a topological subgraph. We intend this result as a demonstration of
the algorithmic use of the Global Structure
Theorem~\ref{theo:structure}: it shows that by combining the
techniques that work on almost-embeddable and on bounded-degree
graphs, we can solve problems on graphs excluding a topological
subgraph. We would like to emphasize that all the algorithmic
techniques in this section are standard: it is the new structure
theorem that allows us to use these standard techniques on a larger
class of graphs. We remark that an $f(k)\cdot n^{f(H)}$ algorithm was
known for \textsc{Partial Dominating Set} on $H$-minor free graphs
\cite{DBLP:journals/jcss/AminiFS11}, but instead of extending this
algorithm, we give here a self-contained presentation of the result on
graphs excluding $H$ as a topological subgraph.

We begin by defining a generalization of \textsc{Partial Dominating
  Set}, which will be convenient for computations on tree
decompositions. We extend the problem by introducing a {\em cost
  function} $\kappa:V(G)\to \{0,1\}$ and {\em value function}
$\nu:V(G)\to \{0,1\}$; now the goal is to find a set $Z\subseteq V(G)$
with $\kappa(Z)\le k$ such that $\nu(N^G[Z])$ is maximizied. (As
usual, $\kappa$ and $\nu$ are extended to sets by $\nu(Z)=\sum_{v\in
  Z}\nu(v)$ and $\kappa(Z)=\sum_{v\in Z}\kappa(v)$.) That is, the vertices with $\kappa(v)=0$ can be used for ``free'' and the domination of a vertex with $\nu(v)=0$  does not increase the objective function.
\begin{defn}\label{def:partdomprofile}
  Let $G$ be a graph and $S\subseteq V(G)$ a set of vertices. The {\em
    $k$-profile} of $G$ with respect to $S$ is a function
  $\pi(z,\kappa,\nu)$, which, for every integer $0 \le z \le k$ and
  functions $\kappa,\nu:V(G)\to \{0,1\}$ that have value 1 on
  $V(G)\setminus S$, gives the maximum of $\nu(N^G[Z])$ taken over
  every $Z\subseteq V(G)$ with $\kappa(Z)\le z$.
\end{defn}
That is, the $k$-profile with respect to $S$ is described by $(k+1)\cdot 2^{|S|}\cdot
2^{|S|}$ integers. Observe that if the $k$-profile with respect to $S$
is known, then it is easy to deterimine the $k$-profile with respect
to some $S'\subseteq S$.

First we show that the $k$-profile can be computed in a bottom-up
manner on a tree decomposition if every bag is small, that is, the
decomposition has bounded width. Then we use a standard layering
argument to compute the $k$-profile on almost-embeddable torsos by
reducing it to the bounded-treewidth case. For this reduction, we need
the fact that almost-embeddable graphs have bounded {\em local treewidth:}
\begin{theo}[\cite{DBLP:journals/combinatorica/Grohe03}]\label{th:localtw}
  For every $p,q,r\in\NN$, there is a constant $\lambda>0$ such that
  the following holds. Let $G$ be a minor of a $(p,q,r,0)$-almost
  embeddable graph, let $x\in V(G)$, and let $N_d[\{x\}] \subseteq
  V(G)$ be the set of vertices at distance at most $d$ from $x$. Then
  $G\big[N_d[\{x\}]\big]$ has treewidth at most $\lambda\cdot d$ for every
  $d\ge 0$.
\end{theo}
Finally, we compute the $k$-profile on almost bounded-degree torsos by
using a standard random coloring technique.
\begin{lem}\label{lem:partialdom}
  Let $(T,\sep,\comp)$ be a tree decomposition of a graph $G$ and $t$
  a node of $T$. Suppose that, for every child $t'$ of $t$, the
  $k$-profile of $G[\cone(t')]$ with respect to $\sep(t')$ is
  known. Then the $k$-profile of $G[\cone(t)]$ with respect to
  $\sep(t)$ can be computed
\begin{enumerate}
\item in time $f(s)\cdot n^{O(1)}$ if $|\bag(t)|\le s$ and $|N_+^T(t)|\le 2$.
\item in time $f(w)\cdot n^{O(1)}$ if a tree decomposition of $\torso(t)$ having width $w$ is given.
\item in time $f(k,p,q,r,s,a)\cdot n^{O(1)}$ if $|\sep(t)|\le a$ and a set $P$ of size at
  most $s$ is given such that $\torso(t)\setminus P$ is almost $(p,q,r,0)$-embeddable.
\item in time $f(k,c,d,a)\cdot n^{O(1)}$ if $|\sep(t)|\le a$ and all but at most $c$ vertices have degree at most $d$ in $\torso(t)$.
\end{enumerate}
\end{lem}
\begin{proof}
  (1) Let us assume that $t$ has exactly two children $t_1$ and $t_2$;
  it is easy to adapt the proof for the simpler cases that $t$ has
  exactly one child or is a leaf. Let $G_0=G[\cone(t)]$. For $i=1,2$, let $G_i=G[\cone(t_i)]$ and let $\pi_i$ be the
  $k$-profile of $G_i$ with respect to $\sep(t_i)$. We claim that
  $\pi(z,\kappa,\nu)$ with respect to $\sep(t)$ is the maximum of
\begin{equation}
\nu(N^{G_0}[Z_0]\cap \bag(t)) + \pi_1(z_1,\kappa_1,\nu_1)+
\pi_2(z_2,\kappa_2,\nu_2)
\label{eq:partialdom}\end{equation}
taken over functions $\kappa_i,\nu_i:\cone(t_i)\to \{0,1\}$ for $i=1,2$ and 
\begin{align*}
z_1,z_2&\ge 0\\
Z_0&\subseteq \bag(t)\\
\kappa(Z_0)+z_1+z_2&\le z\\
\kappa_i(v)=\nu_i(v)&=1 &&\text{for $i=1,2$ and every $v\in \comp(t_i)$}\\
\nu_i(v)&=0 &&\text{for $i=1,2$ and every $v\in N^{G_0}[Z_0]\cap \sep_i(t)$}\\
\nu_1(v)+\nu_2(v)&\le \nu(v)&& \text{for every $v\in \bag(t)$}\\
\kappa_i(v)&=1 &&\text{for $i=1,2$ and every $v\in \bag(t)\setminus Z_0$}\\
\end{align*}
Suppose that $Z\subseteq \cone(t)$ is the set reaching the maximum in
$\pi(z,\kappa,\nu)$; we show that (\ref{eq:partialdom}) can reach
$\nu(N^{G_0}[Z])$. Let $Z_0=Z\cap \bag(t)$ and let 
let $\kappa_i(v)=0$
if and only if $v\in Z_0\cap \sep(t_i)$.
Let $Z_i=Z\cap \cone(t_i)$,
 $z_i=\kappa_i(Z_i)$,
$z_2=\kappa_1(Z_2)$. Observe that $Z_0,Z_1,Z_2$ are not necessarily disjoint,  but $\kappa(Z_0)+z_1+z_2\le \kappa(Z)\le z$ holds, as $\kappa_i(v)=0$ for any vertex that appears in more than one of the sets. Let $\nu_1(v)=0$ if 
 $v\in \sep(t_1)\cap N^{G_0}[Z_0]$ and let $\nu_1(v)=\nu(v)$ otherwise.
Let  $\nu_2(v)=0$ if 
$v\in \sep(t_2)\cap N^{G_0}[Z\cup Z_1]$ and let $\nu_2(v)=\nu(v)$ otherwise. It is easy to verify that every condition is satisfied. Furthermore, $Z_i$ shows that $\pi_i(z_i,\kappa_i,\nu_i)$ is at least $\nu_i(N^{G_i}[Z_i])$. The sum of the three terms in (\ref{eq:partialdom}) is at least $\kappa(N^{G_0}[Z])$: every vertex $v\in N^{G_0}[Z]$ is accounted for in one of the three terms depending on which of $v\in N^{G_0}[Z_0]$, $v\in N^{G_0}[Z_1]\setminus N^{G_0}[Z_0]$, or $v\in N^{G_0}[Z_2]\setminus N^{G_0}[Z_0\cup Z_1]$ holds.

Conversely, consider the values of $z_1$, $z_2$, $Z_0$, $\kappa_i$,
$\nu_i$ that maximize \eqref{eq:partialdom}. Let $Z_i$ be a set that
reaches the maximum in the definition of $\pi_i(z_i,\kappa_i,\nu_i)$.  Let $Z:=Z_0\cup Z_1 \cup
Z_2$. Clearly, $\kappa(Z)\le \kappa(Z_0)+\kappa_1(Z_1)+\kappa_2(Z_2)\le \kappa(Z_0)+z_1+z_2\le  z$ (in the first inequality, we use the fact that if
$\kappa(v)=1$ and $\kappa_i(v)=0$ for some $v\in Z_i$, then $v$ is in $Z_0$ as well). We claim that \eqref{eq:partialdom} is
at most $\nu(N^{G_0}[Z])$. The first term counts the vertices $v\in N^{G_0}[Z_0]\cap \bag(t)$ with $\nu(v)=1$. The second term is $\nu_1(N^{G_1}[Z_1])$, i.e, counts those vertices $v\in N^{G_1}[Z_1]$ that have $\nu_1(v)=1$.
Similarly, the third counts $v\in N^{G_2}[Z_2]$ that have $\nu_2(v)=1$. 
Every vertex $v$ counted this way is in $N^{G_0}[Z]$ and has $\kappa(v)=1$. Furthermore,  every such vertex is counted in at most one of the three terms: this is ensured by the conditions on $\nu_1$ and $\nu_2$.  Thus
the three terms count the sizes of disjoint sets, which means that
their sum is at most $\nu(N^{G_0}[Z])\le
\pi(z,\kappa,\nu)$.

(2)
 The treewidth of $\torso(t)$ is at most $w$ by assumption, thus we
can use standard algorithms to compute in time $f(w)\cdot n^{O(1)}$ a
tree decomposition $(T_t,\bag_t)$ with $|\bag_t(x)|<w$ for
every $x\in V(T_t)$. Since $\sep(t)$ is a clique in $\torso(t)$, there
is an $x\in V(T_t)$ with $\sep(t)\subseteq \bag_t(x)$. Furthermore, if the
children of $t$ in $T$ are $t_1$, $\dots$, $t_m$, then $\sep(t_i)$ is
a clique in $\torso(t)$, hence there is an $x_i\in V(T_t)$ with
$\sep(t_i)\subseteq \bag_t(x_i)$ for every $1\le i \le m$. By standard
transformations, we can assume that $\sep(t)\subseteq \bag_t(r)$ for
the root $r$ of $T_t$, for every child $t_i$ there is a leaf $x_i$ of
$T_t$ with $\sep(t_i)\subseteq \bag_t(x_i)$, and every node of $T_t$
has at most two children (we omit the details).

We modify $(T,\bag)$ to obtain a new tree decomposition $(T',\bag')$
the following way. The tree $T'$ is obtained from $T$ by removing node
$t$, adding every node of $T_t$, letting the parent of $t$ be the
parent of $r$ (the root of $T_t$), and letting $x_i$ be the parent of
$t_i$ for every $1\le i \le m$. We set $\bag'(y)$ to be $\bag(y)$ if
$y\in V(T)\setminus \{t\}$ and to $\bag_t(y)$ otherwise.  It is not
difficult to verify that $(T',\bag')$ is also a tree decomposition of
$G$. We apply statement (1) on every node $t'\in V(T_t)\subseteq
V(T')$ of this tree decomposition in a bottom-up order to compute the
$k$-profile of $G[\gamma'(t')]$ with respect to $\sep'(t')$. The
conditions of (1) hold: whenever we consider a $t'\in V(T_t)$, then we
already know the $k$-profile of each child of $t'$: either it is in
$V(T_t)$ and we know the $k$-profile because of the bottom-up order or
it is $t_i$ and we know the $k$-profile by assumption. When this
procedure reaches $r$ (the root of $T_t$), it computes the $k$-profile
of $G[\gamma'(r)]=G[\gamma(t)]$ with respect to $\sep'(r)=\sep(t)$, as
required.

(3) Let $G'$ be the $(p,q,r,0)$-almost embeddable graph
$\torso(t)\setminus P$ and let $Q$ contain one vertex from each
connected component of $G'$. Let {\em level $L[i]$} contain those
vertices of $G'$ whose distance from $Q$ in $G'$ is exactly $i$. We define
$L[i,j]:=\bigcup_{\ell=i}^j L[\ell]$. We claim that $G'[L[i,j]]$ has
treewidth at most $\lambda\cdot (j-i+1)$ for some $\lambda$ depending
only on $p,q,r$. If $i\le 1$, then this follows immediately from
Theorem~\ref{th:localtw}. If $i\ge 1$, then let $H$ be obtained from
$L[0,j]$ by contracting every edge whose both endpoints are at
distance at most $i-1$ from $Q$. Observe now that $G'[L[i,j]]$ is a
subgraph of $H$ and that every vertex of $H$ is at distance at most
$j-i+1$ from $Q$. As $H$ is a minor of an $(p,q,r,0)$-almost
embeddable graph, Theorem~\ref{th:localtw} implies that the treewidth
of $H$ (and hence of $G'[L[i,j]]$) is at most $\lambda\cdot (j-i+1)$.

We compute $\pi(z,\kappa,\nu)$ with respect to $\sep(t)$ as follows.
Suppose that $Z\subseteq \gamma(t)$ is the set reaching the maximum in
the definition of $\pi(z,\kappa,\nu)$.  We claim that $N^G[Z\setminus
P]$ intersects at most $D:=3(k+a)$ levels. For every vertex $v\in (Z\setminus
P)\cap \bag(t)$, the closed neighborhood of $v$ is fully contained in
at most 3 levels. For every vertex $v\in Z\cap \comp(t')$ for some
child $t'$ of $t$, $N^G(\{v\})$ intersects $\bag(t)$ in a subset of
$\sep(t')$, which induces a clique in $\torso(t)$. Thus $N^G(\{v\})$
intersects at most 2 levels of $G'$. We have $|Z|\le k+a$, because $\kappa(Z)\le z\le
k$ and $\kappa(v)=1$ for all but the at most $a$ vertices $v\in
\sep(t)$. Therefore,  $N^G[Z\setminus P]$ intersects at most
$2(k+a)\le D$ levels.

For $0 \le h \le D$, let $M_h:=\bigcup_{j\ge 0}L[((D+1)j+h]$. Note
that these $D+1$ sets are pairwise disjoint. As $N^G[Z\setminus P]$
intersects at most $D$ levels, there is an $h$ such that $M_h$ is
disjoint from $N^G[Z\setminus P]$. This means that if we obtain $G_h$
from $G[\cone(t)]$ by removing from every $v\in M_h$ the edges incident
to $v$ and {\em not} going to $P$, then this does not change the
value of $\pi(z,\kappa,\nu)$: removing edges cannot increase this
value, and as $M_h$ is disjoint from $N^G[Z\setminus P]$, none of
these edges are induced by $N^G[Z]$, thus it does not decrease
the value either. Therefore, if we denote by $\pi_h$ the $k$-profile
of $G_h$ with respect to $\sep(t)$, then
$\pi(z,\kappa,\nu)=\max_{h=0}^{D}\pi_h(z,\kappa,\nu)$.

$(T,\sep,\comp)$ is a tree decomposition also for $G_h$. Let $(T_h,\sep_h,\comp_h)$ be the tree decomposition of $G_h$ with the
slight modification that $\sep_h(t_j)=\sep(t_j)\setminus M_h$ for
every child $t_j$. This is still a tree decomposition: there are no
edges between $\comp(t_j)$ and $\sep(t_j)\cap M_h$ in $G_h$, as
vertices of $M_h$ have neighbors only in $P$. Furthermore, the
$k$-profile of $G[\cone(t_j)]$ with respect to $\sep_h(t_j)\subseteq \sep(t_j)$ can be
determined from the $k$-profile of $G[\cone(t_j)]$ with respect to
$\sep(t_j)$, which is assumed to be known.

Let us bound the treewidth of $\torso_h(t)$.  Observe that vertices of
$M_h$ are isolated in $\torso_h(t)\setminus P$ and thus every
component of $\torso_h(t)\setminus P$ contains at most $D$ consecutive
levels. Therefore, by our observation above on the treewidth of
$G'[L[i,j]]$, we have that the treewidth of $\torso_h(t)\setminus P$
is at most $\lambda\cdot (D+1)$. The set $P$ can increase treewidth by
at most $s$. Thus the treewidth of $\torso_h(t)$ can be bounded by a
function depending only on $p$, $q$, $r$, $s$, and $a$. This means
that we can use statement (2) to compute the $k$-profile $\pi_h$ for
every $0 \le h \le D$ and deduce the value of $\pi(z,\kappa,\nu)$ with
respect to $\sep(t)$.

(4)
%
Let $P$ be the set of vertices with degree more than $d$ in $\tau(t)$.  Let us
color every vertex of $\bag(t)\setminus P$ red or blue
uniformly and independently at random.  If there is a connected red
component of size larger than $D:=(k+a)(d+1)$ in $\torso(t)$, then recolor every
vertex of this component blue. Let us obtain $G'$ from $G[\cone(t)]$
by removing every edge  with at least one blue endpoint and {\em not} incident to $P$.

It is clear that this transformation cannot increase
$\pi(z,\kappa,\nu)$. We claim that with positive probability
depending only on $k$, $a$, and $d$, this transformation does not decrease
it either. Suppose that $Z\subseteq \cone(t)$ is the set reaching the
maximum in the definition of $\pi(z,\kappa,\nu)$. Let $R:=N^G[Z\setminus P]\cap \bag(t)$ and let $B$ be the
(open) neighborhood of $R$ in $\torso(t)\setminus P$. As there are at most $a$ vertices $v\in \bag(t)$ with $\kappa(v)=0$, 
the size of $R$ is at most $(k+a)(d+1)$ and the size of $B$ is at most $(k+a)(d+1)d$. Thus with positive probability
depending only on $k$ and $d$, $R$ ends up red and $B$ ends up
blue. This means that no vertex of $R$ is recolored blue and therefore none of the edges in $N^G[Z\setminus P]$ is
removed, implying that $\pi(z,\kappa,\nu)$ does not decrease.

Let $(T',\sep',\comp')$ be a tree decomposition of $G'$ obtained from
$(T,\sep,\comp)$ by setting, for every child $t_j$ of $t$,
$\comp'(t_j)=\comp(t_j)$ and letting $\sep'(t_j)$ be $\sep(t_j)$ minus
the blue vertices. Similarly to (3), this remains a tree decomposition
and the $k$-profiles of $G[\cone(t_j)]$ with respect to $\sep'(t_j)$
can be assumed to be known. We claim that the treewidth of
$\torso'(t)$ can be bounded by a function $k$, $c$, $d$, and
$a$. Indeed, every component of $\torso'(t)\setminus P$ has size at
most $D$ (as we recolored every larger red component to blue)
and $P$ can increase treewidth by at most $c$. Thus we can use (2) to
compute the profile of the modified graph and with probability
depending only on $k$, $a$, and $d$, this will give us exactly
$\pi(z,\kappa,\nu)$ with respect to $\sep(t)$. By repeated
application, the error probability can be made an arbitrary small
constant. Furthermore, the algorithm can be derandomized using
standard techniques, see e.g., \cite{DBLP:conf/wg/KneisMRR06}.
\end{proof}
Theorem~\ref{th:partialdom} follows immediately by putting together
Corollary~\ref{cor:structure-emb} and
Lemma~\ref{lem:partialdom}(3--4): in a bottom-up order, for every node
$t$ of the decomposition given by Corollary~\ref{cor:structure-emb},
we can compute the $k$-profile of $G[\cone(t)]$ with respect to
$\sep(t)$, which gives us the value of the optimum solution of
\textsc{Partial Dominating Set}.

\begin{rem}
  Recall that a graph is {\em $d$-degenerate} if every subgraph has a
  vertex of degree at most $d$. A classical result of Mader
  \cite{MR0220616} shows that every graph excluding $H$ as a
  topological subgraph is $d_H$-degenerate for some constant $d_H$
  depending on $H$, thus it is a natural question whether
  Theorem~\ref{th:partialdom} can be generalized to the more general
  class of $d$-degenerate graphs.  However, \textsc{Partial Dominating
    Set} is W[1]-hard parameterized by $k$ and $d$ on $d$-degenerate
  graphs. To see this, note that \textsc{Maximum Independent Set},
  parameterized by the size $k$ of the solution, is W[1]-hard even on
  regular graphs. Let $G$ be an $r$-regular graph ($r\ge 3$) and let
  us subdivide every edge by a new vertex. It is not difficult to see
  that $G$ has an independent set of size $k$ if and only if the new
  graph $G'$ has a set of $k$ vertices whose closed neighborhood has
  size $(r+1)k$. As $G'$ is 2-degenerate, an $f(k,d)\cdot n^{O(1)}$
  time algorithm for \textsc{Partial Dominating Set} on $d$-degenerate
  graphs would imply an $f(k)\cdot n^{O(1)}$ time algorithm for
  \textsc{Maximum Independent Set}. Thus the fixed-parameter
  tractability of \textsc{Partial Dominating Set} on graph excluding
  $H$ as a topological subgraph is not simply a consequence of the
  sparsity/degeneracy of such graphs, but essentially depends on the
  structural properties of this class of graphs.
\end{rem}


%% file: treelike.tex
\section{Invariant Treelike Decompositions}
\label{sec:treel-decomp}

In this section, we relax the notion of tree decomposition to the more liberal
notion of treelike decomposition, first introduced in
\cite{gro08,gro10+a}. The reason is that we want to make our
decompositions invariant under automorphisms of the underlying graph, and
this is not possible for tree decompositions. Treelike decompositions
are based on the axiomatisation of tree decompositions by
\ref{li:t1}--\ref{li:t5}. From now on, a \emph{decomposition} of a
graph $G$ is a
triple $\Delta=(D,\sep,\comp)$, where $D$ is a digraph and
$\sep,\comp:V(D)\mapsto 2^{V(G)}$. For every $t\in V(D)$, we define
sets $\cone(t),\bag(t)\subseteq V(G)$ and a graph $\tau(t)$ as in
\eqref{eq:cone}, \eqref{eq:bag}, and \eqref{eq:torso}. 
Two nodes $t,u\in V(D)$
are \emph{$\Delta$-equivalent} (we write $t\doublesmile u$) if
$\sep(t)=\sep(u)$ and $\comp(t)=\comp(u)$. Note that $t\doublesmile u$
implies $\cone(t)=\cone(u)$, but not $\bag(t)=\bag(u)$ or
$\tau(t)=\tau(u)$. We will occasionally work with several
decompositions at the same time, and in such situations may use an
index ${}^\Delta$, as for example in $\sep^\Delta(t)$ or
$t\doublesmile^\Delta u$, to indicate which decomposition we are
referring to. However, we usually prefer implicit naming conventions
such as the following: If we have a decomposition
$\Delta'=(D',\sep',\comp')$, then we will denote $\cone^{\Delta'}(t)$
by $\cone'(t)$, $\bag^{\Delta'}(t)$
by $\bag'(t)$, et cetera. 

The \emph{width} and
\emph{adhesion} of a decomposition are defined, as for tree decompositions, to
be the maximum size of the bags minus one and the maximum size of the
separators, respectively. A decomposition is \emph{over} a
class $\CA$ of graphs if all its torsos are in $\CA$. 

\begin{defn}\label{def:treelike}
  A \emph{treelike decomposition} of a graph $G$ is a decomposition
  $\Delta=(D,\sep,\comp)$ of $G$ that satisfies the following axioms:
\begin{nlist}{TL}
  \item\label{li:tl1}
    $D$ is acyclic.
  \item\label{li:tl2} For all $t\in V(D)$ it holds that
    $\comp(t)\cap\sep(t)=\emptyset$ and
    $N^G(\comp(t))\subseteq\sep(t)$.

  \item\label{li:tl3} For all $t\in V(D)$ and $u\in N_+^D(t)$ it holds that
    $\comp(u)\subseteq\comp(t)$ and $\cone(u)\subseteq\cone(t)$.
  \item\label{li:tl4} For all $t\in V(D)$ and $u_1,u_2\in N_+^D(t)$, either
    $u_1\doublesmile u_2$ or 
$  \cone(u_1)\cap\cone(u_2)=\sep(u_1)\cap\sep(u_2).$
  \item\label{li:tl5} For every connected component $A$ of $G$ there is a $t\in
    V(D)$ with $\sep(t)=\emptyset$ and $\comp(t)=V(A)$.
\end{nlist}
\end{defn}

\begin{rem}
  Recall the axiomatisation \ref{li:t1}--\ref{li:t5} of tree decompositions.
  Note that \ref{li:t2} coincides with \ref{li:tl2} and \ref{li:t3} coincides
  with \ref{li:tl3}. Moreover, \ref{li:t1} implies \ref{li:tl1} and \ref{li:t4}
  implies \ref{li:tl4}. For connected graphs $G$, \ref{li:t5} coincides with
  \ref{li:tl5}, and thus every tree decomposition of a connected graph is a
  treelike decomposition.
For disconnected graphs, this is not necessarily the case, but it can be shown
that from every treelike decomposition one can construct a tree decomposition
with the same torsos. (See \cite{gro10+a} for details.)  
\end{rem}

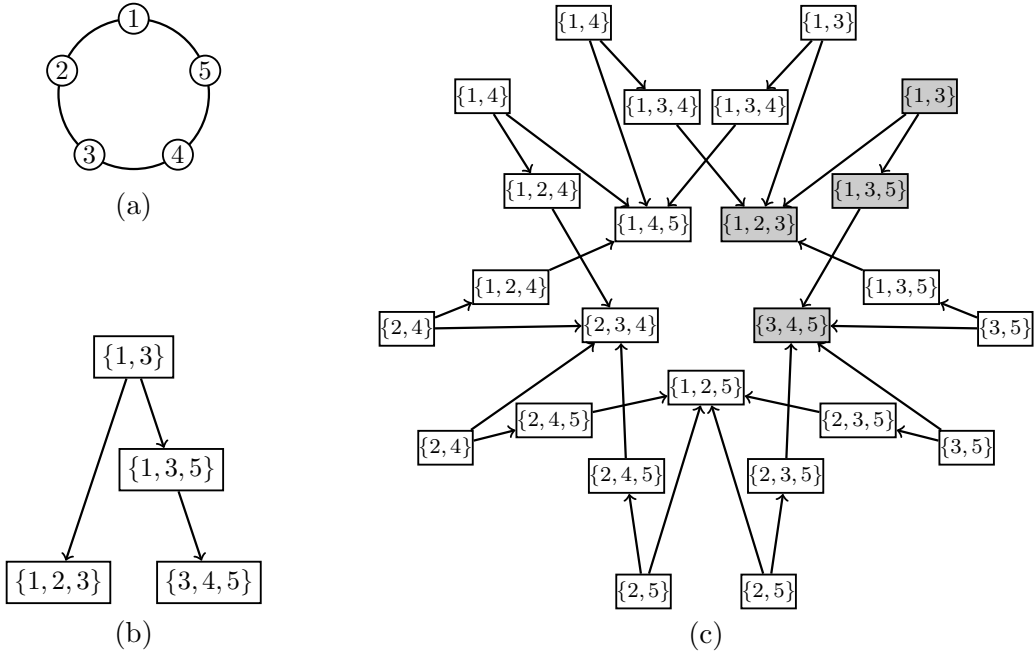
\begin{figure}
  \centering
  \input{c5dec}
  \caption{(a) The cycle $C_5$ with (b) a tree decomposition and (c)
    an automorphism-invariant treelike decomposition}
  \label{fig:c5dec}
\end{figure}

\begin{exa}\label{exa:c5dec}
  Figure~\ref{fig:c5dec}(a) shows the cycle $C_5$.
  Figure~\ref{fig:c5dec}(b) shows a tree decomposition $(T,\bag)$ of 
  $C_5$ of width $2$. Note that this tree decomposition is not
  invariant under automorphisms of $C_5$, in the sense that there is
  an automorphism $f$ of $C_5$ for which we cannot find an automorphism $g$ of
  $T$ such that for all $t\in V(T)$ we have
  $f(\bag(t))=\bag(g(t))$. It is easy to see that there is no tree
  decomposition of $C_5$ of width $2$ that is invariant under
  automorphisms.

  Figure~\ref{fig:c5dec}(b) shows a treelike decomposition
  $(D',\sep',\comp')$ of $C_5$ of width $2$. Actually, the sets
  displayed in
  the nodes are the bags, but we can easily compute the
  separators and components using \eqref{eq:bagsep} and
  \eqref{eq:bagcomp}. For instance, for the grey node $t$ with bag
  $\bag'(t)=\{1,3,5\}$ we have $\sep'(t)=\{1,3\}$ and
  $\comp'(t)=\{4,5\}$. For the sake of completeness, we observe that $\cone'(t)=\{1,3,4,5\}$ and
  $\tau'(t)=K[\{1,3,5\}]$. 

  Note that the ``subdecomposition'' induced by the four grey nodes
  is precisely the tree decomposition shown in
  Figure~\ref{fig:c5dec}(b). The treelike decomposition contains many
  other tree decompositions of $C_5$; actually, it contains all images
  of the decomposition shown in Figure~\ref{fig:c5dec}(b) under
  automorphisms of $C_5$. And indeed, the treelike decomposition is
  invariant under automorphisms.
  \uend
\end{exa}

Example~\ref{exa:c5dec} illustrates how treelike decompositions can be
made ``invariant.'' However, the automorphism invariance of the
example is not sufficient for our purposes, we need a more general
notion of invariance that involves decompositions of more than one graph.

\begin{defn}
  A \emph{decomposition mapping} for a class $\CC$ of graphs is a
  mapping $\Delta$ that associates with each $G\in\CC$ a decomposition
  $\Delta_G=(D_G,\sep_G,\comp_G)$ of $G$. 

  $\Delta$ is \emph{invariant}
    if for all isomorphic graphs $G,G'\in\CC$ and all isomorphisms $f$
    from $G$ to $G'$ there is an isomorphism $g$ from $D_G$ to
    $D_{G'}$ such that for all $t\in V(D_G)$ we have
    $\sep_{G'}(g(t))=f(\sep_G(t))$ and $\comp_{G'}(g(t))=f(\comp_G(t))$.
\end{defn}
We need some additional terminology about decomposition mappings: We
say that a decomposition mapping $\Delta$ for a class $\CC$ is \emph{treelike} if for all $G\in\CC$ the
    decomposition $\Delta_G$ is treelike. It has \emph{adhesion at most $a$} if
    for all $G\in\CC$ the adhesion of $\Delta_G$ is at most $a$, and
    it is \emph{over} a class $\CA$ of graphs if for all $G\in\CC$ the
    decomposition $\Delta_G$ is over $\CA$. We
    say that a class $\CC$ \emph{admits polynomial time
      computable invariant treelike decompositions over $\CA$ (of
      adhesion at most $a$)}
    if there is a polynomial time computable invariant
    treelike decomposition mapping for $\CC$ over $\CA$ (of adhesion $a$).

\begin{rem}
  The decomposition schemes of \cite{gro10+a} yield polynomial time
  computable invariant decomposition mappings.
  \uend
\end{rem}

The main result of the section is the following:
\begin{theo}[Invariant Decomposition Theorem]\label{theo:idec}
  For every graph $H$ there are constants $a,b,c,d,e\in\NN$ and a
  polynomial time computable invariant treelike decomposition mapping
  $\Delta$ of adhesion at most $a$ for the class of graphs $G$ with
  $H\not\preceq_T G$ such that for every $G$ from this class with
  $\Delta_G=:(D,\sep,\comp)$ and every $t\in V(D)$ one of the following
  three conditions is satisfied:
  \begin{eroman}
  \item
    $|\bag(s)|\le b$
\item
    At most $c$ vertices of $\tau(t)$ have degree greater than $d$.
   \item
$K_e\not\preceq\tau(t)$.
\end{eroman}
\end{theo}

\begin{proof}
  We let $k:=|H|$. We choose $c=c^*(k)$, $d=d^*(k)$, $\ell=\ell^*(k)$,
  and
  $m=m^*(k)$ according to Lemma~\ref{lem:conndegstar} and $e=e^*(\ell,m)$
  according to Lemma~\ref{lem:conncliquestar}. We let 
  $a:=3m-3$ and $b:=4m-3$..

  Let $G$ be a graph with $H\not\preceq_T G$.  We shall define a
  decomposition $\Delta_G=(D,\sep,\comp)$ of $G$ of adhesion at most
  $a$ such that every node $t$ satisfies one of (i)--(iii). Then we
  will argue that the decomposition $G\mapsto\Delta_G$ is polynomial
  time computable and invariant.

  There will be three kinds of nodes in $V(D)$: b-nodes (``\underline
  bounded nodes''), 
  d-nodes (``bounded \underline degree nodes''), and e-nodes (``\underline excluded minor nodes''). All nodes are triples
  $t=(A_t,X_t,Y_t)$ satisfying the following conditions:
  \begin{ealph}
  \item\label{ea:idec-a} $A_t$ is a connected induced subgraph of $G$ with
    $|N^G(A_t)|\le a$. To simplify the notation, in the following we let
    $C_t:=G\big[N^G[A_t]\big]$.
  \item \label{ea:idec-b}
    $X_t\subseteq V(C_t)$ such that $N^G(A_t)\subset X_t$ and $|X_t|=\min\{a+1,|C_t|\}$.
    \item\label{ea:idec-c}
      $Y_t\subseteq V(C_t)$ such that $|Y_t|<m$. (Actually, $Y_t$ will
      be empty for d-nodes and e-nodes.)
  \end{ealph}
  Let us call such triples ``nodes'' and let $U$ be the set of all
  nodes (the actual nodes of $D$ will form a subset of $U$). For every
  node $t\in U$ we let $\comp(t):=V(A_t)$, $\sep(t):= N^G(A_t)$, and
  $\cone(t):=V(C_t)$.
  \begin{ealph}[resume]
  \item\label{ea:idec-d} A \emph{b-node} is a node $t\in U$ such that
    for every connected component $A$ of $C_t\setminus
    Y_t$ it holds that $|(V(A)\cap X_t)\cup Y_t|<|X_t|$.
  \end{ealph}
  Let $V_b$ be the set of all b-nodes. Let $U_b$ be the set
  of all nodes $t\in U$ for which there exists a $Y\subseteq V(C_t)$ of size
  $|Y|<m$ such that for every connected component of $A$ of $C_t\setminus Y$ it
  holds that $|(V(A)\cap X_t)\cup Y|<|X_t|$. Note that $V_b\subseteq
  U_b$ and that that for every $t\in
  U\setminus U_b$ the set $X_t$ is $m$-unbreakable in $C_t$.
  \begin{ealph}[resume]
  \item\label{ea:idec-f} An \emph{e-node} is a node $t\in U\setminus
    U_b$ such that $Y_t=\emptyset$ and the algorithm of
    Lemma~\ref{lem:conncliquestar} on $C_t$, $\ell$, $m$, and $X_t$
    returns star decomposition $\Sigma_t:=\Sigma_{X_t}$ of $C_t\cup K[X_t]$.
  \item\label{ea:idec-e} A \emph{d-node} is a node $t\in U\setminus
    U_b$ such that $Y_t=\emptyset$ and the algorithm of
    Lemma~\ref{lem:conncliquestar} on $C_t$, $\ell$, $m$, and $X_t$
    returns an image $I$ of $K_\ell$ in $C_t$ that is $m$-attached to
    $X_t$. In this case, the algorithm of Lemma~\ref{lem:conndegstar}
    applied to $C_t$, $k$, the set $X_t$, and the image $I$
    computes a star decomposition $\Sigma_t:=\Sigma_{X_t}$ of $C_t\cup K[X_t]$
    (since $K_k\not\preceq_T C_t$ by assumption).
 \end{ealph}
 Let $V_d$ and $V_e$ be the sets of d-nodes and e-nodes,
 respectively. Note that the three sets $V_b,V_d,V_e$ are mutually
 disjoint. We let $V(D):=V_b\cup V_d\cup V_e$.  
   \begin{claim}\label{claim:idec1}
     Let $A$ be a (nonempty) connected induced subgraph of $G$ with
     $|N^G(A)|\le a$. Then there is a node $t\in V(D)$ such that
     $A_t=A$.

     \proof 
     Let $C:=G\big[N^G[A]\big]$, and choose an arbitrary $X\subseteq V(C)$
     such that $N^G(A)\subset X$ and $|X|=\min\{a+1,|C|\}$. Clearly,
     such a set $X$ exists, because $A\neq\emptyset$ and $|N^G(A)|\le a$.

     If there is a set $Y\subseteq V(G)$ such that $|Y|<m$ and for every
     connected component $A'$ of $C\setminus Y$ it holds that $|(V(A')\cap
     X)\cup Y|<|X|$, then $(A,X,Y)\in V_b$.

     Suppose there is no such set $Y$. Then $(A,X,\emptyset)\not\in
     U_b$ and thus  $(A,X,\emptyset)\in V_e\cup V_d$.  \uend
 \end{claim}

By \ref{ea:idec-f} and \ref{ea:idec-e}, for all $t\in V_e\cup
 V_d$ we have a star decomposition
 $\Sigma_t=:(T_t,\sep_t,\comp_t)$. Let $s_t$ be the center of $T_t$.
 To define the edge relation $E(D)$, for every node $t\in V(D)$ we
define the set $N_+^D(t)$ of its children in $D$.
\begin{ealph}[resume]
  \item\label{ea:idec-g}
  For $t\in V_b$, we let $N_+^D(t)$ be the set of all $u\in V(D)$ such
  that $A_u$ is a connected component of $C_t\setminus(X_t\cup Y_t)$.
\item\label{ea:idec-h}
  For $t\in V_d\cup V_e$, we let $N_+^D(t)$ be the set of all $u\in V(D)$ such
  that $A_u$ is a
  connected component of $C_t\setminus \bag_t(s_t)$.
\end{ealph}
This completes the definition of the decomposition
$\Delta_G=(D,\sep,\comp)$. 

  

\begin{claim}[resume]\label{claim:idec3}
  $\Delta_G$ is a treelike decomposition of $G$.

  \proof
  It follows immediately from the definitions of $\sep$ and $\comp$
  that $\Delta_G$ satisfies \ref{li:tl2}.

  To verify \ref{li:tl3}, let $tu\in E(D)$.  We have $X_t\subseteq
  \bag_t(s_t)$ (either by \ref{ea:idec-g} or by the statements of 
  Lemmas~\ref{lem:conncliquestar} and
  \ref{lem:conndegstar}). Therefore, by \ref{ea:idec-g} and
  \ref{ea:idec-h} we have
  \begin{equation}
    \label{eq:idec1}
  \comp(u)=V(A_u)\subseteq V(C_t)\setminus \bag_t(s_t) \subseteq
  V(C_t)\setminus X_t\subset V(C_t)\setminus N^G(A_t)=\comp(t).
  \end{equation}
  Moreover, if
  $t\in V_b$, then we have $N^{C_t}(A_u)\subseteq X_t\cup Y_t$. Since
  every vertex of $C_t$ with a neighbor outside $C_t$ is in
  $N^G(A_t)\subseteq X_t$ and we have $V(A_u)\cap X_t=\emptyset$, this
  implies $N^{G}(A_u)\subseteq X_t\cup Y_t$. Hence
  $\cone(u)=V(A_u)\cup N^G(A_u)\subseteq V(C_t)=\cone(t)$. If $t\in
  V_e\cup V_b$ then we have $N^{C_t}(A_u)\subseteq
  \bag_t(s_t)$. Again, every vertex of $C_t$ with a neighbor outside
  $C_t$ is in $N^G(A_t)\subseteq X_t\subseteq\bag_t(s_t)$ and
  $V(A_u)\cap\bag_t(s_t)=\emptyset$.  Hence $\cone(u)=V(A_u)\cup
  N^G(A_u)\subseteq V(C_t)=\cone(t)$.

  Note that in \eqref{eq:idec1} we proved that for all edges $tu\in E(D)$ the inclusion
  $\comp(u)\subset\comp(t)$ is strict. This implies that $D$ is acyclic, that
  is, \ref{li:tl1}.

 To verify \ref{li:tl4}, let $t\in V(D)$ and $u_1,u_2\in N_+^D(t)$. For
 $i=1,2$, we let $C_i:=C_{u_i}$ and $X_i:=X_{u_i}$ and
  $A_i:=A^t_{u_i}$.
  \begin{cs}
    \case1
    $t\in V_b$.\\
    Then by \ref{ea:idec-g}, both $A_1$ and $A_2$ are connected
    components of $C_t\setminus(X_t\cup Y_t)$. Hence either $A_1=A_2$
    or $A_1\cap A_2=\emptyset$. If $A_1=A_2$ then
    $\comp(u_1)=V(A_1)=V(A_2)=\comp(u_2)$ and
    $\sep(u_1)=N^G(A_1)=N^G(A_2)=\sep(u_2)$ and thus $u_1\doublesmile
    u_2$. Suppose that $A_1\cap A_2=\emptyset$.  Note that we also
    have $V(A_1)\cap N^G(A_2)\subseteq V(A_1)\cap (X_t\cup Y_t)=\emptyset$
    and, symmetrically, $V(A_2)\cap N^G(A_1)=\emptyset$. This implies
   $
    \cone(u_1)\cap\cone(u_2)=V(C_1)\cap
    V(C_2)=N^G(A_1)\cap N^G(A_2)=\sep(u_1)\cap\sep(u_2).
   $
   \case2
    $t\in V_e\cup V_d$.\\
    Then both $A_1$ and
    $A_2$ are connected components of $C_t\setminus \bag_t(s_t)$, 
    and we can argue as in Case~1. 
 \end{cs}
  To verify \ref{li:tl5}, let $A$ be a connected component of
  $G$. Then $N^G(A)=\emptyset$, and by
  Claim~\ref{claim:idec1} there is a $t\in V(D)$ such that
  $C_t=A$. For each such $t$, we have
  $\sep(t)=N^G(A)=\emptyset$ and $\comp(t)=V(A)$.
  \uend
\end{claim}

\begin{claim}[resume]\label{claim:idec4}
  Let $t\in V(D)$.
  \begin{enumerate}
 \item If $t\in V_b$ then $\bag(t)=X_t\cup Y_t$.
 \item If $t\in V_e\cup V_d$ then $\bag(t)=\bag_t(s_t)$ and
   $\tau(t)\subseteq\tau_t(s_t)$.
 \end{enumerate}

 \proof Recall that $\bag(t)=\cone(t)\setminus\bigcup_{u\in
   N_+^D(t)}\comp(u)=V(C_t)\setminus\bigcup_{u\in N_+^D(t)}V(A_u)$. 

 To prove (1), suppose that $t\in V_b$. It
 follows from \ref{ea:idec-g} that for all $u\in N_+^D(t)$ we have $V(A_u)\subseteq
 V(C_t)\setminus(X_t\cup Y_t)$. Hence $(X_t\cup
 Y_t)\subseteq\bag(t)$. To prove the converse inclusion, we shall
 prove that for every connected component $A$ of $C_t\setminus(X_t\cup
 Y_t)$ there is a $u\in N_+^D(t)$ with $A_u=A$. By \ref{ea:idec-g} and
 Claim~\ref{claim:idec1}, it suffices to prove that $|N^G(A)|\le
 a$.
To see this, observe first that $N^G(A)=N^{C_t}(A)$, because
 $A\subseteq C_t\setminus X_t\subseteq A_t$ and thus $N^G(A)\subseteq
 N^G[A_t]=V(C_t)$. Now let $A'$ be the connected component of $C_t\setminus
 Y_t$ with $A\subseteq A'$. Then $N^{C_t}(A)\subseteq (V(A')\cap
 X_t)\cup Y_t$. By \ref{ea:idec-d}, we have $|(V(A')\cap X_t)\cup
 Y_t|<|X_t|=a+1$. Thus $|N^G(A)|=|N^{C_t}(A)|\le a$.

 To prove (2), let $t\in V_e\cup V_d$. By \ref{ea:idec-h}, for all $u\in N_+^D(t)$ we have $V(A_u)\subseteq
 V(C_t)\setminus\bag_t(s_t)$. Hence $\bag_t(s_t)\subseteq\bag(t)$. For
 the converse inclusion, let $A$ be a connected component of
 $C_t\setminus\bag_t(s_t)$.  Then
 there is a tip $x$ of $T_t$ such that
 $A$ is a connected component of $C_t[\comp_t(x)]=G[\comp_t(x)]$. We
 have $N^G(A)\subseteq\sep_t(x)$,  and as the adhesion of $\Sigma_t$
 is $<|X_t|$, we have $|N^G(A)|\le|\sep_t(x)|<|X_t|=a+1$. Thus by
 Claim~1 and \ref{ea:idec-h}, there is a $u\in N_+^D(t)$ with $A_u=A$. 

 It remains to prove that $\tau(t)\subseteq\tau_t(s_t)$. First, note
 that for all $u\in N_+^D(t)$ there is an $x\in N_+^{T_t}(s_t)$ such that
 $\sep(u)\subseteq\sep_t(x)$.  Furthermore, $\Sigma_t$ is a
 decomposition of $C_t\cup K[X_t]$, thus $\sep(t)\subseteq X_t$ is a
 clique in $\tau_t(s_t)$.  Let us remark that $\tau_t(s_t)\subseteq
 \tau(t)$ is not necessarily true: $X_t$ is a proper superset of
 $\sep(t)$, thus $X_t$ is a clique $\tau_t(s_t)$, but it is not
 necessarily a clique in $\tau(t)$.
 \uend
\end{claim}

It follows from \ref{ea:idec-a} that the adhesion of $\Delta_G$ is at
most $a$. By Claim~\ref{claim:idec4}(1), every $t\in V_b$
satisfies (i). By Claim~\ref{claim:idec4}(2) and
Lemmas~\ref{lem:conndegstar} and \ref{lem:conncliquestar}, every $t\in
V_d$ satisfies (ii) and every $t\in V_e$ satisfies (iii).

It it easy to see that the decomposition mapping $\Delta$ is
polynomial time computable. Indeed, note first that the set $U$ has
size $O(n^{a+1+3m-2+m-1})$ (here we use $n^{a+1}$ as an upper bound
for the number of connected induced subgraphs $A$ of $G$ with $N^G(A)|\le a$)
and that the set is polynomial time computable. Remember that the
parameters $a,m$ et cetera are all treated as constants depending only
on $H$. The subset $U_b$ is also polynomial time computable, because
to decide whether $t\in U_b$ we can go through all subsets $Y\subseteq
V(C_t)$ of size less than $m$ and see if the condition is
satisfied. Now it follows from Lemmas~\ref{lem:conncliquestar} and
\ref{lem:conndegstar} that the sets $V_e$ and $V_d$ are polynomial
time computable. Hence $V(D)$ is polynomial time computable. Since for
nodes $t\in V_e\cup V_d$ the star decomposition $\Sigma_t$ is
polynomial time computable (again by Lemmas~\ref{lem:conncliquestar}
and \ref{lem:conndegstar}), the edge relation $E(D)$ is polynomial
time computable as well. Since the mappings $\sep$ and $\comp$ are
almost trivially polynomial time computable, this shows that $\Delta$
is polynomial time computable.

It remains to prove that $\Delta$ is
invariant. To prove this, we take isomorphic graphs $G,G'$ with $H\not\preceq G,G'$ and let $f$ be an
isomorphism from $G$ to $G'$. Let
$\Delta_G=(D,\sep,\comp)$ and $\Delta_{G'}=(D',\sep',\comp')$. 
We define the sets $U,U_b,V_b,V_e,V_d$ for $G$ as above and let
$U',U_b',V_b',V_e',V_d'$ be the corresponding sets for $G'$. We denote the constituents of a node
$t'\in U'$ by $(A'_{t'},X'_{t'},Y'_{t'})$ and let
$C'_{t'}:=G'\big[N^{G'}[A'_t]\big]$. For $t'\in V_e\cup V_d$ we denote the
star decomposition of $C'_{t'}$ (obtained as above) by
$\Sigma'_{t'}$. The isomorphism $f$ has a natural extension to subsets
of $V(G)$, tuples of subsets, and similar objects defined in terms of
$V(G)$. We denote this extension by $f^*$. As $f$ is an
isomorphism, we obviously have $f^*(U)=U'$, $f^*(U_b)=U_b'$, and
$f^*(V_b)=V_b'$. Moreover, for every $t\in U$ we have
$f^*(A_t)=A'_{f^*(t)}$, $f^*(C_t)=C'_{f^*(t)}$, et cetera. It follows from the
invariance conditions of Lemmas~\ref{lem:conncliquestar} and
\ref{lem:conndegstar} that $f^*(V_e)=V'_e$ and $f^*(V_d)=V'_d$ and that
for every $t\in V_e\cup V_d$ there is an isomorphism $g_t$ from $T_t$
to $T'_t$ such that $f^*(\sep_t(x))=\sep_{f^*(t)}(g_t(x))$ and
$f^*(\comp_t(x))=\comp_{f^*(t)}(g_t(x))$ for all $x\in
V(T_t)$. But this implies that $N_+^D(f^*(t))=\{ f^*(u)\mid u\in N_+^D(t)\}$.
As $f^*(C_t)=C'_{f^*(t)}$ and $f^*(X_t)=X'_{f^*(t)}$ and $f^*(Y_t)=Y'_{f^*(t)}$,
we also have $N_+^D(f^*(t))=\{ f^*(u)\mid u\in N_+^D(t)\}$ for all $t\in V_b$.
Hence the restriction of $f^*$ to $V(D)$ is an isomorphism from $D$ to $D'$. As
$f^*(A_t)=A'_{f^*(t)}$ for all $t\in V(D)$, we have
$f(\comp(t))=\comp'(f^*(t))$ and thus
$f(\sep(t))=f(N^G(\comp(t)))=N^{G'}(f(\comp(t)))=N^{G'}(\comp'(f^*(t)))=\sep'(f^*(t))$. 
This proves that $\Delta$ is invariant.
\end{proof}


%% file: c5dec.tex
\begin{tikzpicture}
 [
 line width=0.3mm,
 vertex/.style={draw,fill=white,circle,line width=0.25mm,inner sep=0mm,minimum
   size=4mm},
 tvertex/.style={draw,line width=0.25mm,inner sep=1mm},
 every edge/.style={draw,->}
 ]
 \useasboundingbox (-1,-7.5) (1,1);
 \small
 \draw circle (1cm);
\node[vertex] (1) at (90:1cm) {$1$};
 \node[vertex] (2) at (162:1cm) {$2$};
 \node[vertex] (3) at (234:1cm) {$3$};
 \node[vertex] (4) at (306:1cm) {$4$};
 \node[vertex] (5) at (18:1cm) {$5$};

 \path (0,-1.5) node {\normalsize(a)};

 \node[tvertex] (13) at (-0,-3.5) {$\{1,3\}$};

\node[tvertex] (123) at (-1,-6.5) {$\{1,2,3\}$};
 \node[tvertex] (135) at (0.5,-5) {$\{1,3,5\}$};
 \node[tvertex] (345) at (1,-6.5) {$\{3,4,5\}$};
 \path (13) edge (123) (13) edge (135) (135) edge (345);

 \path (0,-7.2) node {\normalsize(b)};

\end{tikzpicture}
\hspace{2cm}
  \begin{tikzpicture}
  [
  line width=0.3mm,
  vertex/.style={draw,line width=0.25mm,inner sep=0mm,minimum
    size=4.5mm},
  every edge/.style={draw,->}
  ]
  \scriptsize
  \foreach \name/\b/\bb/\alpha in {
    b/1/3/66, c/1/4/114, d/1/4/138, e/2/4/186, f/2/4/210, g/2/5/258, h/2/5/282,
    i/3/5/330, j/3/5/354
  }
  \node[vertex] (\name) at (\alpha:4cm) { $\{\b,\bb\}$ };
  \node[vertex,fill=black!20] (a) at (42:4cm) {$\{1,3\}$};
  
  \foreach \name/\b/\bb/\bbb/\alpha in {
    B/1/3/4/77, C/1/3/4/103, D/1/2/4/147, E/1/2/4/177, F/2/4/5/219, G/2/4/5/246,
    H/2/3/5/294, I/2/3/5/321, J/1/3/5/3    
  }
  \node[vertex] (\name) at (\alpha:2.6cm) { $\{\b,\bb,\bbb\}$ };
    \node[vertex,fill=black!20] (A) at (33:2.6cm) {$\{1,3,5\}$};

  \foreach \name/\b/\bb/\bbb/\alpha in {
    145/1/4/5/126, 234/2/3/4/198, 125/1/2/5/270
  }
  \node[vertex] (\name) at (\alpha:1.2cm) { $\{\b,\bb,\bbb\}$ };
 \node[vertex,fill=black!20] (123) at (54:1.2cm) {$\{1,2,3\}$};
 \node[vertex,fill=black!20] (345) at (342:1.2cm) {$\{3,4,5\}$};

 \foreach \v/\w in { 
   a/123, a/A, A/345, b/123, b/B, B/145, c/145, c/C, C/123, d/145,
   d/D, D/234, e/234, e/E, E/145, f/234, f/F, F/125, g/125, g/G,
   G/234, h/125, h/H, H/345, i/345, i/I, I/125, j/345, j/J, J/123}
  \path (\v) edge (\w);

 \path (0,-4.5) node {\normalsize(c)};
\end{tikzpicture}


%% file: canon.tex
\section{Canonization}
\label{sec:canonization}

A \emph{canonisation mapping} $\Kc$ for a class $\CC$ of graphs is a
mapping that associates with each graph $G\in\CC$ a graph $\Kc(G)\cong
G$ such that for all $G,H\in\CC$ we have $G\cong
H\iff\Kc(G)=\Kc(H)$. That is, $\Kc(G)$ and $\Kc(H)$ are not only
isomorphic, but they are actually the same graph on the same set of
vertices. Thus the isomorphism of $G$ and $H$ can be tested simply by
comparing $\Kc(G)$ and $\Kc(H)$. A \emph{canonisation algorithm}
computes a canonisation mapping. Without loss of generality we may
always assume a canonisation mapping $\Kc$ to map graphs $G$ to graphs
$\Kc(G)$ with vertex set $V(\Kc(G))=[n]$, where $n:=|G|$. We say that
a class $\CC$ of graphs \emph{admits polynomial time canonisation} if
there is a polynomial time algorithm that computes a canonisation
mapping for $\CC$.

\begin{fact}[Babai and Luks~\cite{babluk83}]\label{fact:babluk}
  For every $d\in\NN$ the class of all graphs of maximum degree at
  most $d$ admits polynomial time canonisation.
\end{fact}

\begin{fact}[Ponomarenko~\cite{pon88}]\label{fact:pon}
  For every graph $H$ the class of all graphs excluding $H$ as a minor
  admits polynomial time canonisation.
\end{fact}

An alternative proof of Fact~\ref{fact:pon} can be found in \cite{gro10+a}.

Our goal in this section is to prove a ``Lifting Lemma'' that allows us to
lift a canonisation from the torsos of a treelike decomposition of a graph to
the whole graph. To be able to prove such a lemma, we need to work with more
general structures than graphs and a stronger notion of canonisation.

We often denote tuples $(v_1,\ldots,v_k)$ by $\vec v$. For $\vec
v=(v_1,\ldots,v_k)$, by $\tilde v$ we denote the set
$\{v_1,\ldots,v_k\}$. 
A \emph{vocabulary} is a finite set of relation symbols, each of which
has a prescribed arity in $\NN$. (Note that we admit $0$-ary relation
symbols. For every set $S$ the set $S^0$ just consists of the empty tuple.) Let $\lambda$ be a vocabulary. A
\emph{weighted $\lambda$-structure} $A$ consists of a \emph{universe}
(or \emph{vertex set}) $V(A)$ and for each $k$-ary relation symbol
$R\in\lambda$ a mapping $R^A:V(A)^k\to\NN$. A (plain)
$\lambda$-structure is a weighted $\lambda$-structure $A$ with
$\operatorname{range}(R^A)\subseteq\{0,1\}$ for all
$R\in\lambda$. We usually identify a function $R^A:V(A)^k\to\{0,1\}$ with
the relation $R(A):=\{\vec v\in V(A)^k\mid R^A(\vec v)=1\}$ and view a
plain structure as a finite set (the universe) together with a collection of relations
on this universe. For example, graphs and digraphs may be viewed as plain
$\{E\}$-structures, where $E$ is a binary relation symbol. Graphs with
multiple edges may be viewed as weighted $\{E\}$-structures. 

Let $\lambda,\mu$ be vocabularies with $\lambda\subseteq\mu$, and let
$A$ be a weighted $\lambda$-structure and $B$ a weighted
$\mu$-structure. Then $A$ is the \emph{$\lambda$-restriction} of $B$
if $V(A)=V(B)$ and $R^A=R^B$ for all symbols
$R\in\lambda$. Conversely, $B$ is a \emph{$\mu$-expansion} of $A$ if
$A$ is the $\lambda$-restriction of $B$. For every $W\subseteq V(A)$, we
define the \emph{induced substructure} $A[W]$ to be the weighted
$\lambda$-structure with universe $V(A[W]):=W$, relations
$R^{A[W]}:=R^A\restriction_{W^k}$ for all $k$-ary $R\in\lambda$. We
let $A\setminus W:=A[V(A)\setminus W]$. If $f$ is an injective mapping with
domain $V(A)$, we let $f(A)$ be the weighted $\lambda$-structure with
universe $V(f(A)):=f(V(A))$ and mappings $R^{f(A)}$ defined by
$R^{f(A)}(f(\bar a)):=R^A(\bar a)$. If $A$ and $B$ are weighted
$\lambda$-structures such that for all $k$-ary $R\in\lambda$ and all
$\vec v\in V(A)^k\cap V(B)^k$ we have $R^A(\vec v)=R^B(\vec v)$, then
we define the union $A\cup B$ to be the weighted $\lambda$-structure
with $V(A\cup B):=V(A)\cup V(B)$ and
\[
R^{A\cup B}(\vec a):=
\begin{cases}
  R^A(\vec a)&\text{if }\vec a\in V(A)^k,\\
  R^B(\vec a)&\text{if }\vec a\in V(B)^k,\\
  0&\text{otherwise}.
\end{cases}
\]
for all $k$-ary relation symbols $R\in\lambda$ and $\vec a\in V(A\cup
B)^k$.

The \emph{Gaifman graph} of a weighted $\lambda$-structure
$A$ is the graph $G_A$ with vertex set $V(G_A):=V(A)$ and edge set
\[
E(G_A):=\Big\{vw\in\binom{V(A)}{2}\Bigmid \exists k\text{-ary }R\in\lambda,\vec
v\in V(A)^k\text{ with }R^A(\vec v)>0\text{ and }v,w\in\tilde v\Big\}.
\]
An \emph{isomorphism} from a weighted $\lambda$-structure $A$ to a
weighted $\lambda$-structure $B$ is a bijective mapping $f:V(A)\to
V(B)$ such that for all $k$-ary $R\in\lambda$ and all $\vec v\in
V(A)^k$ we have $R^A(\vec v)=R^B(f(\vec v))$. Canonisation mappings
and algorithms for weighted structures are defined in the obvious
way. We say that a class $\CC$ of graphs \emph{admits polynomial time
  strong canonisation} if for every vocabulary $\lambda$ there is a
polynomial time computable canonisation mapping for the class of all
weighted $\lambda$-structures with Gaifman graph in $\CC$.

\begin{lem}\label{lem:scdeg}
  For all $d\in\NN$ the class of all graphs of maximum degree at most
  $d$ admits polynomial time strong canonisation.
\end{lem}

\begin{proof}
  We derive this from Fact~\ref{fact:babluk} by a simple gadget
  construction. 

  We first define auxiliary graphs $N_n$ for all $n\in\NN$: Let
  $b_{\ell-1}\ldots b_0$ be the binary representation of
  $n=\sum_{i=0}^{\ell-1}b_i2^i$. Then $N_n$ has vertices
  $v_0,\ldots,v_{\ell-1}$ and a vertex $w_i$ for every $i\in[0,\ell-1]$
  such that $b_i=1$. It has edges between $v_{i-1}$ and $v_i$ for all
  $i\in[\ell-1]$ and between $v_i$ and $w_i$ for all $i\in[0,\ell-1]$
  such that $b_i=1$. We call $v_0$ the ``anchor'' of $N_n$.

  Next, we define further auxiliary graphs $M^m_n$ for all $m,n\in\NN$
  (see Figure~\ref{fig:gadget}): We take a copy of $N_m$, a copy of
  $N_n$, a triangle $T$ with vertex set $x,y,z$ and vertices
  $w_1,w_2,w_3,w_4$. We add edges between $x$ and the anchor of
  $N_m$, between $y$ and the anchor of $N_n$, and additional edges
  $yw_1,zw_2,zw_3,zw_4$. We call $z$ the anchor of $M^m_n$. Note that
  $z$ is the only vertex of degree $\ge 5$ in $M^m_n$, and $y$ is the
  only vertex of degree $4$ and $x$ is the only vertex of degree $3$
  that is contained in a triangle. Thus whenever we have an
  isomorphism $f$ between a copy of $M^m_n$ and a copy of
  $M^{m'}_{n'}$, we must have $m=m'$ and $n=n'$, and $f$ maps the
  anchor of the first copy to the anchor of the second copy.
\begin{figure}
\begin{center}
{\small \def\svgwidth{0.5\linewidth}%
\executeiffilenewer{gadget.svg}{gadget.pdf}%
{inkscape -z -D --file=gadget.svg %
--export-pdf=gadget.pdf --export-latex}%
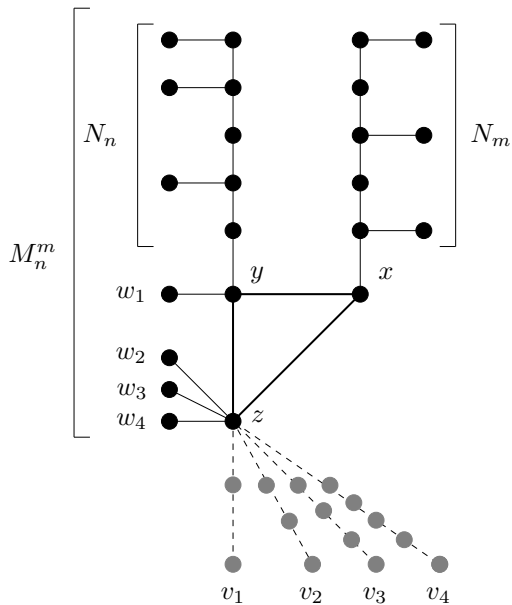%
}
\end{center}
\caption{The auxiliary graph $M_n^m$ in the proof of Lemma~\ref{lem:scdeg}.}\label{fig:gadget}

\end{figure}

  For some $d\in\NN$, let $\CD_d$ be the class of all graphs of
  maximum degree at most $d$. Let
  $\lambda=\{R_1,\ldots,R_m\}$ be a vocabulary,
  where $R_i$ is $k_i$-ary. Let $A$ be a weighted
  $\lambda$-structure with $G_A\in\CD_d$. 

  We define a graph $H_A$ as follows: We start by taking the graph
  with vertex set $V(A)$ and no edges.  For every $i\in[m]$ and for
  every tuple $\vec v=(v_1,\ldots,v_{k_i})\in V(A)^{k_i}$ with
  $r:=R_i^A(\vec v)>0$, we add a fresh copy $M_{i,\vec v}$ of $M^i_r$ and for
  $1\le j\le k_i$ a path of length $j+1$ from the anchor of $M_{i,\vec v}$ to
  $v_j$.

  Clearly, for every vertex $x\in V(H_A)\setminus V(A)$ the degree of
  $x$ in $H_A$ is at most $\max\{k_i+5\mid i\in[m]\}$. Observe that
  for every $v\in V(A)$, and every $i\in[m]$ there are at most
  $(d+1)^{k_i}$ tuples $\vec v\in V(A)^{k_i}$ with $R_i^A(\vec v)>0$
  and $v\in\tilde v$. This means that the degree of $v$ in $H_A$ is at
  most $\sum_{i\in [m]}(d+1)^{k_i}$. Thus we have proved:
  \begin{ealph}
  \item The maximum degree of $H_A$ is at most
    $k:=\max\{k_i+5,\sum_{i\in [m]}(d+1)^{k_i}\}$.
  \end{ealph}
  Now let $A,B$ be weighted $\lambda$-structures. Observe that the only
  triangles in $H_A$ and $H_B$ appear within the gadgets $N^m_n$. Hence every
  isomorphism from $H_A$ to $H_B$ must map $V(A)$ to $V(B)$, and it must map
  gadgets with matching parameters onto one another. It follows that:
  \begin{ealph}[resume]
    \item
    For all weighted $\lambda$-structures $A,B$ we have $A\cong B\iff H_A\cong
    H_B$.
  \end{ealph}
  Finally, it is easy to verify the following two algorithmic claims:
  \begin{eroman}
    \item There is a polynomial time computable mapping $\mathfrak a$
      from the class of weighted $\lambda$-structures to the class of
      graphs such that $\mathfrak a(A)\cong H_A$ for every weighted
      $\lambda$-structure $A$.
    \item There is a polynomial time computable mapping $\mathfrak b$
      from the class of graphs to the class of weighted
      $\lambda$-structures such that $\mathfrak b(H_A)\cong A$ for
      every weighted $\lambda$-structure $A$.
  \end{eroman}
  Now let $\mathfrak c$ be a polynomial time computable canonisation
  mapping for the class of all graphs of maximum degree at most
  $k$. Then $\mathfrak b\circ\mathfrak c\circ\mathfrak a$ is a
  polynomial time computable canonisation mapping for the class of all
  all weighted $\lambda$-structures $A$ with $G_A\in\CD_d$.
\end{proof}

\begin{lem}\label{lem:scmin}
  For every graph $H$, the class of graphs $G$ with $H\not\preceq G$ 
  admits polynomial time strong canonisation.
\end{lem}

\begin{proof}
  It suffices to prove that this is true for every $H=K_k$ with $k\ge
  4$.  This follows from Fact~\ref{fact:pon} using the same gadget
  construction as the proof of Lemma~\ref{lem:scdeg}, observing that
  for every weighted structure $A$ and every $k\ge 4$ we have
  $K_{k-1}\not\preceq G_A\implies K_k\not\preceq H_A$. To see this
  last implication, suppose that there is a $K_k$-minor image in
  $H_A$. Clearly, it has to be contained in the 2-connected component
  of $H_A$ not containing any of $M_{n}^m$ gadgets. It is easy to see
  that at most one branch set of the $K_k$-minor image can be disjoint
  from $V(A)$; let us consider the $K_{k-1}$-minor image where every
  branch set intersects $V(A)$. Now it can be verified that if two
  branch sets touch, then they have to contain vertices of $V(A)$ that
  are adjacent in $G_A$.
\end{proof}

We define the \emph{lexicographical order}
  $\lex^\lambda$ on weighted $\lambda$-structures $A$
  with $V(A)\subseteq\NN$. Let $\lambda=\{R_1,\ldots,R_m\}$, where $R_i$ is
  $k_i$-ary. The order $\lex^\lambda$ actually not only depends on the
  set $\lambda$, but on the order in which the relations are
  listed. Hence we fix this order. We first review the
  lexicographical order on tuples and sets of integers:
\begin{itemize}
\item\label{ea:ll1} For tuples $\vec x=(x_1,\ldots,x_k)\in\NN^k$, $\vec
  y=(y_1,\ldots,y_\ell)\in\NN^\ell$ we let $\vec x\slex\vec y$ if and only if
  either there exists an $i\le\min\{k,\ell\}$ such that $x_i<y_i$ and
  $x_j=y_j$ for $1\le j<i$ or $k<\ell$ and $x_i=y_i$ for all $i\le k$. 
\item\label{ea:ll2} For sets $X,Y\subseteq\NN$ we let $X\slex Y$ if and only if there exists an $i\in Y\setminus X$ such that for all $j<i$ it holds that
  $j\in X\iff j\in Y$.
\end{itemize}
Now let $A,B$ be weighted $\lambda$-structures with $V(A),V(B)\subseteq\NN$. Then we let
$A\slex^\lambda B$ if one of the following conditions is satisfied:
\begin{itemize}
\item\label{ea:ll3} $V(A)\slex V(B)$. Note that if both $V(A)$ and $V(B)$ are initial
  segments of the positive integers then this just means $|A|<|B|$.

\item\label{ea:ll4} $V(A)=V(B)=:V$ and there is an $i\in[m]$ and a tuple $\vec a\in V^{k_i}$
  such that $R_j^A=R_j^B$ for $1\le j<i$, and $R_i^A(\vec a)<R_i^B(\vec
  a)$, and $R_i^A(\vec b)=R_i^B(\vec b)$ for all $\vec b\in V^{k_i}$ with
  $\vec b\slex\vec a$.
\end{itemize}
We let $A\lex^\lambda B$ if $A\slex^\lambda B$ or $A=B$.
Note that $\lex^\lambda$ is indeed a linear order on the class of weighted
$\lambda$-structures whose universe is a set of natural numbers and that given
$A,B$, it can be decided in polynomial time whether $A\lex^\lambda B$.

We want to prove that not only the class of bounded-degree graphs
admits polynomial-time strong canonization, but also the more general
class containing those graphs that can be made bounded-degree by the
removal of $k$ vertices. We deduce this from a more general statement
about enlarging graphs by $k$ additional vertices.  For every class $\CC$
of graphs and every $k\in\NN$, we let
\[
\CN_k(\CC):=\{G\mid\exists X\subseteq V(G):\;|X|\le k\text{ and }G\setminus
X\in\CC\}.
\]
We call the graphs in $\CN_k(\CC)$ \emph{$k$-enlargements} of the graphs in
$\CC$. 

\begin{lem}
  Let $\CC$ be a class of graphs that is decidable in polynomial time
  and admits polynomial time strong
  canonisation. Then for every $k\in\NN$ the class $\CN_k(\CC)$ admits
  polynomial time strong canonisation.
\end{lem}

\begin{proof}
  It suffices to prove that $\CN_1(\CC)$ is polynomial time decidable
  and admits polynomial time strong canonisation. The full assertion follows
  by induction.

  We first observe that $\CN_1(\CC)$ is polynomial time decidable. To
  prove that it admits polynomial time strong canonisation, let
  $\lambda$ be a vocabulary. For every $k$-ary relation symbol
  $R\in\lambda$ and every set $S\subset [k]$ we introduce a fresh
  $(k-|S|)$-ary relation symbol
  $R_S$, and we let $\lambda^*$ be the extension of $\lambda$ by all
  these new relation symbols. For every weighted $\lambda$-structure
  $A$ and every $x\in V(A)$ we let $A_x$ be the $\lambda^*$-expansion of
  $A\setminus\{x\}$ defined as follows: For every $k$-ary
  $R\in\lambda$ and $S=\{i_1,\ldots,i_\ell\}\subseteq[k]$, where
  $i_1<\ldots<i_\ell$, we let 
  \[
  R_S^{A_x}(v_1,\ldots,v_{k-\ell}):=R^A(v_1,\ldots,v_{i_1-1},x,v_{i_1},\ldots,v_{i_2-2},x,v_{i_2-1},\;\ldots\;,v_{i_\ell-\ell},x,v_{i_{\ell}+1-\ell},\ldots,v_{k-\ell}).
  \]
  Then clearly for all weighted $\lambda$-structures $B$ and all
  $y\in V(B)$, there is
  an isomorphism $f$ from $A$ to $B$ with $f(x)=y$ if and only if
  $A_x$ and $B_y$ are isomorphic. Furthermore, we can compute
  $A_x$ from $A$ and $x$ and conversely $A$ from $A_x$ and $x$ in
  polynomial time. 

  Now let $\Kc^*$ be a polynomial time computable canonisation mapping
  for the class of all weighted $\lambda^*$-structures $A^*$ with
  $G_{A^*}\in \CC$. We shall define a polynomial time computable
  canonisation mapping $\Kc$ for the class of all
  weighted $\lambda$-structures with Gaifman graph in $\CN_1(\CC)$.
  Let $A$ be a weighted $\lambda$-structure with
  $G_A\in\CN_1(\CC)$. If $G_A\in\CC$, we simply let $\Kc(A)$ by the
  $\lambda$-restriction of $\Kc^*(A^*)$, where $A^*$ is the
  $\lambda^*$-expansion of $A$ with $R_S^{A^*}\equiv0$ for all
  $R_S\in\lambda^*\setminus\lambda$. 

  In the following, we assume that $G_A\not\in\CC$. Let $X$ be the set
  of all $x\in V(A)$ such that $G_{A\setminus\{x\}}\in\CC$. Note that
  $X$ is nonempty, because $G_A\in\CN_1(\CC)\setminus\CC$, and
  polynomial time computable, because $\CC$ is polynomial time
  decidable. As usual, we assume that $\Kc^*$ maps every structure
  $A^*$ to a structure whose universe is an initial segment of
  the positive integers. For every $x\in X$, we let $C_x:=\Kc^*(A_x)$. Let $C^*$ be
  lexicographically minimal among all $C_x$ for $x\in X$. Let
  $n:=|A|$; then $V(C^*)=[n-1]$. We let
  $\Kc(A)$ be the structure $C$ with universe $[n]$ such that $C_n=C^*$.

  It is easy to see that the mapping $\Kc$ has the desired properties.
\end{proof}

\begin{cor}\label{cor:scdeg}
  For all $c,d\in\NN$ the class of all graphs $G$ such that at most $c$
  vertices of $G$ have degree greater than $d$ admits polynomial time strong
  canonisation.
\end{cor}

The main result of the section is the following lemma:
\begin{lem}[Lifting Lemma]\label{lem:lift}
  Let $\CA,\CC$ be two classes of graphs and $a\in\NN$. Suppose that
  $\CA$ admits polynomial time strong canonisation and that $\CC$
  admits polynomial time computable invariant treelike decompositions
  over $\CA$ of adhesion $a$.

  Then $\CC$ admits polynomial time strong canonisation.
\end{lem}

\begin{proof}
  We shall describe a polynomial time computable a canonisation
  mapping $\Kc$ for the class of all weighted $\lambda$-structures
  with Gaifman graph in $\CC$. As another small piece of terminology
  used in the proof, we say that a tuple $(x_1,\ldots,x_{k})$ is an
  \emph{enumeration} of a finite set $S$ (or \emph{enumerates} $S$) if
  $k=|S|$ and $S=\{x_1,\ldots,x_k\}$.

Let $P_1,\ldots,P_a,Q_1,\ldots,Q_a\not\in\lambda$ be fresh relation symbols,
where $P_i$ and $Q_i$ are $i$-ary for all $i\in[a]$.  Let
$\mu:=\lambda\cup\{P_1,\ldots,P_a,Q_1,\ldots,Q_a\}$. Let
$\mathfrak a$ be a polynomial time computable canonisation mapping on
the class of all weighted $\mu$-structures whose Gaifman graph
is in $\CA$.  Such a mapping $\Ka$ exists by the assumption that $\CA$
admits polynomial time strong canonisation.  Let $\Delta$ be a
polynomial time computable invariant treelike decomposition mapping
for $\CC$ over $\CA$ of adhesion at most $a$.

To explain our canonisation mapping $\Kc$, let us fix a
$\lambda$-structure $C$ with Gaifman graph $G_C\in\CC$. Without loss
of generality we may assume that $G_C$ is connected. Let
$\Delta_{G_C}=(D,\sep,\comp)$. For every $t\in V(D)$ we let
$C_t:=C[\cone(t)]$ and $s_t:=|\sep(t)|$. Note that $0\le s_t\le a$.
By induction on $D$, starting from the
leaves, for every node $t\in V(D)$ and every enumeration $\vec
x$ of $\sep(t)$ we define a 
copy $C^*_{t,\vec x}$ of $C_t$ and a mapping $g_{t,\vec x}:\sep(t)\to
V(C_{t,\vec x}^*)$ with the following properties:
\begin{ealph}[resume]
\item\label{ea:ll13} $V(C_{t,\vec x}^*)$ is an initial segment of the
  positive integers.
\item\label{ea:ll14} There is an isomorphism $f$ from $C_t$ to
  $C_{t,\vec x}^*$ such that $g_{t,\vec x}\subseteq f$.
\end{ealph}
Let $t\in V(D)$, and let $\vec x=(x_1,\ldots,x_{s_t})$ be an
enumeration of $\sep(t)$. Let $u_1,\ldots,u_m$ be the children of $t$ in
$D$. For every $i\in[m]$, let $C_i:=C_{u_i}$ and $n_i:=|\comp(u_i)|$
and $s_i:=s_{u_i}$. Note that $|C_i|=n_i+s_i$.
Recall from Section~\ref{sec:treel-decomp} that $t\doublesmile^{\Delta} u$ means that 
$\sep(t)=\sep(u)$ and $\comp(t)=\comp(u)$ hold in the treelike decomposition $\Delta$. 
 For all $i,j\in[m]$, let
$i\doublesmile j\defiff u_i\doublesmile^{\Delta_{G_C}} u_j$.  For every
$i\in[m]$ and every tuple $\vec y$ that enumerates $\sep(u_i)$, let
$C_{i,\vec y}^*:=C_{u_i,\vec y}^*$ and $g_{i,\vec y}:=g_{u_i,\vec
  y}$. Then $C_{i,\vec y}^*$ and $g_{i,\vec y}$ satisfy
\ref{ea:ll13} and \ref{ea:ll14}. 

  Let $\CY$ be the set of all $\vec y\in \bag(t)^{\le a}$ that enumerate
  $\sep(u_i)$ for some $i\in[m]$. For each $\vec y\in\CY$, let
  $M_{\vec y}$ be the set of all $i\in[m]$ with $\sep(u_i)=\tilde
  y$. Let $\preceq_{\vec y}$ be a linear order on $M_{\vec y}$
  such that for all $i,j\in M_{\vec y}$:
  \begin{ealph}[resume]
  \item\label{ea:ll30}  If $C_{i,\vec y}^*\slex^\lambda C_{j,\vec
      y}^*$ then $i\prec_{\vec y}j$.
  \item\label{ea:ll31}  If $C_{i,\vec y}^*=C_{j,\vec y}^*$ and
    $g_{i,\vec y}(\vec y)\slex g_{j,\vec y}(\vec y)$ then $i\prec_{\vec y}j$.
  \end{ealph}
  Note that conditions \ref{ea:ll30} and \ref{ea:ll31} do not
  determine a linear order on $M_{\vec y}$, since there may be distinct
  $i,j\in M_{\vec y}$ such that $C_{i,\vec y}^*=C_{j,\vec y}^*$ and
  $g_{i,\vec y}(\vec y)=g_{j,\vec y}(\vec y)$. If this is the case, decide arbitrarily
  whether $i\prec_{\vec y}j$ or $j\prec_{\vec y}i$. No matter how we
  decide, the resulting structure $C_{t,\vec x}$ and mapping
  $g_{t,\vec x}$ will be the same.

  Note that for every $\doublesmile$-equivalence class $K$, either
  $K\cap M_{\vec y}=\emptyset$ or $K\subseteq M_{\vec y}$.  Let
  $N_{\vec y}$ be the system of representatives for the
  $\doublesmile$-equivalence classes in $M_{\vec y}$ that contains the
  $\preceq_{\vec y}$-smallest element of each class. Let $i_0$
  be the minimal element of $N_{\vec y}$. We define
  $D_{\vec y}^*$ to be the structure obtained in the following three
  steps:
  \begin{ealph}[resume]
  \item For each $i\in
    N_{\vec y}$, we take a copy $C_{i,\vec y}^{**}$ of $C_{i,\vec
      y}^*$ and shift the universes of these copies in such a way
    that they are disjoint intervals of
    nonnegative integers arranged in the order given by
    $\preceq_{\vec y}$.
  \item We take the union of all the $C_{i,\vec y}^{**}$. Then for each
    $i\in N_{\vec y}$ we identify the tuple $g_{i,\vec y}(\vec y)$
    with the tuple $g_{i_0,\vec y}(\vec y)$.
  \item
    We shrink the universe so that it becomes an initial segment
    of the positive integers.
  \end{ealph}
  Then $D_{\vec y}^*$ is an isomorphic copy of the union $D_{\vec y}$
  of all structures $C_{i,\vec y}$ for $i\in N_{\vec y}$. Let $\vec
  y^*:=g_{i_0,\vec y}(\vec y)$. Observe that $D_{\vec y}^*$ and $\vec
  y^*$ indeed do not depend on the order $\preceq_{\vec y}$, as long as
  it satisfies \ref{ea:ll30} and \ref{ea:ll31}.


  Let $\rho$ be the unique mapping from $\CY$ to an initial segment of
  the positive integers such that $\rho(\vec y)\le\rho(\vec z)$ if and only
  if one of the following two conditions is satisfied:
  \begin{ealph}[resume]
  \item\label{ea:ll32}  $D_{\vec y}^*\slex^\lambda D_{\vec
      z}^*$.
  \item\label{ea:ll33}  $D_{\vec y}^*= D_{\vec
      z}^*$ and $\vec y^*\lex \vec z^*$.
  \end{ealph}
  Then $\rho(\vec y)=\rho(\vec z)$ if and only if $D_{\vec y}^*= D_{\vec z}^*$ and
  $\vec y^*=\vec z^*$.  Let $r:=\max\{\rho(\vec y)\mid \vec y\in\CY\}$.

  We let $A_{t,\vec
    x}$ be the $\mu$-expansion of $C[\bag(t)]$ defined as
  follows:
  \begin{ealph}[resume]
  \item For all $i\in[a]$ we define $P_i(A_{t,\vec x})$ by
    \[
    P_i^{A_{t,\vec x}}(\vec y):=
    \begin{cases}
      1&\text{if }i=s_t\text{ and }\vec y=\vec x,\\
      0&\text{otherwise},
    \end{cases}
    \]
    for all $\vec y\in\bag(t)^i$.
  \item For all $i\in[a]$ we define $Q_i(A_{t,\vec x})$ by
    \[
    Q_i^{A_{t,\vec x}}(\vec y):=
    \begin{cases}
      \rho(\vec y)&\text{if }\vec y\in\CY,\\
      0&\text{otherwise},
    \end{cases}
    \]
    for all $\vec y\in\bag(t)^i$.
  \end{ealph}
  Observe that the Gaifman graph of $A_{t,\vec x}$ is
  $\tau(t)$, because the sets $\tilde x=\sep(t)$ and $\tilde
  y=\sep(u_i)$ for all $\vec y\in\CY$ are cliques in $\tau(t)$. Hence the
  canonisation mapping $\Ka$ is applicable to 
  $A_{t,\vec x}$. Let $A_{t,\vec x}^*:=\Ka(A_{t,\vec x})$.

  We define the structure $C_{t,\vec x}^*$ as follows: We take the
  disjoint union of $A_{t,\vec x}^*$ with copies of the structures
  $D_{\vec y}^*$ for $\vec y\in\CY$. These copies are chosen such
  that their universes are consecutive intervals of positive integers
  and such that $D_{\vec y}^*$ comes before $D_{\vec z}^*$ if
  $\rho(\vec y)<\rho(\vec z)$. Let $C^1$ be the resulting
  structure. Then for each tuple $\vec z\in V(A_{t,\vec x}^*)$ with
  $q:=Q_{|\vec z|}^{A_{t,\vec x}^*}(\vec z)\in[1,r]$ we choose a tuple $\vec y\in\CY$
  with $\rho(\vec y)=q$ and identify the copy of $\vec
  y^*$ in the copy of $D_{\vec y}^*$ in $C^1$ with $\vec z$. Of course there may be
  several $\vec z$ with $Q_{|\vec z|}^{A_{t,\vec x}^*}(\vec z)=q$, say, $\vec
  z_1\slex\vec z_2\slex\ldots\slex\vec z_\ell$. Then there are $\vec
  y_1,\ldots,\vec y_\ell\in\CY$ with $\rho(\vec y_j)=q$. For all
  these, the structures $D_{\vec y_j}^*$ are isomorphic and their
  copies appear consecutively in $C^1$. We identify $\vec
  z_j$ with the copy of $\vec y_j^*$ in the copy of $D_{\vec
    y_j}^*$. After doing all these identifications, we shrink the
  universe of the structure so that it is an initial segment of the
  positive integers. Let $C^2$ be the resulting $\mu$-structure,
  and let $C_{t,\vec x}^*$ be the $\lambda$-restriction of $C^2$. If
  $s_t=0$ (and thus $\vec x$ is the empty tuple) we let $g_{t,\vec x}$
  be the empty mapping. Otherwise, there is a unique tuple $\vec
  x^*=(x_1^*,\ldots,x_{s_t}^*)\in P_{s_t}(A_{t,\vec x}^*)$. We
  define $g_{t,\vec x}$ by letting $g_{t,\vec x}(x_i):=x_i^*$ for all
  all $i\in[s_t]$.

  To define the canonical copy $\Kc(C)$ of $C$, we let $M\subseteq V(D)$
  be the set of all nodes $t$ with $\sep(t)=\emptyset$ and
  $\cone(t)=V(C)$. Such nodes exist by \ref{li:tl5}, because $G_C$ is
  connected. We look at the set $\CM$ of all structures $C_t^*$ for
  nodes $t\in T$. (Here we write $C_t^*$ instead
  of $C_{t,()}^*$, omitting the empty tuple enumerating
  $\sep(t)=\emptyset$.) By \ref{ea:ll14}, all structures in $\CM$ are
  isomorphic to $C$. We let $\Kc(C)$ be the $\lex^\lambda$-minimal
  structure in $\CM$.

  It is important to note that our construction is ``invariant'', that
  is, completely determined by the structure $C$ and the decomposition
  ${\Delta_{G_C}}$. The only freedom we have during the construction is in
  the exact order of the children $u_1,\ldots,u_m$ of $t$, but we have
  already noted that conditions \ref{ea:ll30} and \ref{ea:ll31} restrict
  the choices we can make in such a way that they do no longer matter
  because the resulting structures will be isomorphic.  The invariance
  of $\Delta$ implies that if $f$ is an isomorphism from $C$ to a
  $\lambda$-structure $C'$ then, letting $\Delta_{G_C'}=(D',\sep',\comp')$, there
  is an isomorphism $g$ from $D$ to $D'$ such that for all $t\in V(D)$
  and all enumerations $\vec x$ of $\sep(t)$ the restriction of $f$ to
  $\cone(t)$ is an isomorphism from $C_{t,\vec x}$ to $C_{g(t),f(\vec
    x)}$. By the invariance of our construction, it follows that
  $C_{t,\vec x}^*={C'}_{g(t),f(\vec x)}^*$ and $g_{t,\vec x}(\vec x)=
  g_{g(t),f(\vec x)}(f(\vec x))$. This implies that $\CM=\CM'$, where
  $\CM'$ is defined from $C'$ and $\Delta_{G_C'}$ in the same way as
  $\CM$ is defined from $C$ and $\Delta_{G_C}$, and thus
  $\Kc(C)=\Kc(C')$.

  As the decomposition mapping $\Delta$ is polynomial time computable, the
  canonization mapping $\Kc$ is polynomial time computable as well.
\end{proof}

Now we are ready to prove the main algorithmic result of the paper (which proves Theorem~\ref{th:introiso} in the introduction):
\begin{theo}
  For every graph $H$, the class of graphs excluding $H$ as
  topological subgraph admits polynomial time strong canonisation.
\end{theo}

\begin{proof}
  Choose the constants $a,b,c,d,e$ as in the Invariant Decomposition
  Theorem~\ref{theo:idec}. Let $\CA_1$ be the class of all graphs $G$
  with $|V(G)|\le b$, and let $\CA_2$ be the class of all graphs $G$
  with $K_e\not\preceq G$, and let $\CA_3$ be the class of all graphs
  $G$ such that at most $c$ vertices of $G$ have degree higher than
  $d$. The class $\CA_1$ trivially admits polynomial time strong
  canonisation. The class $\CA_2$ admits polynomial time strong
  canonisation by Lemma~\ref{lem:scmin}. The class $\CA_3$ admits
  polynomial time strong canonisation by Corollary~\ref{cor:scdeg}. As
  $\CA_1$ and $\CA_3$ are polynomial time decidable ($\CA_2$ is as
  well, but we do not need this), it follows that the class
  $\CA:=\CA_1\cup\CA_2\cup\CA_2$ admits polynomial time strong
  canonisation as well. That is, the canonization algorithm for $\CA$
  uses the algorithm for $\CA_1$ if $|V(G)|\le b$; otherwise it uses
  the algorithm for $\CA_3$ if there are at most $c$ vertices having
  degree higher than $d$; otherwise it uses the algorithm for $\CA_2$.

  By the Invariant Decomposition Theorem~\ref{theo:idec} the class of
  $H$-topological subgraph free graphs admits polynomial time
  invariant treelike decompositions over $\CA$ of adhesion $a$. Hence
  by the Lifting Lemma~\ref{lem:lift}, the class of graphs excluding
  $H$ as topological subgraph admits polynomial time strong
  canonisation.
\end{proof}


%% file: gadget.pdf_tex
\begingroup%
  \makeatletter%
  \providecommand\color[2][]{%
    \errmessage{(Inkscape) Color is used for the text in Inkscape, but the package 'color.sty' is not loaded}%
    \renewcommand\color[2][]{}%
  }%
  \providecommand\transparent[1]{%
    \errmessage{(Inkscape) Transparency is used (non-zero) for the text in Inkscape, but the package 'transparent.sty' is not loaded}%
    \renewcommand\transparent[1]{}%
  }%
  \providecommand\rotatebox[2]{#2}%
  \ifx\svgwidth\undefined%
    \setlength{\unitlength}{624.8004924bp}%
    \ifx\svgscale\undefined%
      \relax%
    \else%
      \setlength{\unitlength}{\unitlength * \real{\svgscale}}%
    \fi%
  \else%
    \setlength{\unitlength}{\svgwidth}%
  \fi%
  \global\let\svgwidth\undefined%
  \global\let\svgscale\undefined%
  \makeatother%
  \begin{picture}(1,1.0768316)%
    \put(0,0){\includegraphics[width=\unitlength]{gadget.pdf}}%
    \put(0.51463364,0.6152528){\color[rgb]{0,0,0}\makebox(0,0)[lb]{\smash{$y$}}}%
    \put(0.72132819,0.61342362){\color[rgb]{0,0,0}\makebox(0,0)[lb]{\smash{$x$}}}%
    \put(0.5164628,0.38295005){\color[rgb]{0,0,0}\makebox(0,0)[lb]{\smash{$z$}}}%
    \put(0.34818051,0.58415717){\color[rgb]{0,0,0}\makebox(0,0)[rb]{\smash{$w_1$}}}%
    \put(0.34818051,0.48538275){\color[rgb]{0,0,0}\makebox(0,0)[rb]{\smash{$w_2$}}}%
    \put(0.34964382,0.42684977){\color[rgb]{0,0,0}\makebox(0,0)[rb]{\smash{$w_3$}}}%
    \put(0.34891216,0.37929178){\color[rgb]{0,0,0}\makebox(0,0)[rb]{\smash{$w_4$}}}%
    \put(0.48719631,0.09760186){\color[rgb]{0,0,0}\makebox(0,0)[b]{\smash{$v_1$}}}%
    \put(0.61340804,0.09760186){\color[rgb]{0,0,0}\makebox(0,0)[b]{\smash{$v_2$}}}%
    \put(0.71584074,0.09760186){\color[rgb]{0,0,0}\makebox(0,0)[b]{\smash{$v_3$}}}%
    \put(0.81827344,0.09760186){\color[rgb]{0,0,0}\makebox(0,0)[b]{\smash{$v_4$}}}%
    \put(0.30245162,0.83475142){\color[rgb]{0,0,0}\makebox(0,0)[rb]{\smash{$N_n$}}}%
    \put(0.86948978,0.83475142){\color[rgb]{0,0,0}\makebox(0,0)[lb]{\smash{$N_m$}}}%
    \put(0.18172737,0.65000675){\color[rgb]{1,1,1}\makebox(0,0)[rb]{\smash{$M_n^m$}}}%
    \put(0.20367723,0.64086098){\color[rgb]{0,0,0}\makebox(0,0)[rb]{\smash{$M_n^m$}}}%
  \end{picture}%
\endgroup%

%% file: conclusions.tex
\section{Conclusions}
Our first main result is a structure theorem for graphs with excluded topological subgraphs, stating that they have tree decompositions into torsos with excluded minors and torsos were all but a bounded number of vertices have bounded degree. This is, in some sense, the best one can expect, because all classes of graphs with excluded minors and all classes of bounded degree (even bounded degree up to a bounded number of exceptional vertices) exclude topological subgraphs. Our proof of this theorem is self-contained, and it does not involve any enourmous hidden constants (something we feel is worth mentioning in the context of graph structure theory). The algorithmic version of the theorem, though, does depend on a minor test, which can be carried out in cubic time, but only with astronomical constant factors which make the algorithms completely impractical. It is an interesting open question whether minor tests can be avoided in the algorithms for computing our decompositions. Just like Robertson and Seymour's structure theorem for graphs with excluded minors, we expect our structure theorem to have many algorithmic applications. As a case in point, we show that the \textsc{Partial Dominating
  Set} problem is fixed-parameter tractable on graph classes with excluded topological subgraphs.

Our second main result is a polynomial time isomorphism test for graph classes with excluded topological subgraphs. Such classes form a natural common generalisation of classes of bounded degree and classes with excluded minors, both of which were known to have polynomial isomorphism tests.
To prove this result, we need a generalisation of our structure theorem which gives invariant treelike decompositions instead of tree decompositions. Treelike decompositions were introduced in \cite{gro08} in the context of descriptive complexity theory as a logically definable substitute for tree decompositions. Here we show them to be useful in a purely algorithmic context as well. We develop a new, fairly generic machinery that, starting from the ``local structure lemmas,'', which form the core of our structure theorem, first yields the ``global'' invariant structure theorem and then, by a generic ``lifting lemma'', allows us to lift canonisation algorithms from the torsos of the decomposed graphs to the whole graphs. The same machinery could be applied to even richer classes of graphs, provided it can be proved that they have a ``local structure'' that admits polynomial time canonisation.

Let us remark that it is unlikely that
our isomorphism test could be generalized to all classes of graphs
with bounded expansion, as the isomorphism problem on such a class can be as hard 
as on general graphs. To see this, consider two graphs $G_1$ and $G_2$
on $n$-vertices and let us obtain $G'_1$ and $G'_2$ by subdividing
each edge with $n$ new vertices. Now $G'_1$ and $G'_2$ have bounded
expansion and they are isomorphic if and only if $G_1$ and $G_2$ are.

Let us point out that the exponent of the running time of our
isomorphism test for graphs excluding $H$ as a topological subgraph
depends on the graph $H$. It is an obvious question whether this can
be improved to $f(H)\cdot n^{O(1)}$, i.e., the problem is
fixed-parameter tractable parameterized by the excluded graph
$H$. However, it is a significant open question already for graphs of
maximum degree $k$ if the known $n^{O(k)}$ time algorithms for graph
isomorphism can be improved to $f(k)\cdot n^{O(1)}$ time. On the othe
hand, very recently Lokshtanov et
al.~\cite{DBLP:journals/corr/LokshtanovPPS14} presented an $f(k)\cdot
n^{O(1)}$ time algorithm for graphs of treewidth at most $k$, giving
some hope for fixed-parameter tractability results for more general
classes.
